\documentclass[twocolumn]{aastex62}

\begin{document}

\title{\bf A Reexamination of Phosphorus and Chlorine Depletions in the Diffuse Interstellar Medium\footnote{Based on observations made with the NASA/ESA \emph{Hubble Space Telescope} and the \emph{Far Ultraviolet Spectroscopic Explorer} mission, obtained from the MAST data archive at the Space Telescope Science Institute. STScI is operated by the Association of Universities for Research in Astronomy, Inc., under NASA contract NAS5-26555.}}

\correspondingauthor{Adam M.~Ritchey}

\author{Adam M.~Ritchey}
\affiliation{Eureka Scientific, 2452 Delmer Street, Suite 100, Oakland, CA 96402, USA}
\email{ritchey.astro@gmail.com}

\author{J.~M.~Brown}
\affiliation{Department of Physics and Astronomy, University of Toledo, Toledo, OH 43606, USA}

\author{S.~R.~Federman}
\affiliation{Department of Physics and Astronomy, University of Toledo, Toledo, OH 43606, USA}

\author{Paule Sonnentrucker}
\affiliation{European Space Agency Office, Space Telescope Science Institute, 3700 San Martin Dr., Baltimore, MD 21218, USA}

\begin{abstract}
We present a comprehensive examination of interstellar P and Cl abundances based on an analysis of archival spectra acquired with the Space Telescope Imaging Spectrograph of the Hubble Space Telescope and the Far Ultraviolet Spectroscopic Explorer. Column densities of P~{\sc ii}, Cl~{\sc i}, and Cl~{\sc ii} are determined for a combined sample of 107 sight lines probing diffuse atomic and molecular gas in the local Galactic interstellar medium (ISM). We reevaluate the nearly linear relationship between the column densities of Cl~{\sc i} and H$_2$, which arises from the rapid conversion of Cl$^+$ to Cl$^0$ in regions where H$_2$ is abundant. Using the observed total gas-phase P and Cl abundances, we derive depletion parameters for these elements, adopting the methodology of Jenkins. We find that both P and Cl are essentially undepleted along sight lines showing the lowest overall depletions. Increasingly severe depletions of P are seen along molecule-rich sight lines. In contrast, gas-phase Cl abundances show no systematic variation with molecular hydrogen fraction. However, enhanced Cl (and P) depletion rates are found for a subset of sight lines showing elevated levels of Cl ionization. An analysis of neutral chlorine fractions yields estimates for the amount of atomic hydrogen associated with the H$_2$-bearing gas in each direction. These results indicate that the molecular fraction in the H$_2$-bearing gas is at least 10\% for all sight lines with $\log N({\rm H}_2)\gtrsim18$ and that the gas is essentially fully molecular at $\log N({\rm H}_2)\approx21$.
\end{abstract}


\section{INTRODUCTION\label{sec:intro}}
Dust plays an important role in the evolution of galaxies due to the impact it has on the physics and chemistry of the interstellar medium (ISM). However, questions remain regarding the dominant dust production mechanisms, both in the present day Milky Way and in galaxies at high redshift. Efficient dust production occurs in the extended outer atmospheres of evolved stars \citep[e.g.,][]{g16} and in the cooling ejecta of supernovae \citep[SNe; e.g.,][]{sc15}. However, the timescale for dust production by stars is longer than the timescale for destruction of the grains by SN shocks \citep{ds80,j94,s15}. Moreover, the large observed depletions of refractory elements in cold, diffuse clouds \citep[where $\sim$95\% of the Mg and Si atoms and $\sim$99\% of the Fe atoms are locked up in grains; e.g.,][]{j09} are inconsistent with the levels of condensation expected for material returned to the ISM by evolved stars and SNe \citep[e.g.,][]{d16}. These considerations have generally been interpreted as indications that dust grains must grow through the accretion of refractory elements in cold diffuse interstellar clouds.

Direct constraints on the growth of dust grains in the diffuse ISM come from detailed knowledge of the depletions of different elements from the gas-phase and how those depletions change with changing environmental conditions. \citet{j09} compiled information on interstellar depletions for 17 elements along 243 sight lines sampling diffuse atomic and molecular gas in the solar vicinity. \citet{j09} developed a unified framework for examining gas-phase element depletions predicated on the empirical observation that, while different elements exhibit different degrees of depletion, the depletions of most elements tend to increase in a systematic way as the overall strength of depletions increases from one sight line to the next. \citet{r18} extended that framework to include several rare neutron-capture elements (and the light element B) yielding a more complete picture of gas-phase depletions for elements with low-to-moderate condensation temperatures.

To quantify the overall strength of the depletions seen in a given direction, \citet{j09} defined the sight line depletion strength factor $F_*$. Sight lines showing strong depletions (such as those that characterize relatively dense and/or cold H$_2$-rich gas in the Galactic disk where grain growth occurs via cold accretion) have depletion factors near $F_*=1$, while sight lines showing only very modest depletions (characteristic of warm and/or low-density H$_2$-poor gas where substantial grain growth has not yet occurred or where the grains have been destroyed or disrupted by shocks) have values of $F_*$ closer to zero. The observed depletions at $F_*=0$ are probably the best representation that we have of the composition of grain cores that emerge from evolved low and high-mass stars and core-collapse SNe.

Most elements with low-to-moderate condensation temperatures ($T_C\lesssim800$~K) have depletions at $F_*=0$ that are consistent with zero \citep[see Figure~21 in][]{r18}. For elements with higher condensation temperatures, a fairly well-defined trend emerges of increasing depletion with increasing $T_C$, particularly for Ga, Ge, Cu, Mn, Cr, Fe, Ni, and Ti. \citet{r18} suggest that this trend constitutes a dust condensation sequence and that the temperature where the sequence seems to terminate ($\sim$800~K) is related to the dust formation temperature. This idea is corroborated by observations of dust shells around asymptotic giant branch (AGB) stars, which are found to have temperatures in the range 800 to 1100~K \citep{g13}. However, while the trend of increasing depletion with increasing $T_C$ seems convincing for the elements mentioned above, there are also glaring exceptions to the trend for the elements Si and Mg but especially for Cl, As, and P. These latter three elements appear to have gas-phase abundances at $F_*=0$ that are significantly higher than the corresponding solar values, which seems particularly unusual since the elements have moderate-to-high condensation temperatures.

One possible factor that could be influencing the unusual Cl and P depletions is that the column density measurements for the dominant singly-ionized species may not be very accurate. All of the Cl~{\sc ii} measurements (and over half of the P~{\sc ii} results) compiled by \citet{j09} are from observations obtained by the Copernicus satellite \citep[e.g.,][]{j86}. More precise measurements of interstellar column densities may be derived from spectra acquired using the higher resolution and/or higher sensitivity instruments of the Hubble Space Telescope (HST) and Far Ultraviolet Spectroscopic Explorer (FUSE).

An even more important factor in the case of P depletions is that the transition oscillator strengths ($f$-values) used by \citet{j09} and others to derive P~{\sc ii} column densities are in need of revision. \citet{j09} adopted theoretical P~{\sc ii} $f$-values originally obtained by \citet{h88}. However, more precise experimentally-determined $f$-values for the important $\lambda1152$ and $\lambda1301$ transitions are now available \citep{f07,b18}. These new experimental results indicate that the P~{\sc ii} column densities obtained from the $\lambda1152$ and $\lambda1301$ transitions should be revised downward by 0.045 dex and 0.188 dex, respectively, in agreement with recent theoretical calculations \citep{t03,ff06}. Furthermore, these more recent theoretical efforts suggest that a downward revision of $\sim$0.4 dex is required for column densities derived using the older \citet{h88} value for P~{\sc ii}~$\lambda1532$ \citep[e.g.,][]{l05,j09}.

A major contributing factor to the unusual Cl depletions reported by \citet{j09} is that he considered only Cl~{\sc ii} column densities when deriving total Cl abundances, ignoring any contribution from Cl~{\sc i}. However, because Cl$^+$ reacts rapidly with H$_2$ to form HCl$^+$, which leads (eventually) to Cl$^0$ and H$^0$, Cl will be predominantly neutral in regions where H$_2$ is abundant \citep{j74,jy78}. Indeed, there are a number of sight lines where previous studies indicate that $N({\rm Cl~\textsc{i}})>N({\rm Cl~\textsc{ii}})$ \citep[e.g.,][]{hb84,j86,m12} and these sight lines tend to have high molecular hydrogen fractions. Since these same sight lines will likely have high values of $F_*$ as well, the depletion trend for Cl calculated by \citet{j09} is too steep and the extrapolated value of [Cl/H] at $F_*=0$ is too high.

In this investigation, we seek to present a definitive analysis of interstellar P and Cl depletions based on high-quality archival HST and FUSE data. The selection of sight lines and the steps involved in processing the archival data are described in Section~\ref{sec:observations}. The procedures used to obtain column densities of P~{\sc ii}, Cl~{\sc i}, and Cl~{\sc ii} for the sight lines in our survey are explained in Section~\ref{sec:results}. In Section~\ref{subsec:correlations}, we reexamine the relationships that exist between various Cl and H species. In Section~\ref{subsec:depletions}, the column densities derived in Section~\ref{sec:results} are used to evaluate the depletion trends for P and Cl, with some additional insights discussed in Section~\ref{subsec:depletions2}. In Section~\ref{subsec:fractions}, we use the observed neutral chlorine fractions to estimate the amount of atomic hydrogen associated with the molecular gas in each direction. The implications of our results are discussed in Section~\ref{sec:discussion}. Our summary and conclusions are presented in Section~\ref{sec:conclusions}. Two appendices compile a set of chlorine measurements from Copernicus observations and present an analysis of fluorine depletions from the literature.

\section{OBSERVATIONS AND DATA PROCESSING\label{sec:observations}}
Since the ionization potential of P~{\sc i} (10.5 eV) is below that of neutral hydrogen (13.6 eV), most P in neutral, diffuse clouds is singly ionized. The same would be true of Cl, since its first ionization potential is 13.0 eV, except that Cl$^+$ reacts exothermically with H$_2$, which leads to the conversion of Cl$^+$ to Cl$^0$ in regions where H$_2$ is optically thick \citep[e.g.,][]{nw09}. Thus, while P~{\sc ii} column densities are sufficient for studies of interstellar P abundances, column densities of both Cl~{\sc i} and Cl~{\sc ii} are required for an evaluation of total interstellar Cl abundances.

Most previous studies of interstellar P have relied on either the P~{\sc ii} $\lambda1152$ transition or the $\lambda1301$ transition. However, the former is often strongly saturated and the latter may be blended with the wing of the nearby strong O~{\sc i} $\lambda1302$ absorption feature. Fewer studies make use of the weak P~{\sc ii} $\lambda1532$ absorption line, which is unblended and typically unaffected by saturation, yet still strong enough to yield reliable measurements for most sight lines. Similarly, while the Cl~{\sc i} $\lambda1347$ line is almost always present in stellar spectra probing diffuse molecular gas, this feature is very often affected by saturation in the line core, making reliable column density determinations difficult. The nearby Cl~{\sc i} line at 1379.5~\AA{} is much weaker and yet (again) is often strong enough to be reliably measured.

Since the goal of our survey is to accurately deduce the depletion characteristics of Cl and P, minimizing the associated uncertainties, we use measurements of the P~{\sc ii} $\lambda1532$ and Cl~{\sc i} $\lambda1379$ absorption features, from high or medium resolution STIS spectra, wherever possible. We include measurements of the P~{\sc ii} $\lambda1301$ line only in cases where there is no apparent blending with O~{\sc i} $\lambda1302$. The strong Cl~{\sc i} $\lambda1347$ line is included in our survey. However, in most cases, a weaker Cl~{\sc i} line (typically $\lambda1379$) is used to help constrain the total column density. For some sight lines, we use observations of the Cl~{\sc i} $\lambda1004$, $\lambda1094$, or $\lambda1097$ transition, available from FUSE data, to constrain the total Cl~{\sc i} column density (see Section~\ref{subsec:chlorine}). The column density of Cl~{\sc ii} is derived from observations of the Cl~{\sc ii} line at 1071.0~\AA{}, available from FUSE spectra.

\startlongtable
\begin{deluxetable*}{lcccccccc}
\tablecolumns{9}
\tabletypesize{\small}
\tablecaption{Stellar Data for the Combined Sample\label{tab:sample}}
\tablehead{ \colhead{Star} & \colhead{Name} & \colhead{Sp.~Type} & \colhead{$V$} & \colhead{$E(\bv)$} & \colhead{$l$} & \colhead{$b$} & \colhead{$d$\tablenotemark{a}} & \colhead{$z$} \\
\colhead{} & \colhead{} & \colhead{} & \colhead{(mag)} & \colhead{(mag)} & \colhead{(deg)} & \colhead{(deg)} & \colhead{(kpc)} & \colhead{(kpc)} }
\startdata
HD~108        & \ldots & O8fp            & 7.40   & 0.43   & 117.93   & $+1.25$   & $1.93^{+0.08}_{-0.09}$   & $+0.042$ \\
HD~1383       & \ldots & B1II            & 7.63   & 0.47   & 119.02   & $-0.89$   & $2.50^{+0.18}_{-0.13}$   & $-0.039$ \\
HD~3827       & \ldots & B0.7V          & 7.95   & 0.02   & 120.79   &$-23.23$   & $1.67^{+0.14}_{-0.15}$   & $-0.658$ \\
HD~12323      & \ldots & ON9.2V             & 8.92   & 0.23   & 132.91   & $-5.87$   & $2.44^{+0.23}_{-0.22}$   & $-0.250$ \\
HD~13268      & \ldots & ON8.5IIIn           & 8.18   & 0.36   & 133.96   & $-4.99$   & $1.77^{+0.09}_{-0.07}$   & $-0.154$ \\
HD~13745      & V354~Per & O9.7IIn         & 7.90   & 0.46   & 134.58   & $-4.96$   & $2.28^{+0.15}_{-0.09}$   & $-0.197$ \\
HD~14434      & \ldots & O5.5Vnfp        & 8.49   & 0.48   & 135.08   & $-3.82$   & $2.24^{+0.10}_{-0.09}$   & $-0.150$ \\
HD~15137      & \ldots & O9.5II-IIIn     & 7.86   & 0.35   & 137.46   & $-7.58$   & $2.05^{+0.17}_{-0.13}$   & $-0.271$ \\
HD~23478      & \ldots & B3IV            & 6.67   & 0.27   & 160.76   &$-17.42$   & $0.249^{+0.004}_{-0.003}$   & $-0.074$ \\
HD~24190      & \ldots & B2Vn            & 7.45   & 0.30   & 160.39   &$-15.18$   & $0.375^{+0.005}_{-0.005}$   & $-0.098$ \\
HD~24534      & X~Per & O9.5III         & 6.72   & 0.59   & 163.08   &$-17.14$   & $0.596^{+0.017}_{-0.014}$   & $-0.176$ \\
HD~25443      & \ldots & B0.5III         & 6.77   & 0.54   & 143.68   & $+7.35$   & $1.13^{+0.07}_{-0.06}$   & $+0.145$ \\
HD~37903      & \ldots & B1.5V         & 7.83   & 0.35   & 206.85   &$-16.54$   & $0.394^{+0.003}_{-0.004}$   & $-0.112$ \\
HD~41161      & \ldots & O8Vn            & 6.76   & 0.21   & 164.97   &$+12.89$   & $1.43^{+0.12}_{-0.10}$   & $+0.319$ \\
HD~46223      & \ldots & O4Vf            & 7.28   & 0.54   & 206.44   & $-2.07$   & $1.39^{+0.05}_{-0.06}$   & $-0.050$ \\
HD~52266      & \ldots & O9.5IIIn           & 7.23   & 0.29   & 219.13   & $-0.68$   & $1.35^{+0.06}_{-0.07}$   & $-0.016$ \\
HD~53975      & \ldots & O7.5Vz           & 6.50   & 0.21   & 225.68   & $-2.32$   & $1.12^{+0.07}_{-0.06}$   & $-0.045$ \\
HD~63005      & \ldots & O7Vf            & 9.13   & 0.27   & 242.47   & $-0.93$   & $3.72^{+0.60}_{-0.50}$   & $-0.060$ \\
HD~66788      & \ldots & O8V             & 9.83   & 0.22   & 245.43   & $+2.05$   & $3.31^{+0.32}_{-0.30}$   & $+0.119$ \\
HD~72754      & FY~Vel & B2I:pe          & 8.88   & 0.36   & 266.83   & $-5.82$   & $1.62^{+0.05}_{-0.05}$   & $-0.165$ \\
HD~73882      & NX~Vel & O8.5IV          & 7.19   & 0.70   & 260.18   & $+0.64$   & $0.737^{+0.031}_{-0.030}$   & $+0.008$ \\
HD~75309      & \ldots & B1IIp           & 7.84   & 0.29   & 265.86   & $-1.90$   & $1.82^{+0.12}_{-0.13}$   & $-0.060$ \\
HD~79186      & GX Vel & B5Ia            & 5.00   & 0.40   & 267.36   & $+2.25$   & $1.81^{+0.34}_{-0.23}$   & $+0.071$ \\
HD~88115      & \ldots & B1.5Iin         & 9.36   & 0.16   & 285.32   & $-5.53$   & $2.53^{+0.16}_{-0.17}$   & $-0.243$ \\
HD~89137      & \ldots & ON9.7IIn       & 7.97   & 0.27   & 279.69   & $+4.45$   & $2.44^{+0.20}_{-0.14}$   & $+0.189$ \\
HD~90087      & \ldots & O9.2III          & 8.92   & 0.28   & 285.16   & $-2.13$   & $2.19^{+0.11}_{-0.12}$   & $-0.082$ \\
HD~91597      & \ldots & B1IIIne         & 9.92   & 0.30   & 286.86   & $-2.37$   & $3.90^{+0.26}_{-0.24}$   & $-0.161$ \\
HD~91651      & \ldots & ON9.5IIIn           & 9.52   & 0.28   & 286.55   & $-1.72$   & $1.85^{+0.07}_{-0.07}$   & $-0.055$ \\
HD~91824      & \ldots & O7Vfz             & 8.14   & 0.24   & 285.70   & $+0.07$   & $1.83^{+0.08}_{-0.08}$   & $+0.002$ \\
HD~91983      & \ldots & B1III           & 8.55   & 0.29   & 285.88   & $+0.05$   & $2.40^{+0.15}_{-0.14}$   & $+0.002$ \\
HD~92554      & \ldots & O9.5IIn         &10.15   & 0.39   & 287.60   & $-2.02$   & $4.04^{+0.31}_{-0.28}$   & $-0.142$ \\
HD~93129      & \ldots & O2If*            & 6.90   & 0.48   & 287.41   & $-0.57$   & $2.41^{+0.10}_{-0.10}$   & $-0.024$ \\
HD~93205      & V560~Car & O3.5V            & 7.75   & 0.38   & 287.57   & $-0.71$   & $2.25^{+0.12}_{-0.11}$   & $-0.028$ \\
HD~93222      & \ldots & O7IIIf          & 8.10   & 0.36   & 287.74   & $-1.02$   & $2.41^{+0.14}_{-0.15}$   & $-0.043$ \\
HD~93843      & \ldots & O5IIIf          & 7.33   & 0.27   & 288.24   & $-0.90$   & $2.37^{+0.19}_{-0.14}$   & $-0.037$ \\
HD~94493      & \ldots & B1Ib            & 7.59   & 0.23   & 289.01   & $-1.18$   & $2.15^{+0.14}_{-0.12}$   & $-0.044$ \\
HD~97175      & \ldots & B0.5III         & 9.20   & 0.18   & 294.53   & $-9.17$   & $2.48^{+0.18}_{-0.14}$   & $-0.396$ \\
HD~99857      & \ldots & B0.5Ib          & 7.49   & 0.35   & 294.78   & $-4.94$   & $1.80^{+0.08}_{-0.07}$   & $-0.155$ \\
HD~99890      & \ldots & B0IIIn          & 9.26   & 0.24   & 291.75   & $+4.43$   & $2.53^{+0.18}_{-0.17}$   & $+0.195$ \\
HD~100199     & \ldots & B0IIIn         & 8.17   & 0.30   & 293.94   & $-1.49$   & $1.73^{+0.14}_{-0.11}$   & $-0.045$ \\
HD~101190     & \ldots & O6IVf            & 7.33   & 0.37   & 294.78   & $-1.49$   & $2.36^{+0.13}_{-0.12}$   & $-0.061$ \\
HD~103779     & \ldots & B0.5Iab         & 7.22   & 0.21   & 296.85   & $-1.02$   & $2.02^{+0.10}_{-0.08}$   & $-0.036$ \\
HD~104705     & DF~Cru & B0Ib            & 9.11   & 0.23   & 297.45   & $-0.34$   & $1.94^{+0.16}_{-0.15}$   & $-0.011$ \\
HD~108639     & \ldots & B0.2III         & 8.57   & 0.37   & 300.22   & $+1.95$   & $1.98^{+0.10}_{-0.11}$   & $+0.067$ \\
HD~109399     & \ldots & B0.7II          & 7.67   & 0.21   & 301.71   & $-9.88$   & $2.33^{+0.13}_{-0.14}$   & $-0.400$ \\
HD~114886     & \ldots & O9III          & 6.89   & 0.40   & 305.52   & $-0.83$   & $1.83^{+0.95}_{-0.78}$   & $-0.026$ \\
HD~115455     & \ldots & O8IIIf         & 7.97   & 0.47   & 306.06   & $+0.22$   & $1.80^{+0.09}_{-0.07}$   & $+0.007$ \\
HD~116852     & \ldots & O8.5II-IIIf           & 8.47   & 0.21   & 304.88   &$-16.13$   & $3.42^{+0.39}_{-0.32}$   & $-0.949$ \\
HD~121968     & \ldots & B1V             &10.26   & 0.07   & 333.97   &$+55.84$   & $3.94^{+0.91}_{-0.63}$   & $+3.26$ \\
HD~122879     & \ldots & B0Ia            & 6.50   & 0.36   & 312.26   & $+1.79$   & $2.22^{+0.16}_{-0.13}$   & $+0.069$ \\
HD~124314     & \ldots & O6IVnf           & 6.64   & 0.53   & 312.67   & $-0.42$   & $1.61^{+0.11}_{-0.10}$   & $-0.012$ \\
HD~137595     & \ldots & B3Vn            & 7.49   & 0.25   & 336.72   &$+18.86$   & $0.751^{+0.021}_{-0.022}$   & $+0.243$ \\
HD~144965     & \ldots & B2Vne           & 7.11   & 0.35   & 339.04   & $+8.42$   & $0.258^{+0.002}_{-0.002}$   & $+0.038$ \\
HD~147683     & V760~Sco & B4V+B4V         & 7.05   & 0.39   & 344.86   &$+10.09$   & $0.290^{+0.002}_{-0.002}$   & $+0.051$ \\
HD~147888     & $\rho$~Oph~D & B3V             & 6.74   & 0.47   & 353.65   &$+17.71$   & $0.124^{+0.006}_{-0.006}$   & $+0.038$ \\
HD~147933     & $\rho$~Oph~A & B2V             & 5.05   & 0.45   & 353.69   &$+17.69$   & $0.137^{+0.003}_{-0.003}$   & $+0.042$ \\
HD~148422     & \ldots & B1Ia            & 8.65   & 0.29   & 329.92   & $-5.60$   & $4.18^{+0.52}_{-0.31}$   & $-0.408$ \\
HD~148937     & \ldots & O6f?p            & 6.71   & 0.65   & 336.37   & $-0.22$   & $1.15^{+0.03}_{-0.03}$   & $-0.004$ \\
HD~151805     & \ldots & B1Ib            & 8.86   & 0.43   & 343.20   & $+1.59$   & $1.51^{+0.07}_{-0.06}$   & $+0.042$ \\
HD~152590     & V1297~Sco & O7.5Vz             & 9.29   & 0.48   & 344.84   & $+1.83$   & $1.68^{+0.08}_{-0.06}$   & $+0.053$ \\
HD~156359     & \ldots & B0Ia            & 9.72   & 0.14   & 328.68   &$-14.52$   & $8.98^{+3.14}_{-2.11}$   & $-2.25$ \\
HD~157857     & \ldots & O6.5IIf        & 7.78   & 0.43   &  12.97   &$+13.31$   & $2.22^{+0.20}_{-0.13}$   & $+0.511$ \\
HD~163522     & \ldots & B1Ia            & 8.43   & 0.19   & 349.57   & $-9.09$   & $4.01^{+0.56}_{-0.47}$   & $-0.633$ \\
HD~165246     & \ldots & O8Vn            & 7.60   & 0.38   &   6.40   & $-1.56$   & $1.19^{+0.04}_{-0.05}$   & $-0.032$ \\
HD~165955     & \ldots & B3Vn            & 9.59   & 0.15   & 357.41   & $-7.43$   & $1.47^{+0.12}_{-0.09}$   & $-0.190$ \\
HD~167402     & \ldots & B0Ib     & 9.03   & 0.21   &   2.26   & $-6.39$   & $4.94^{+0.83}_{-0.73}$   & $-0.550$ \\
HD~168076     & \ldots & O4IIIf     & 8.25   & 0.78   &  16.94   & $+0.84$   & $1.65^{+1.51}_{-0.67}$   & $+0.024$ \\
HD~168941     & \ldots & O9.5IVp         & 9.37   & 0.24   &   5.82   & $-6.31$   & $4.00^{+0.60}_{-0.53}$   & $-0.440$ \\
HD~170740     & \ldots & B2IV-V          & 5.72   & 0.48   &  21.06   & $-0.53$   & $0.225^{+0.005}_{-0.005}$   & $-0.002$ \\
HD~177989     & \ldots & B0III           & 9.34   & 0.23   &  17.81   &$-11.88$   & $2.41^{+0.20}_{-0.19}$   & $-0.496$ \\
HD~178487     & \ldots & B0Ib          & 8.69   & 0.35   &  25.78   & $-8.56$   & $2.85^{+0.36}_{-0.25}$   & $-0.425$ \\
HD~179407     & \ldots & B0.5Ib          & 9.44   & 0.28   &  24.02   &$-10.40$   & $4.44^{+0.69}_{-0.47}$   & $-0.802$ \\
HD~185418     & \ldots & B0.5V           & 7.49   & 0.50   &  53.60   & $-2.17$   & $0.692^{+0.010}_{-0.009}$   & $-0.026$ \\
HD~190918     & V1676~Cyg & O9.7Iab+WN4     & 6.75   & 0.45   &  72.65   & $+2.07$   & $1.78^{+0.07}_{-0.07}$   & $+0.064$ \\
HD~191877     & \ldots & B1Ib            & 6.27   & 0.21   &  61.57   & $-6.45$   & $1.73^{+0.11}_{-0.13}$   & $-0.194$ \\
HD~192035     & RX~Cyg & B0III-IVn       & 8.22   & 0.34   &  83.33   & $+7.76$   & $1.65^{+0.06}_{-0.06}$   & $+0.223$ \\
HD~192639     & \ldots & O7.5Iab           & 7.11   & 0.66   & 74.90    & $+1.48$   & $1.81^{+0.07}_{-0.06}$   & $+0.047$ \\
HD~195455     & \ldots & B0.5III         & 9.20   & 0.10   &  20.27   &$-32.14$   & $2.35^{+0.35}_{-0.24}$   & $-1.25$ \\
HD~195965     & \ldots & B0V             & 6.97   & 0.25   &  85.71   & $+5.00$   & $0.790^{+0.023}_{-0.025}$   & $+0.069$ \\
HD~198478     & 55~Cyg & B3Ia            & 4.86   & 0.57   &  85.75   & $+1.49$   & $1.84^{+0.35}_{-0.22}$   & $+0.048$ \\
HD~198781     & \ldots & B0.5V           & 6.45   & 0.35   &  99.94   &$+12.61$   & $0.915^{+0.022}_{-0.027}$   & $+0.200$ \\
HD~201345     & \ldots & ON9.2IV             & 7.76   & 0.15   &  78.44   & $-9.54$   & $1.83^{+0.15}_{-0.11}$   & $-0.303$ \\
HD~202347     & \ldots & B1.5V           & 7.50   & 0.17   &  88.22   & $-2.08$   & $0.764^{+0.023}_{-0.019}$   & $-0.028$ \\
HD~203374     & \ldots & B0IVpe           & 6.67   & 0.53   & 100.51   & $+8.62$   & $2.04^{+1.42}_{-0.76}$   & $+0.306$ \\
HD~203938     & \ldots & B0.5IV           & 7.08   & 0.74   &  90.56   & $-2.23$   & $2.96^{+2.26}_{-1.35}$   & $-0.115$ \\
HD~206267     & \ldots & O6Vf           & 5.62   & 0.53   &  99.29   & $+3.74$   & $0.790^{+0.172}_{-0.112}$   & $+0.052$ \\
HD~206773     & \ldots & B0Vpe        & 6.87   & 0.45   &  99.80   & $+3.62$   & $0.888^{+0.016}_{-0.014}$   & $+0.056$ \\
HD~207198     & \ldots & O8.5II       & 5.94   & 0.62   & 103.14   & $+6.99$   & $0.978^{+0.034}_{-0.027}$   & $+0.119$ \\
HD~207308     & \ldots & B0.5V   & 7.49   & 0.53   & 103.11   & $+6.82$   & $0.906^{+0.017}_{-0.013}$   & $+0.108$ \\
HD~207538     & \ldots & O9.7IV           & 7.30   & 0.64   & 101.60   & $+4.67$   & $0.830^{+0.013}_{-0.013}$   & $+0.068$ \\
HD~208440     & \ldots & B1V             & 7.91   & 0.28   & 104.03   & $+6.44$   & $0.877^{+0.019}_{-0.018}$   & $+0.098$ \\
HD~209339     & \ldots & O9.7IV            & 8.51   & 0.36   & 104.58   & $+5.87$   & $0.936^{+0.028}_{-0.024}$   & $+0.096$ \\
HD~210809     & \ldots & O9Iab           & 7.56   & 0.31   &  99.85   & $-3.13$   & $3.66^{+0.52}_{-0.34}$   & $-0.200$ \\
HD~210839     & $\lambda$~Cep & O6.5Infp          & 5.05   & 0.57   & 103.83   & $+2.61$   & $0.833^{+0.066}_{-0.049}$   & $+0.038$ \\
HD~212791     & V408~Lac & B3ne            & 8.02   & 0.17   & 101.64   & $-4.30$   & $0.893^{+0.019}_{-0.016}$   & $-0.067$ \\
HD~218915     & \ldots & O9.2Iab        & 7.20   & 0.30   & 108.06   & $-6.89$   & $2.97^{+0.33}_{-0.25}$   & $-0.357$ \\
HD~219188     & \ldots & B0.5IIIn        & 7.06   & 0.13   &  83.03   &$-50.17$   & $2.10^{+0.25}_{-0.27}$   & $-1.61$ \\
HD~220057     & \ldots & B3IV            & 6.94   & 0.23   & 112.13   & $+0.21$   & $0.385^{+0.004}_{-0.004}$   & $+0.001$ \\
HD~224151     & V373~Cas & B0.5II-III      & 6.00   & 0.44   & 115.44   & $-4.64$   & $1.89^{+0.13}_{-0.10}$   & $-0.153$ \\
HDE~232522    & \ldots & B1II            & 8.70   & 0.27   & 130.70   & $-6.71$   & $3.46^{+0.41}_{-0.44}$   & $-0.404$ \\
HDE~303308    & \ldots & O4.5Vfc            & 8.17   & 0.45   & 287.59   & $-0.61$   & $2.17^{+0.09}_{-0.10}$   & $-0.023$ \\
HDE~308813    & \ldots & O9.7IVn           & 9.73   & 0.34   & 294.79   & $-1.61$   & $2.43^{+0.11}_{-0.09}$   & $-0.068$ \\
BD$+$35~4258  & \ldots & B0.5Vn          & 9.46   & 0.25   &  77.19   & $-4.74$   & $2.21^{+0.14}_{-0.11}$   & $-0.182$ \\
BD$+$53~2820  & \ldots & B0IVn          & 9.96   & 0.29   & 101.24   & $-1.69$   & $3.57^{+0.25}_{-0.17}$   & $-0.105$ \\
CPD$-$59~2603 & V572~Car & O7Vnz            & 8.81   & 0.46   & 287.59   & $-0.69$   & $2.63^{+0.16}_{-0.14}$   & $-0.032$ \\
CPD$-$59~4552 & \ldots & B1III           & 8.24   & 0.38   & 303.22   & $+2.47$   & $2.04^{+0.07}_{-0.06}$   & $+0.088$ \\
CPD$-$69~1743 & \ldots & B0.5IIIn        & 9.64   & 0.30   & 303.71   & $-7.35$   & $3.20^{+0.19}_{-0.16}$   & $-0.410$ \\
\enddata
\tablenotetext{a}{Distances are based on Gaia EDR3 parallax measurements \citep{bj21}.}
\end{deluxetable*}

\subsection{Processing of the STIS Data\label{subsec:stis}}
Archival STIS spectra acquired using either the medium-resolution echelle grating (E140M) or the high-resolution grating (E140H) at central wavelength settings that cover the relevant P~{\sc ii} and/or Cl~{\sc i} lines were obtained from the Mikulski Archive for Space Telescopes (MAST). We focus only on sight lines with reliable column densities of H~{\sc i} and H$_2$ published in the literature \citep[e.g.,][]{j19}. An additional requirement for the Cl analysis is that each sight line must have FUSE observations available.\footnote{An exception is made in the case of HD~147933 ($\rho$~Oph~A), which was not observed with FUSE but does have high-resolution STIS spectra covering the Cl~{\sc i}~$\lambda1347$ transition. \citet{m12} obtained Cl~{\sc i} and Cl~{\sc ii} column densities for this sight line using Copernicus observations. However, the total Cl abundance derived from their analysis is unusually low. We therefore reexamine the Copernicus data toward $\rho$~Oph~A, in conjunction with the newer high-resolution STIS spectra, in order to re-evaluate the Cl~{\sc i} and Cl~{\sc ii} column densities in this direction.} The final P and Cl samples differ from one another slightly due to the constraints placed on measuring the various absorption features (as discussed above). Basic information regarding the background stars associated with the 107 sight lines that constitute the final combined sample is presented in Table~\ref{tab:sample}. The wavelengths and adopted oscillator strengths for the interstellar lines of interest to our survey are provided in Table~\ref{tab:lines}.

After downloading the pipeline-processed archival files from MAST, the STIS data were further reduced in a manner analogous to that described in \citet{r11,r18}. Multiple exposures of a given target acquired with the same echelle grating were co-added to increase the signal-to-noise (S/N) ratio in the final spectrum. When a feature of interest appeared in adjacent echelle orders with sufficient continua on both sides of the line, the overlapping portions of the two orders were averaged together. Portions of the co-added spectrum surrounding interstellar lines of interest (typically 2~\AA{} wide) were cut from the data. These smaller spectral segments were then normalized by fitting the continuum regions with a low-order polynomial function.

\begin{deluxetable}{lccc}
\tablecolumns{4}
\tabletypesize{\small}
\tablecaption{Atomic Data\label{tab:lines}}
\tablehead{ \colhead{Species} & \colhead{$\lambda_0$} & \colhead{$f$} & \colhead{Ref.} \\
\colhead{} & \colhead{(\AA{})} & \colhead{} & \colhead{} }
\startdata
P~{\sc ii}   & 1152.818 & 0.272\phn\phn & 1 \\
  & 1301.874 & 0.0196\phn & 2 \\
  & 1532.533 & 0.00737 & 3 \\
Cl~{\sc i}   & 1004.678 & 0.0473\phn & 4 \\
  & 1094.769 & 0.0385\phn & 4 \\
  & 1097.369 & 0.0088\phn & 5 \\
  & 1347.240 & 0.153\phn\phn & 5 \\
  & 1379.528 & 0.00269 & 6 \\
Cl~{\sc ii}   & 1071.036 & 0.0142\phn & 7 \\
\enddata
\tablerefs{(1) \citet{f07}, (2) \citet{b18}, (3) this work, (4) \citet{a19}, (5) \citet{s93}, (6) \citet{oh13}, (7) \citet{s05}.}
\end{deluxetable}

In most cases, the process of normalizing the continuum was straightforward. The P~{\sc ii} $\lambda1532$ line and the Cl~{\sc i} $\lambda1347$ and $\lambda1379$ lines are relatively isolated and are easily distinguished from the underlying stellar spectra. However, as already mentioned, the P~{\sc ii} $\lambda1301$ line is positioned very close to the strong O~{\sc i} $\lambda1302$ resonance feature. (The difference in wavelength between the two transitions amounts to 68~km~s$^{-1}$.) Thus, great care had to be taken in choosing an appropriate normalization for the P~{\sc ii} $\lambda1301$ line. Any cases where the P~{\sc ii} line was inextricably blended with O~{\sc i} were rejected. We considered the spectrum to be ``blended'' if there was no obvious continuum region between the P~{\sc ii} feature and the core of the O~{\sc i} line. Of the 107 sight lines in our combined sample, only 48 were deemed to have unblended P~{\sc ii} $\lambda1301$ absorption lines. Furthermore, in many of the ``unblended'' cases, the P~{\sc ii} feature is superimposed onto the gently sloping damping wing of the O~{\sc i} line. Thus, in these cases, the continuum being fit is not the stellar continuum but the Lorentzian damping wing of the O~{\sc i} resonance line.

\subsection{Processing of the FUSE Data\label{subsec:fuse}}
All FUSE exposures for the sight lines in our survey were downloaded from the MAST archive. For each detector segment, multiple exposures of a given target were cross-correlated in wavelength space and then co-added by taking the weighted mean of the measured intensities.\footnote{The registration and co-addition of FUSE data was accomplished using IDL routines within the LTOOLS package, available at the following URL: \url{https://archive.stsci.edu/fuse/analysis/idl_tools.html}.} The overlapping portions of different detector segments that covered lines of interest to our survey were then cross-correlated and co-added in the same manner. In this way, we produced high S/N ratio spectra for the Cl~{\sc ii} $\lambda1071$ line and also, in some cases, for the Cl~{\sc i} $\lambda1004$, $\lambda1094$, and $\lambda1097$ transitions.

A major complication when dealing with FUSE spectra are the numerous absorption features arising from electronic transitions within the Lyman and Werner bands of H$_2$. Most concerning for this investigation is that the $R$(4) line of the H$_2$ (3$-$0) Lyman band at 1070.9~\AA{} is very close to the Cl~{\sc ii} line at 1071.0~\AA{}. (The velocity separation between the two lines is only 38~km~s$^{-1}$.) The difficulty in distinguishing the two features is compounded by the relatively low resolution of the FUSE spectra ($\Delta v\approx18$~km~s$^{-1}$). The stellar continuum surrounding the H$_2$ lines in the vicinity of the Cl~{\sc ii} $\lambda1071$ feature is easy to identify in most cases. The same continuum fitting procedure used for the STIS data provided adequate solutions for the normalization of the FUSE spectra. A more specialized procedure was required to deblend the absorption features associated with the H$_2$ and Cl~{\sc ii} lines near 1071~\AA{}. This procedure is described in more detail in Section~\ref{subsubsec:ionized}.

\section{RESULTS ON COLUMN DENSITIES\label{sec:results}}
In order to obtain column densities for P~{\sc ii}, Cl~{\sc i}, and Cl~{\sc ii} for the sight lines in our survey, we employed the technique of multi-component Voigt profile fitting using the code ISMOD \citep{s08}. The profile fitting routine treats the column density $N$, Doppler broadening parameter $b$, and velocity $v$ of each component as a free parameter and seeks to minimize the root mean square (rms) deviations between the observed spectrum and the synthetic one. In the following subsections we describe in more detail the specific procedures used to fit the P~{\sc ii}, Cl~{\sc i}, and Cl~{\sc ii} lines, each of which presented us with a unique challenge.

\begin{figure}
\centering
\includegraphics[width=0.44\textwidth]{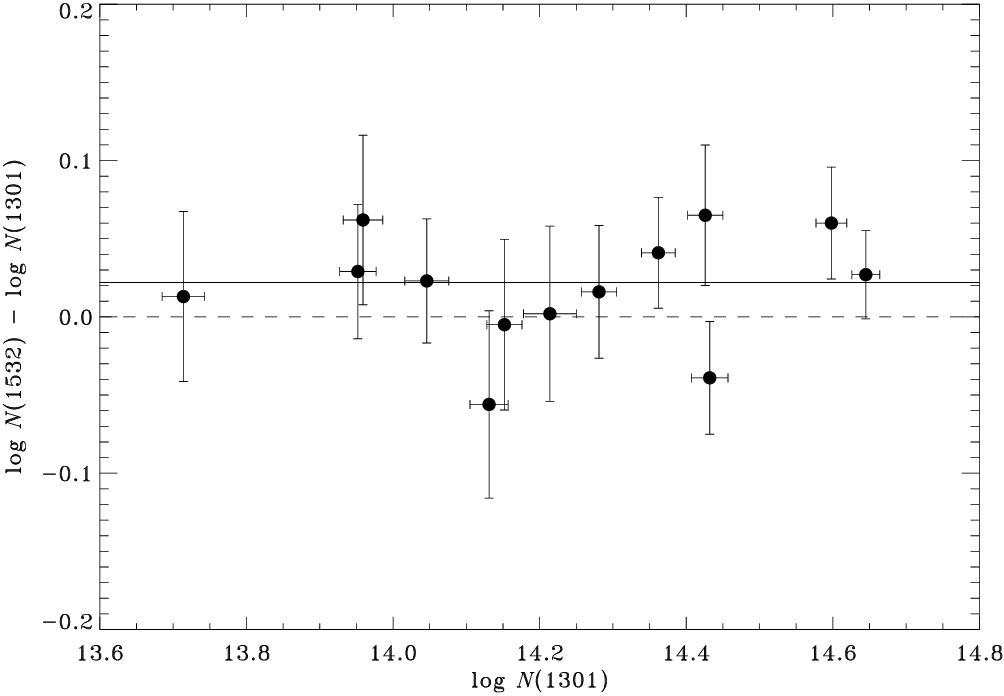}
\caption{Logarithmic difference in the column densities obtained from the P~{\sc ii}~$\lambda1301$ and $\lambda1532$ lines when the theoretical $f$-value from \citet{ff06} is adopted for the $\lambda1532$ transition and the experimental $f$-value from \citet{b18} is adopted for P~{\sc ii}~$\lambda1301$. The weighted mean of the difference (0.022 dex, as indicated by the solid horizontal line) yields an empirical $f$-value for the P~{\sc ii}~$\lambda1532$ transition of 0.00737.\label{fig:phos_diff}}
\end{figure}

\subsection{Phosphorus Column Densities\label{subsec:phosphorus}}
For our analysis of P~{\sc ii} column densities, we adopt the experimentally determined oscillator strength for the P~{\sc ii} $\lambda1301$ transition recently published by \citet{b18}. Their $f$-value for P~{\sc ii} $\lambda1301$ indicates that a downward revision of 0.188 dex should be applied to column densities derived using the older \citet{h88} $f$-value \citep[which is the one recommended by][]{m03}. The \citet{b18} $f$-value for P~{\sc ii} $\lambda1301$ is consistent with more recent theoretical determinations \citep{t03,ff06}, which indicate that the $f$-value for P~{\sc ii} $\lambda1532$ is also in need of substantial revision. The theoretical calculations of \citet{t03} and \citet{ff06} suggest that a reduction in column density of $\sim$0.4 dex is needed for results based on the \citet{h88} $f$-value for P~{\sc ii} $\lambda1532$. Since no recent experimental result is available for P~{\sc ii} $\lambda1532$, we derived an empirical $f$-value for this transition based on high-resolution STIS spectra.

\startlongtable
\begin{deluxetable*}{lccccc}
\tablecolumns{6}
\tabletypesize{\small}
\tablecaption{Phosphorus Column Densities\label{tab:phosphorus}}
\tablehead{ \colhead{Star} & \colhead{$W_{\lambda}$(1301)} & \colhead{log~$N$(1301)} & \colhead{$W_{\lambda}$(1532)} & \colhead{log~$N$(1532)} & \colhead{log~$N$(P~{\sc ii})\tablenotemark{a}} \\
\colhead{} & \colhead{(m\AA{})} & \colhead{} & \colhead{(m\AA{})} & \colhead{} & \colhead{} }
\startdata
HD~108        & \ldots         & \ldots           & $32.6\pm2.1$   & $14.49\pm0.03$   & $14.49\pm0.03$ \\
HD~1383       & \ldots         & \ldots           & $40.0\pm3.1$   & $14.56\pm0.04$   & $14.56\pm0.04$ \\
HD~12323      & \ldots         & \ldots           & $28.4\pm2.0$   & $14.45\pm0.04$   & $14.45\pm0.04$ \\
HD~13268      & \ldots         & \ldots           & $39.9\pm4.5$   & $14.52\pm0.05$   & $14.52\pm0.05$ \\
HD~13745      & \ldots         & \ldots           & $44.1\pm5.1$   & $14.53\pm0.05$   & $14.53\pm0.05$ \\
HD~14434      & \ldots         & \ldots           & $52.2\pm4.6$   & $14.75\pm0.04$   & $14.75\pm0.04$ \\
HD~15137      & \ldots         & \ldots           & $31.7\pm1.9$   & $14.44\pm0.03$   & $14.44\pm0.03$ \\
HD~23478      & $18.8\pm0.5$   & $13.95\pm0.02$   & $11.9\pm1.0$   & $13.96\pm0.04$   & $13.95\pm0.02$ \\
HD~24190      & $28.1\pm0.4$   & $14.15\pm0.02$   & $17.2\pm2.0$   & $14.13\pm0.05$   & $14.15\pm0.02$ \\
HD~24534      & $19.6\pm0.2$   & $14.05\pm0.03$   & $13.3\pm0.7$   & $14.05\pm0.03$   & $14.05\pm0.02$ \\
HD~25443      & $64.5\pm1.8$   & $14.70\pm0.03$   & \ldots         & \ldots           & $14.70\pm0.03$ \\
HD~37903      & $16.9\pm0.5$   & $13.88\pm0.02$   & \ldots         & \ldots           & $13.88\pm0.02$ \\
HD~41161      & $49.9\pm1.2$   & $14.35\pm0.02$   & \ldots         & \ldots           & $14.35\pm0.02$ \\
HD~46223      & \ldots         & \ldots           & $44.1\pm1.0$   & $14.61\pm0.03$   & $14.61\pm0.03$ \\
HD~52266      & $50.3\pm0.6$   & $14.46\pm0.03$   & \ldots         & \ldots           & $14.46\pm0.03$ \\
HD~53975      & $41.8\pm0.8$   & $14.27\pm0.02$   & \ldots         & \ldots           & $14.27\pm0.02$ \\
HD~63005      & \ldots         & \ldots           & $27.3\pm2.2$   & $14.47\pm0.04$   & $14.47\pm0.04$ \\
HD~66788      & \ldots         & \ldots           & $31.6\pm3.3$   & $14.39\pm0.04$   & $14.39\pm0.04$ \\
HD~72754      & $34.8\pm2.2$   & $14.33\pm0.04$   & \ldots         & \ldots           & $14.33\pm0.04$ \\
HD~73882      & $33.9\pm1.6$   & $14.22\pm0.03$   & \ldots         & \ldots           & $14.22\pm0.03$ \\
HD~79186      & $51.6\pm2.4$   & $14.46\pm0.03$   & \ldots         & \ldots           & $14.46\pm0.03$ \\
HD~88115\tablenotemark{b}     & $33.3\pm1.8$   & $14.15\pm0.03$   & \ldots         & \ldots           & $14.15\pm0.03$ \\
     & $29.2\pm2.8$   & $14.23\pm0.05$   & $16.1\pm4.2$   & $14.12\pm0.10$   & $14.20\pm0.05$ \\
HD~89137      & $40.2\pm1.8$   & $14.27\pm0.03$   & \ldots         & \ldots           & $14.27\pm0.03$ \\
HD~90087      & $45.3\pm0.8$   & $14.36\pm0.02$   & $30.1\pm1.7$   & $14.38\pm0.03$   & $14.37\pm0.02$ \\
HD~91597      & \ldots         & \ldots           & $47.5\pm4.1$   & $14.67\pm0.04$   & $14.67\pm0.04$ \\
HD~91651      & \ldots         & \ldots           & $35.3\pm3.3$   & $14.41\pm0.04$   & $14.41\pm0.04$ \\
HD~91824      & $50.6\pm0.9$   & $14.38\pm0.02$   & \ldots         & \ldots           & $14.38\pm0.02$ \\
HD~91983      & $55.5\pm2.1$   & $14.44\pm0.02$   & \ldots         & \ldots           & $14.44\pm0.02$ \\
HD~92554      & \ldots         & \ldots           & $44.2\pm6.7$   & $14.57\pm0.06$   & $14.57\pm0.06$ \\
HD~93222      & \ldots         & \ldots           & $49.8\pm1.3$   & $14.59\pm0.01$   & $14.59\pm0.01$ \\
HD~93843      & \ldots         & \ldots           & $24.7\pm2.3$   & $14.30\pm0.04$   & $14.30\pm0.04$ \\
HD~94493      & $48.6\pm1.8$   & $14.35\pm0.02$   & \ldots         & \ldots           & $14.35\pm0.02$ \\
HD~97175      & $28.8\pm2.2$   & $14.19\pm0.04$   & \ldots         & \ldots           & $14.19\pm0.04$ \\
HD~99857      & \ldots         & \ldots           & $37.7\pm2.4$   & $14.54\pm0.04$   & $14.54\pm0.04$ \\
HD~99890      & \ldots         & \ldots           & $30.1\pm3.9$   & $14.41\pm0.05$   & $14.41\pm0.05$ \\
HD~100199     & \ldots         & \ldots           & $29.9\pm2.8$   & $14.39\pm0.04$   & $14.39\pm0.04$ \\
HD~101190\tablenotemark{b}    & \ldots         & \ldots           & $25.0\pm2.9$   & $14.31\pm0.05$   & $14.31\pm0.05$ \\
    & \ldots         & \ldots           & $24.7\pm0.8$   & $14.31\pm0.02$   & $14.31\pm0.02$ \\
HD~103779     & \ldots         & \ldots           & $29.3\pm1.8$   & $14.36\pm0.03$   & $14.36\pm0.03$ \\
HD~104705     & \ldots         & \ldots           & $28.6\pm0.9$   & $14.34\pm0.02$   & $14.34\pm0.02$ \\
HD~108639     & \ldots         & \ldots           & $46.5\pm2.3$   & $14.57\pm0.02$   & $14.57\pm0.02$ \\
HD~114886     & \ldots         & \ldots           & $38.0\pm2.4$   & $14.48\pm0.03$   & $14.48\pm0.03$ \\
HD~116852     & \ldots         & \ldots           & $12.2\pm2.0$   & $14.06\pm0.07$   & $14.06\pm0.07$ \\
HD~121968     & $12.4\pm0.6$   & $13.71\pm0.03$   & $ 7.1\pm0.8$   & $13.71\pm0.05$   & $13.71\pm0.02$ \\
HD~122879     & \ldots         & \ldots           & $45.6\pm2.0$   & $14.63\pm0.03$   & $14.63\pm0.03$ \\
HD~124314     & $73.2\pm1.3$   & $14.60\pm0.02$   & $51.9\pm3.2$   & $14.64\pm0.03$   & $14.61\pm0.02$ \\
HD~137595     & $29.0\pm1.0$   & $14.13\pm0.03$   & $15.2\pm2.0$   & $14.05\pm0.05$   & $14.11\pm0.02$ \\
HD~144965     & $18.1\pm1.0$   & $13.95\pm0.04$   & \ldots         & \ldots           & $13.95\pm0.04$ \\
HD~147683     & $31.6\pm1.3$   & $14.21\pm0.04$   & $19.7\pm1.9$   & $14.19\pm0.04$   & $14.21\pm0.03$ \\
HD~147888     & $24.7\pm0.3$   & $14.26\pm0.05$   & \ldots         & \ldots           & $14.26\pm0.05$ \\
HD~147933     & $25.6\pm0.5$   & $14.32\pm0.05$   & \ldots         & \ldots           & $14.32\pm0.05$ \\
HD~148422     & \ldots         & \ldots           & $48.9\pm3.9$   & $14.68\pm0.04$   & $14.68\pm0.04$ \\
HD~148937     & $81.2\pm1.3$   & $14.65\pm0.02$   & $54.5\pm2.5$   & $14.65\pm0.02$   & $14.65\pm0.01$ \\
HD~151805     & \ldots         & \ldots           & $58.8\pm5.7$   & $14.70\pm0.04$   & $14.70\pm0.04$ \\
HD~152590     & \ldots         & \ldots           & $28.4\pm2.4$   & $14.45\pm0.05$   & $14.45\pm0.05$ \\
HD~157857     & $43.6\pm1.1$   & $14.42\pm0.03$   & \ldots         & \ldots           & $14.42\pm0.03$ \\
HD~163522     & \ldots         & \ldots           & $31.9\pm2.9$   & $14.40\pm0.04$   & $14.40\pm0.04$ \\
HD~165955     & \ldots         & \ldots           & $20.4\pm2.5$   & $14.21\pm0.05$   & $14.21\pm0.05$ \\
HD~167402     & \ldots         & \ldots           & $30.8\pm3.2$   & $14.38\pm0.04$   & $14.38\pm0.04$ \\
HD~168941     & \ldots         & \ldots           & $19.5\pm2.7$   & $14.18\pm0.06$   & $14.18\pm0.06$ \\
HD~170740     & $28.4\pm1.4$   & $14.36\pm0.05$   & \ldots         & \ldots           & $14.36\pm0.05$ \\
HD~177989     & \ldots         & \ldots           & $17.0\pm0.6$   & $14.10\pm0.02$   & $14.10\pm0.02$ \\
HD~178487     & \ldots         & \ldots           & $42.2\pm3.0$   & $14.55\pm0.03$   & $14.55\pm0.03$ \\
HD~179407     & \ldots         & \ldots           & $37.6\pm4.0$   & $14.54\pm0.05$   & $14.54\pm0.05$ \\
HD~185418     & $43.3\pm0.7$   & $14.43\pm0.04$   & \ldots         & \ldots           & $14.43\pm0.04$ \\
HD~190918     & \ldots         & \ldots           & $50.5\pm3.0$   & $14.60\pm0.03$   & $14.60\pm0.03$ \\
HD~191877     & $41.6\pm1.2$   & $14.30\pm0.02$   & \ldots         & \ldots           & $14.30\pm0.02$ \\
HD~192035     & $50.9\pm1.8$   & $14.57\pm0.05$   & $33.5\pm3.3$   & $14.50\pm0.05$   & $14.53\pm0.04$ \\
HD~195455     & $14.4\pm1.3$   & $13.73\pm0.04$   & \ldots         & \ldots           & $13.73\pm0.04$ \\
HD~195965     & $39.9\pm1.6$   & $14.28\pm0.02$   & $24.6\pm2.0$   & $14.27\pm0.03$   & $14.28\pm0.02$ \\
HD~198478     & $57.1\pm2.5$   & $14.78\pm0.06$   & \ldots         & \ldots           & $14.78\pm0.06$ \\
HD~198781     & $38.4\pm1.1$   & $14.24\pm0.02$   & \ldots         & \ldots           & $14.24\pm0.02$ \\
HD~201345     & \ldots         & \ldots           & $18.6\pm2.2$   & $14.14\pm0.05$   & $14.14\pm0.05$ \\
HD~202347     & $21.2\pm1.1$   & $13.96\pm0.03$   & $13.6\pm1.5$   & $14.00\pm0.05$   & $13.97\pm0.02$ \\
HD~203374\tablenotemark{b}    & $57.6\pm1.1$   & $14.47\pm0.02$   & \ldots         & \ldots           & $14.47\pm0.02$ \\
    & $54.9\pm1.5$   & $14.49\pm0.03$   & $35.6\pm2.7$   & $14.47\pm0.04$   & $14.48\pm0.02$ \\
HD~203938     & $47.3\pm2.2$   & $14.57\pm0.05$   & \ldots         & \ldots           & $14.57\pm0.05$ \\
HD~206267     & $57.9\pm1.4$   & $14.52\pm0.03$   & \ldots         & \ldots           & $14.52\pm0.03$ \\
HD~206773     & $46.0\pm1.0$   & $14.40\pm0.03$   & \ldots         & \ldots           & $14.40\pm0.03$ \\
HD~207308     & $41.9\pm1.3$   & $14.44\pm0.05$   & $29.4\pm1.3$   & $14.43\pm0.03$   & $14.43\pm0.03$ \\
HD~207538     & $54.9\pm0.9$   & $14.56\pm0.04$   & $36.5\pm1.4$   & $14.52\pm0.03$   & $14.53\pm0.02$ \\
HD~209339     & $48.4\pm1.1$   & $14.43\pm0.03$   & $29.2\pm1.6$   & $14.37\pm0.03$   & $14.40\pm0.02$ \\
HD~210809     & \ldots         & \ldots           & $43.3\pm4.7$   & $14.60\pm0.05$   & $14.60\pm0.05$ \\
HD~210839     & $59.2\pm1.0$   & $14.49\pm0.03$   & \ldots         & \ldots           & $14.49\pm0.03$ \\
HD~218915     & \ldots         & \ldots           & $25.9\pm1.1$   & $14.28\pm0.02$   & $14.28\pm0.02$ \\
HD~219188     & $19.2\pm0.4$   & $13.91\pm0.02$   & \ldots         & \ldots           & $13.91\pm0.02$ \\
HD~220057     & $26.4\pm1.2$   & $14.21\pm0.05$   & \ldots         & \ldots           & $14.21\pm0.05$ \\
HD~224151     & $53.4\pm1.8$   & $14.43\pm0.02$   & $36.8\pm3.2$   & $14.47\pm0.04$   & $14.44\pm0.02$ \\
HDE~308813    & \ldots         & \ldots           & $27.8\pm3.0$   & $14.35\pm0.05$   & $14.35\pm0.05$ \\
BD$+$35~4258    & \ldots         & \ldots           & $29.7\pm3.9$   & $14.35\pm0.05$   & $14.35\pm0.05$ \\
BD$+$53~2820    & \ldots         & \ldots           & $37.5\pm5.2$   & $14.58\pm0.06$   & $14.58\pm0.06$ \\
CPD$-$59~2603\tablenotemark{b}  & \ldots         & \ldots           & $53.4\pm1.9$   & $14.64\pm0.02$   & $14.64\pm0.02$ \\
  & \ldots         & \ldots           & $51.7\pm2.2$   & $14.78\pm0.04$   & $14.78\pm0.04$ \\
CPD$-$59~4552   & $60.4\pm2.5$   & $14.51\pm0.03$   & \ldots         & \ldots           & $14.51\pm0.03$ \\
CPD$-$69~1743   & \ldots         & \ldots           & $35.5\pm4.7$   & $14.50\pm0.06$   & $14.50\pm0.06$ \\
\enddata
\tablenotetext{a}{Final P~{\sc ii} column density. In cases where both P~{\sc ii} lines are measured, the final column density is obtained from the error weighted mean of the two results.}
\tablenotetext{b}{First line gives results from E140H data; second line gives results from E140M data.}
\end{deluxetable*}

There are thirteen sight lines in our sample that have high-resolution STIS spectra covering both P~{\sc ii} transitions ($\lambda1301$ and $\lambda1532$) and have a $\lambda1301$ profile that is not significantly blended with O~{\sc i} $\lambda1302$. For these sight lines, we first fit the P~{\sc ii} $\lambda1301$ profile using the $f$-value from \citet{b18}. We then adopted the component structure found from the $\lambda1301$ line to fit the $\lambda1532$ profile using the $f$-value from \citet{ff06}, which is $f(1532)=0.00701$. The weighted mean of the (logarithmic) differences in the column densities derived from the $\lambda1301$ and $\lambda1532$ transitions is $0.022\pm0.012$ dex (see Figure~\ref{fig:phos_diff}). From this result, we obtain an empirical $f$-value for the P~{\sc ii} $\lambda1532$ transition of 0.00737, very similar to the theoretical results of \citet{t03} and \citet{ff06}.

Our empirical $f$-value is adopted for the remaining sight lines in our sample with observations of P~{\sc ii} $\lambda1532$. (For the sight lines used in the derivation of the empirical $f$-value, the column densities obtained from the $\lambda1532$ transition were adjusted downward by 0.022 dex.) The equivalent widths and column densities derived from profile synthesis fits to the P~{\sc ii} $\lambda1301$ and $\lambda1532$ transitions are presented in Table~\ref{tab:phosphorus} for the 92 sight lines in our final P sample. Errors in equivalent width account for uncertainties due to noise in the spectra and uncertainties in continuum placement. The column density uncertainties include an additional term that depends on the degree of saturation in the absorption profile. In cases where both P~{\sc ii} transitions were analyzed, final P~{\sc ii} column densities were obtained from a weighted mean of the individual results.

For four of the sight lines in our sample (HD~88115, HD~101190, HD~203374, and CPD$-$59~2603), we were able to independently derive P~{\sc ii} column densities from medium-resolution and high-resolution STIS spectra. Both results are included in Table~\ref{tab:phosphorus}. In three out of the four cases, the two independent results agree with one another within the uncertainties. However, for CPD$-$59~2603, the P~{\sc ii} column densities derived from fits to the E140H and E140M spectra disagree at the $3.1\sigma$ level. Since the equivalent widths from the two fits are consistent with one another, the difference in column density is likely related to uncertainties regarding the degree of saturation in the line profile. Indeed, the fit to the medium resolution spectrum results in a smaller $b$-value for the dominant absorption component (1.3 versus 2.4 km~s$^{-1}$) yielding a higher column density. The $b$-values are better constrained in fits to the higher resolution spectra. Thus, for the depletion analysis, we adopt the results from the E140H data for each of the four sight lines discussed here.

\begin{figure*}
\centering
\includegraphics[width=0.44\textwidth]{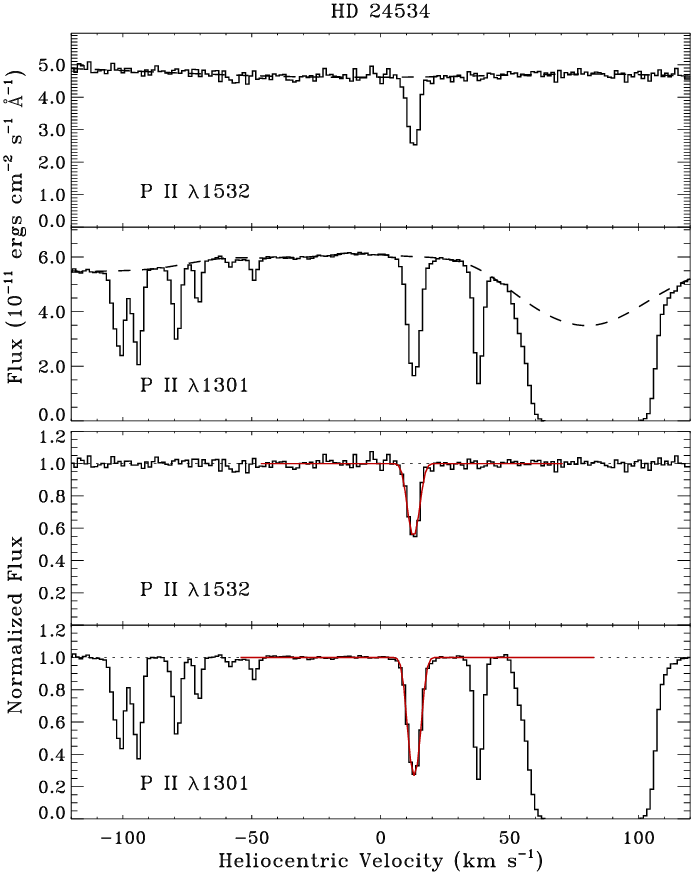}
\includegraphics[width=0.44\textwidth]{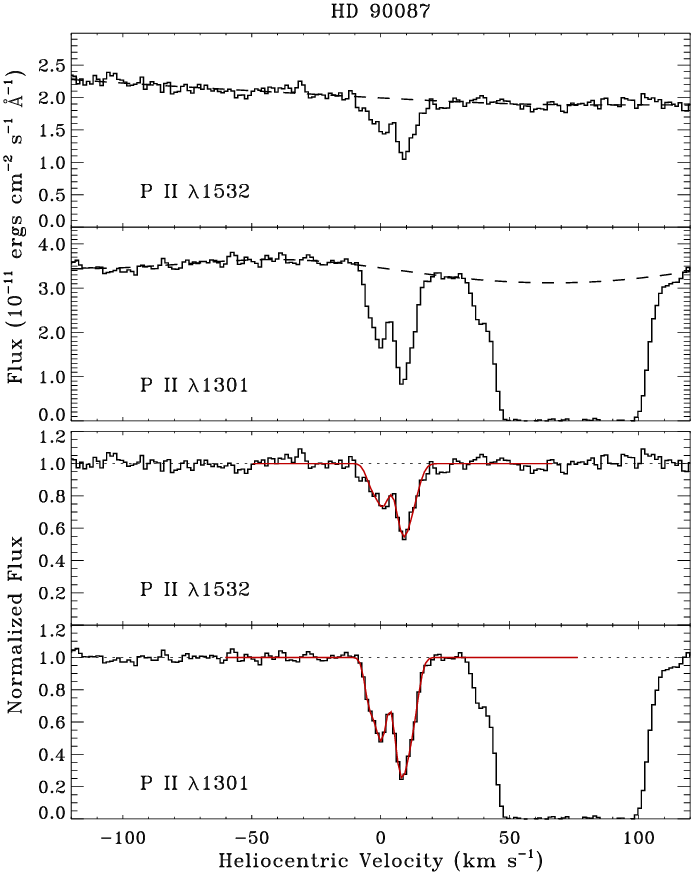}
\caption{High-resolution STIS spectra in the vicinity of the P~{\sc ii}~$\lambda1532$ and $\lambda1301$ lines toward HD~24534 (left panels) and HD~90087 (right panels). Upper panels show the unnormalized spectra with dashed lines indicating the adopted continuum fits. Lower panels show the normalized spectra with solid red lines representing fits to the P~{\sc ii} absorption profiles.\label{fig:phos_fit1}}
\end{figure*}

\begin{figure*}
\centering
\includegraphics[width=0.44\textwidth]{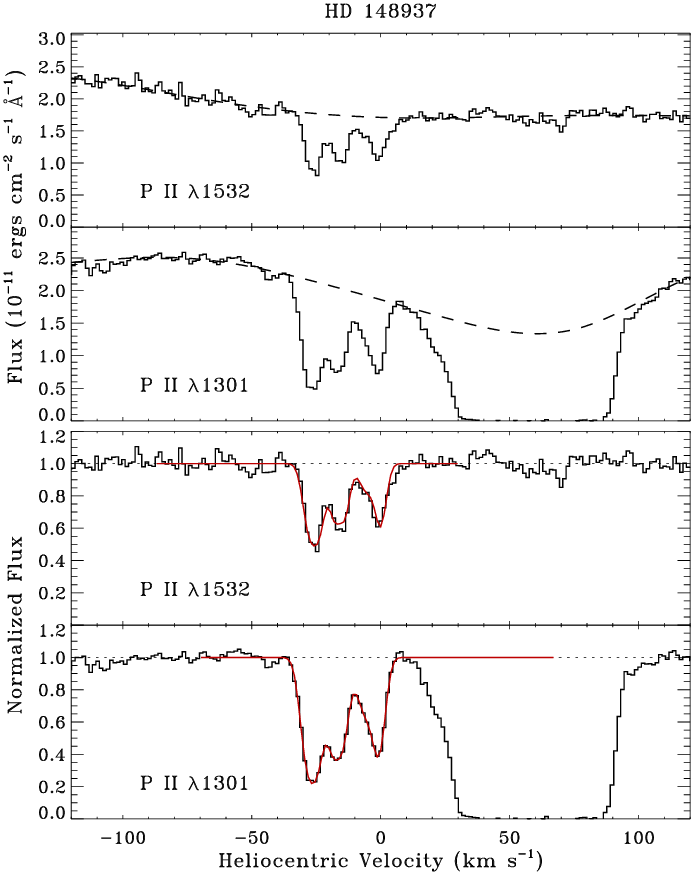}
\includegraphics[width=0.44\textwidth]{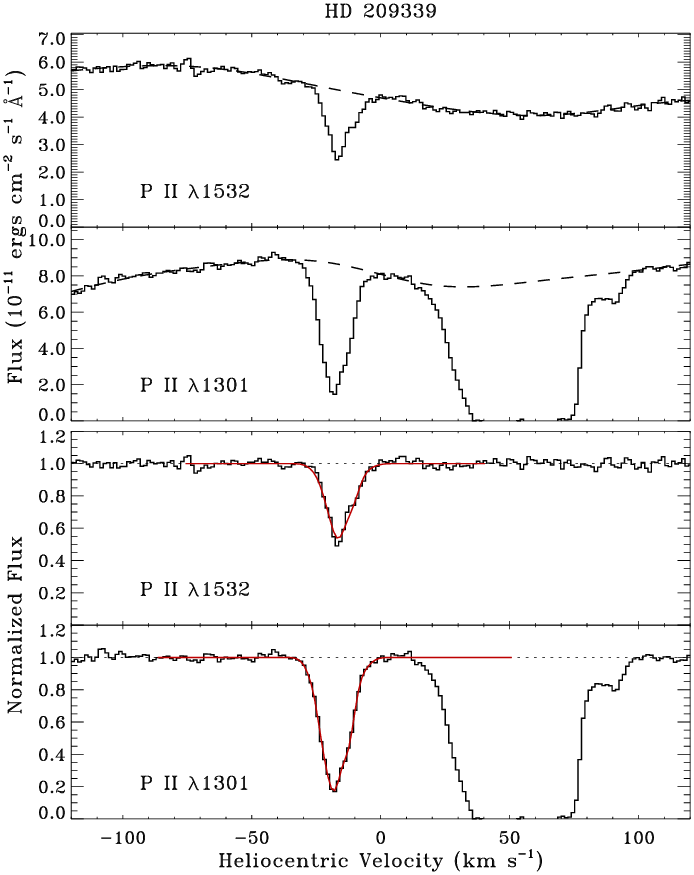}
\caption{Same as Figure~2 except toward HD~148937 (left panels) and HD~209339 (right panels).\label{fig:phos_fit2}}
\end{figure*}

Examples of our fits to the P~{\sc ii} $\lambda1532$ and $\lambda1301$ lines are presented in Figures~\ref{fig:phos_fit1} and \ref{fig:phos_fit2}. In each case, we provide both the unnormalized spectrum with the adopted continuum fit and the normalized spectrum with the derived profile synthesis fit. For each of the examples shown, the component structure obtained from the fit to the stronger P~{\sc ii} $\lambda1301$ feature was adopted in the fit to the $\lambda1532$ line (as described above).

In Table~\ref{tab:comparison}, we present a comparison between the P~{\sc ii} column densities that we obtain and the column densities reported in the literature for sight lines studied previously by \citet{l05} and \citet{c06}. All of the previous results on $N$(P~{\sc ii}) given in Table~\ref{tab:comparison} have been adjusted to be consistent with the set of $f$-values adopted in this investigation (see Table~\ref{tab:lines}). Also note that the P~{\sc ii} column densities from \citet{l05} shown in Table~\ref{tab:comparison} are the ones that those authors obtain from STIS observations of the P~{\sc ii} $\lambda1532$ line. The \citet{c06} column densities were derived from STIS observations of the P~{\sc ii} $\lambda1301$ line. Our values for $\log N({\rm P~\textsc{ii}})$ shown in Table~\ref{tab:comparison} refer to the final P~{\sc ii} column densities from Table~\ref{tab:phosphorus}. In some cases, these final values were obtained from the weighted mean of the results from the $\lambda1301$ and $\lambda1532$ lines.

\begin{deluxetable}{lccc}
\tablecolumns{4}
\tabletypesize{\small}
\tablecaption{Comparison with Previous Studies: P~{\sc ii}\label{tab:comparison}}
\tablehead{ \colhead{Star} & \multicolumn{3}{c}{log~$N$(P~{\sc ii})} \\
\cline{2-4} \\
\colhead{} & \colhead{This Work} & \colhead{Previous Result\tablenotemark{a}} & \colhead{Ref.} }
\startdata
HD~24534    & $14.05\pm0.02$   & $14.03\pm0.05$   & 1 \\
HD~37903    & $13.88\pm0.02$   & $13.91\pm0.04$   & 2 \\
HD~72754    & $14.33\pm0.04$   & $14.27\pm0.04$   & 2 \\
HD~79186    & $14.46\pm0.03$   & $14.55\pm0.06$   & 2 \\
HD~91824    & $14.38\pm0.02$   & $14.38\pm0.03$   & 2 \\
HD~91983    & $14.44\pm0.02$   & $14.44\pm0.04$   & 2 \\
HD~93222    & $14.59\pm0.01$   & $14.43\pm0.01$   & 1 \\
HD~99857    & $14.54\pm0.04$   & $14.25\pm0.02$   & 1 \\
HD~104705   & $14.34\pm0.02$   & $14.15\pm0.04$   & 1 \\
HD~121968   & $13.71\pm0.02$   & $13.63\pm0.09$   & 1 \\
HD~124314   & $14.61\pm0.02$   & $14.57\pm0.24$   & 1 \\
HD~152590   & $14.45\pm0.05$   & $14.46\pm0.02$   & 2 \\
HD~157857   & $14.42\pm0.03$   & $14.43\pm0.04$   & 2 \\
HD~177989   & $14.10\pm0.02$   & $14.08\pm0.07$   & 1 \\
HD~185418   & $14.43\pm0.04$   & $14.45\pm0.03$   & 2 \\
HD~198478   & $14.78\pm0.06$   & $14.50\pm0.03$   & 2 \\
HD~198781   & $14.24\pm0.02$   & $14.24\pm0.04$   & 2 \\
HD~201345   & $14.14\pm0.05$   & $14.23\pm0.06$   & 2 \\
HD~202347   & $13.97\pm0.02$   & $13.94\pm0.07$   & 1 \\
HD~206773   & $14.40\pm0.03$   & $14.42\pm0.04$   & 2 \\
HD~218915   & $14.28\pm0.02$   & $14.14\pm0.09$   & 1 \\
HD~220057   & $14.21\pm0.05$   & $14.15\pm0.03$   & 2 \\
HD~224151   & $14.44\pm0.02$   & $14.48\pm0.09$   & 1 \\
\enddata
\tablenotetext{a}{All previous results have been adjusted to be consistent with the set of $f$-values adopted in this investigation (see Table 2).}
\tablerefs{(1) \citet{l05}, (2) \citet{c06}.}
\end{deluxetable}

Most of the previous results on $N$(P~{\sc ii}) for the sight lines shown in Table~\ref{tab:comparison} are consistent with our values at approximately the $1\sigma$ level. For HD~198478, however, the \citet{c06} value is lower than ours by $4.3\sigma$. \citet{c06} report an equivalent width for the $\lambda1301$ line toward HD~198478 ($47.0\pm1.4$) that is lower than our equivalent width ($57.1\pm2.5$). However, the difference in equivalent width is considerably smaller than the difference in column density (0.28 dex). The discrepancy seems to be caused, therefore, by a combination of continuum placement uncertainties and differences in the optical depth correction (via the $b$-value) for the dominant P~{\sc ii} absorption component.

Similarly, there are three cases where the P~{\sc ii} column density from \citet{l05} is significantly lower than our value. The P~{\sc ii} column densities reported by \citet{l05} for the sight lines to HD~93222, HD~99857, and HD~104705 are lower than our values by $8.9\sigma$, $6.9\sigma$, and $4.3\sigma$, respectively. While \citet{l05} do not provide equivalent widths with their column density determinations, the continua surrounding the P~{\sc ii} $\lambda1532$ lines in these directions are fairly easy to discern. One exception is that, in our fit to the $\lambda1532$ line toward HD~99857, we include very weak components at negative velocities, which are seen in the $\lambda1301$ line but are difficult to discern at $\lambda1532$. \citet{l05} apparently assumed that these features were part of the continuum. In general, we typically include many more velocity components in our profile synthesis fits than \citet{l05} do in theirs. For HD~93222, HD~99857, and HD~104705, we include 6, 8, and 5 components, respectively, while \citet{l05} include 3, 1, and 3 components. Generally speaking, more components will typically result in smaller $b$-values and potentially larger column densities. However, this does not seem to explain the significant discrepancies in the P~{\sc ii} column densities noted for these three sight lines. If we simply integrate the apparent optical depth profiles of the $\lambda1532$ lines, a procedure which should yield a lower limit to the true P~{\sc ii} column density \citep[see, e.g.,][]{ss91}, we find values of $\log N({\rm P~\textsc{ii}})=14.58\pm0.02$ toward HD~93222, $14.47\pm0.02$ toward HD~99857, and $14.34\pm0.02$ toward HD~104705. These values are consistent with the results we obtain from profile fitting but are considerably higher than the column densities reported by \citet{l05}.

\subsection{Chlorine Column Densities\label{subsec:chlorine}}
\subsubsection{Neutral Chlorine\label{subsubsec:neutral}}
Neutral chlorine column densities were obtained (in most cases) from simultaneous profile synthesis fits to the Cl~{\sc i} $\lambda1347$ line from STIS spectra and one other weaker line. If STIS observations were available covering the Cl~{\sc i} $\lambda1379$ feature, then this line served as the weaker line in the simultaneous fit. Along many sight lines, the strong Cl~{\sc i} $\lambda1347$ feature shows weak absorption components displaced in velocity relative to the main, optically-thick absorption component. In such cases, the $\lambda1379$ line will typically only show absorption from the main component but the line is optically thin (i.e., on the linear portion of the curve of growth). A simultaneous fit to both features therefore allows us to probe the full velocity structure of the Cl~{\sc i} absorption and also provides us with an accurate value for the total column density along the line of sight.

\startlongtable
\begin{deluxetable*}{lccccccc}
\tablecolumns{8}
\tabletypesize{\small}
\tablecaption{Chlorine Column Densities\label{tab:chlorine}}
\tablehead{ \colhead{Star} & \colhead{$W_{\lambda}$(1347)} & \colhead{Line\tablenotemark{a}} & \colhead{$W_{\lambda}$\tablenotemark{a}} & \colhead{log~$N$(Cl~{\sc i})} & \colhead{$W_{\lambda}$(1071)} & \colhead{log~$N$(Cl~{\sc ii})} & \colhead{log~$N$(Cl$_{\rm tot}$)} \\
\colhead{} & \colhead{(m\AA{})} & \colhead{} & \colhead{(m\AA{})} & \colhead{} & \colhead{(m\AA{})} & \colhead{} & \colhead{} }
\startdata
HD~108        & $116.5\pm0.7$\phn  & 1379  & $12.5\pm1.6$\phn  & $14.47\pm0.04$  & $20.1\pm3.4$  & $14.28\pm0.07$  & $14.69\pm0.04$ \\
HD~1383       & $121.9\pm0.9$\phn  & 1379  &  $8.4\pm2.4$  & $14.31\pm0.05$  & \ldots        & \ldots          & \ldots         \\
HD~3827       &   \phn$8.8\pm1.0$  &       & \ldots        & $12.57\pm0.04$  &  \phn$7.0\pm1.9$  & $13.72\pm0.11$  & $13.75\pm0.10$ \\
HD~12323      &  $79.1\pm1.4$  & 1379  &  $6.2\pm1.4$  & $14.17\pm0.05$  & $10.1\pm2.9$  & $13.93\pm0.11$  & $14.36\pm0.05$ \\
HD~13268      & $132.9\pm2.0$\phn  & 1379  & $12.5\pm2.5$\phn  & $14.48\pm0.05$  & \ldots        & \ldots          & \ldots         \\
HD~13745      & $122.8\pm2.3$\phn  & 1379  &  $8.1\pm2.3$  & $14.28\pm0.04$  & \ldots        & \ldots          & \ldots         \\
HD~14434      & $124.3\pm3.1$\phn  & 1379  & $16.9\pm3.5$\phn  & $14.64\pm0.05$  & \ldots        & \ldots          & \ldots         \\
HD~15137      & $107.5\pm0.8$\phn  & 1097  & $12.0\pm6.4$\phn  & $14.14\pm0.04$  & $13.9\pm5.6$  & $14.06\pm0.15$  & $14.40\pm0.07$ \\
HD~23478      &  $35.0\pm0.2$  & 1379  &  $8.0\pm0.4$  & $14.33\pm0.02$  & $<9.5$        & $<13.82$          & \ldots         \\
HD~24190      &  $54.8\pm0.4$  & 1379  &  $7.0\pm0.9$  & $14.25\pm0.04$  &  \phn$9.4\pm1.7$  & $13.87\pm0.07$  & $14.40\pm0.04$ \\
HD~24534      &  $31.9\pm0.1$  & 1379  & $10.9\pm0.1$\phn  & $14.51\pm0.02$  &  \phn$4.6\pm2.3$  & $13.55\pm0.17$  & $14.55\pm0.03$ \\
HD~37903      &  $25.5\pm0.3$  & 1097  &  $4.7\pm3.1$  & $13.79\pm0.08$  & $12.4\pm3.9$  & $14.04\pm0.12$  & $14.23\pm0.09$ \\
HD~41161      &  $94.8\pm0.7$  & 1004  & $28.2\pm2.9$\phn  & $13.92\pm0.02$  & $15.1\pm4.5$  & $14.06\pm0.11$  & $14.29\pm0.07$ \\
HD~46223      &  $89.5\pm0.5$  & 1379  & $13.3\pm0.6$\phn  & $14.53\pm0.02$  & $20.1\pm4.8$  & $14.23\pm0.09$  & $14.71\pm0.04$ \\
HD~52266      &  $56.4\pm0.6$  & 1004  & $19.2\pm9.1$\phn  & $13.79\pm0.04$  & \phn$18.8\pm10.6$ & $14.20\pm0.19$  & $14.35\pm0.15$ \\
HD~53975      &  $40.7\pm0.4$  &       & \ldots        & $13.46\pm0.02$  & $12.4\pm7.2$  & $13.97\pm0.20$  & $14.09\pm0.16$ \\
HD~63005      &  $65.7\pm1.1$  & 1379  &  $4.7\pm1.8$  & $14.04\pm0.06$  & $12.5\pm3.6$  & $14.06\pm0.11$  & $14.35\pm0.07$ \\
HD~66788      &  $67.9\pm1.4$  & 1379  &  $2.2\pm1.7$  & $13.70\pm0.03$  & $17.8\pm5.0$  & $14.15\pm0.11$  & $14.28\pm0.08$ \\
HD~72754      &  $39.8\pm0.8$  &       & \ldots        & $13.74\pm0.05$  & \ldots        & \ldots          & \ldots         \\
HD~73882      &  $55.9\pm0.7$  & 1379  &  $13.6\pm1.2$\phn  & $14.62\pm0.04$  & $20.1\pm7.5$  & $14.25\pm0.14$  & $14.77\pm0.05$ \\
HD~75309      &  $53.3\pm0.6$  & 1097  &  $6.2\pm3.8$  & $13.86\pm0.05$  & $12.8\pm3.1$  & $14.02\pm0.10$  & $14.25\pm0.06$ \\
HD~88115      &  $24.6\pm0.7$  &       & \ldots        & $13.19\pm0.04$  & $12.8\pm1.2$  & $14.09\pm0.04$\tablenotemark{b}  & $14.14\pm0.04$ \\
HD~89137      &  $49.9\pm0.6$  & 1094  & $19.9\pm3.1$\phn  & $13.90\pm0.05$  & $14.9\pm3.3$  & $14.07\pm0.09$  & $14.29\pm0.06$ \\
HD~90087      &  $58.9\pm0.6$  &       & \ldots        & $13.65\pm0.04$  & $20.3\pm1.6$  & $14.23\pm0.03$  & $14.34\pm0.03$ \\
HD~91597      &  $46.2\pm2.2$  & 1379  & $<6.4$\tablenotemark{c} & $13.45\pm0.03$  & $31.3\pm3.0$  & $14.50\pm0.04$  & $14.54\pm0.04$ \\
HD~91651      &  $52.1\pm1.4$  &       & \ldots        & $13.52\pm0.02$  & \ldots        & \ldots          & \ldots         \\
HD~91824      &  $67.7\pm1.2$  &       & \ldots        & $13.80\pm0.05$  & $15.4\pm3.2$  & $14.07\pm0.08$  & $14.26\pm0.06$ \\
HD~91983      &  $64.4\pm1.0$  &       & \ldots        & $13.79\pm0.05$  & $17.1\pm5.6$  & $14.13\pm0.12$  & $14.29\pm0.09$ \\
HD~92554      &  $51.8\pm2.8$  & 1379  & $<9.6$\tablenotemark{c} & $13.51\pm0.03$  & $46.4\pm7.1$  & $14.70\pm0.06$  & $14.72\pm0.06$ \\
HD~93129      &  $96.9\pm0.3$  & 1379  & $10.0\pm0.6$\phn  & $14.39\pm0.02$  & \ldots        & \ldots          & \ldots         \\
HD~93205      &  $66.0\pm0.7$  & 1379  &  $4.2\pm1.3$  & $13.98\pm0.06$  & \ldots        & \ldots          & \ldots         \\
HD~93222      &  $48.6\pm0.6$  & 1379  &  $2.0\pm0.8$  & $13.65\pm0.04$  & \ldots        & \ldots          & \ldots         \\
HD~93843      &  $56.5\pm0.8$  &       & \ldots        & $13.55\pm0.03$  & \ldots        & \ldots          & \ldots         \\
HD~94493      &  $52.6\pm0.8$  &       & \ldots        & $13.54\pm0.03$  & $22.5\pm2.8$  & $14.26\pm0.05$  & $14.33\pm0.04$ \\
HD~97175      &  $27.7\pm0.9$  &       & \ldots        & $13.41\pm0.04$  & $13.4\pm4.6$  & $14.08\pm0.13$  & $14.16\pm0.11$ \\
HD~99857      &  $68.1\pm0.5$  & 1097  &  $8.2\pm4.9$  & $13.99\pm0.06$  & $16.7\pm3.2$  & $14.32\pm0.08$\tablenotemark{b}  & $14.49\pm0.06$ \\
HD~99890      &  $70.7\pm0.2$  & 1379  &  $2.3\pm0.3$  & $13.70\pm0.03$  & $20.6\pm3.4$  & $14.21\pm0.07$  & $14.33\pm0.05$ \\
HD~100199     &  $70.7\pm1.1$  & 1379  & $13.0\pm1.5$\phn  & $14.54\pm0.04$  & \ldots        & \ldots          & \ldots         \\
HD~101190     &  $66.2\pm0.5$  & 1379  & $10.6\pm0.4$\phn  & $14.45\pm0.02$  & \ldots        & \ldots          & \ldots         \\
HD~104705     &  $53.4\pm0.5$  & 1379  &  $4.5\pm0.7$  & $14.04\pm0.05$  & \ldots        & \ldots          & \ldots         \\
HD~108639     &  $75.5\pm0.7$  & 1379  &  $4.1\pm1.0$  & $13.98\pm0.05$  & \ldots        & \ldots          & \ldots         \\
HD~109399     &  $53.7\pm0.9$  &       & \ldots        & $13.56\pm0.04$  & $15.4\pm3.2$  & $14.11\pm0.08$  & $14.22\pm0.07$ \\
HD~114886     & $106.9\pm0.9$\phn  & 1379  &  $7.8\pm1.3$  & $14.26\pm0.05$  & \ldots        & \ldots          & \ldots         \\
HD~115455     &  $91.5\pm1.1$  &       & \ldots        & $13.89\pm0.03$  & \ldots        & \ldots          & \ldots         \\
HD~116852     &  $34.1\pm0.3$  &       & \ldots        & $13.44\pm0.04$  &  \phn$7.4\pm1.1$  & $13.97\pm0.07$\tablenotemark{b}  & $14.08\pm0.06$ \\
HD~121968     &   \phn$6.9\pm0.3$  & 1379  & $<0.7$\tablenotemark{c} & $12.57\pm0.03$  & $10.7\pm1.7$  & $13.96\pm0.07$  & $13.98\pm0.06$ \\
HD~137595     &  $59.3\pm0.7$  & 1379  &  $8.2\pm0.9$  & $14.35\pm0.04$  &  \phn$4.6\pm2.1$  & $13.53\pm0.16$  & $14.41\pm0.04$ \\
HD~144965     &  $40.4\pm0.7$  & 1094  & $13.2\pm3.1$\phn  & $13.63\pm0.04$  & $<6.8$        & $<13.67$          & \ldots         \\
HD~147683     &  $51.1\pm0.5$  & 1379  & $10.3\pm0.7$\phn  & $14.46\pm0.03$  & $10.4\pm7.8$  & $13.92\pm0.24$  & $14.57\pm0.07$ \\
HD~147888     &  $25.8\pm0.3$  & 1097  &  $5.3\pm3.4$  & $13.85\pm0.08$  & $19.2\pm4.0$  & $14.43\pm0.09$  & $14.53\pm0.08$ \\
HD~147933\tablenotemark{d}     &  $26.8\pm0.2$  & 1097  &  $4.1\pm3.7$  & $13.71\pm0.08$  & $15.6\pm3.8$  & $14.31\pm0.10$  & $14.41\pm0.09$ \\
HD~148422     &  $84.9\pm1.6$  & 1379  &  $7.2\pm1.9$  & $14.23\pm0.05$  & \ldots        & \ldots          & \ldots         \\
HD~148937     & $120.8\pm0.6$\phn  & 1379  & $13.2\pm1.3$\phn  & $14.51\pm0.03$  & \ldots        & \ldots          & \ldots         \\
HD~151805     &  $81.0\pm1.9$  & 1379  &  $6.6\pm2.4$  & $14.20\pm0.05$  & \ldots        & \ldots          & \ldots         \\
HD~152590     &  $71.5\pm0.3$  & 1379  &  $6.7\pm1.7$  & $14.22\pm0.05$  & $18.3\pm4.3$  & $14.19\pm0.09$  & $14.51\pm0.05$ \\
HD~156359     &   \phn$7.4\pm1.1$  & 1379  & $<4.1$\tablenotemark{c} & $12.59\pm0.06$  &  \phn$9.7\pm3.0$  & $13.93\pm0.12$  & $13.95\pm0.11$ \\
HD~157857     &  $76.6\pm0.6$  & 1097  &  $9.4\pm5.3$  & $14.04\pm0.05$  &  \phn$7.0\pm4.0$  & $13.72\pm0.20$  & $14.21\pm0.08$ \\
HD~163522     &  $52.2\pm1.9$  & 1379  & $<6.2$\tablenotemark{c} & $13.51\pm0.03$  & \ldots        & \ldots          & \ldots         \\
HD~165246     &  $39.2\pm0.3$  & 1094  & $18.4\pm4.1$\phn  & $13.99\pm0.06$  & $20.7\pm1.6$  & $14.32\pm0.04$\tablenotemark{b}  & $14.49\pm0.03$ \\
HD~165955     &  $30.2\pm1.5$  & 1379  & $<4.8$\tablenotemark{c} & $13.23\pm0.03$  & $15.5\pm2.7$  & $14.13\pm0.07$  & $14.18\pm0.06$ \\
HD~167402     &  $35.4\pm1.0$  & 1379  &  $5.3\pm1.3$  & $14.13\pm0.07$  & $12.7\pm4.5$  & $13.98\pm0.13$  & $14.37\pm0.07$ \\
HD~168076     &  $128.7\pm1.8$\phn  & 1379  &  $18.0\pm2.5$\phn  & $14.72\pm0.04$  & \phn$41.8\pm17.1$  & $14.60\pm0.15$  & $14.97\pm0.07$ \\
HD~168941     &  $44.5\pm1.0$  & 1379  &  $7.1\pm1.1$  & $14.27\pm0.05$  & $11.3\pm2.9$  & $13.95\pm0.10$  & $14.44\pm0.05$ \\
HD~177989     &  $56.9\pm0.6$  & 1379  &  $4.8\pm0.5$  & $14.09\pm0.04$  &  \phn$3.7\pm2.5$  & $13.42\pm0.22$  & $14.18\pm0.06$ \\
HD~178487     &  $71.9\pm1.5$  & 1379  &  $8.7\pm1.3$  & $14.34\pm0.05$  & $<16.7$        & $<14.07$          & \ldots         \\
HD~179407     &  $72.6\pm1.8$  & 1379  &  $8.0\pm2.6$  & $14.30\pm0.06$  & $<21.5$        & $<14.17$          & \ldots         \\
HD~185418     &  $67.5\pm0.4$  & 1097  & $16.9\pm2.4$\phn  & $14.38\pm0.04$  & $<14.7$        & $<14.01$          & \ldots         \\
HD~190918     & $115.2\pm1.3$\phn  & 1379  &  $7.6\pm1.4$  & $14.25\pm0.04$  & $29.6\pm3.5$  & $14.38\pm0.05$  & $14.62\pm0.03$ \\
HD~191877     &  $69.1\pm0.5$  & 1097  &  $6.2\pm0.8$  & $13.85\pm0.03$  &  \phn$8.1\pm1.1$  & $13.78\pm0.06$  & $14.12\pm0.03$ \\
HD~192035     &  $62.0\pm2.3$  & 1379  &  $8.4\pm1.5$  & $14.40\pm0.06$  & $10.6\pm4.0$  & $13.93\pm0.14$  & $14.52\pm0.06$ \\
HD~192639     &  $67.6\pm0.4$  & 1097  & $14.3\pm1.7$\phn  & $14.27\pm0.04$  & $18.0\pm3.6$  & $14.22\pm0.08$  & $14.54\pm0.04$ \\
HD~195455     &   \phn$9.6\pm0.8$  &       & \ldots        & $12.62\pm0.03$  &  \phn$4.3\pm3.7$  & $13.49\pm0.27$  & $13.55\pm0.24$ \\
HD~195965     &  $67.8\pm0.9$  & 1379  &  $5.1\pm1.4$  & $14.08\pm0.06$  &  \phn$5.4\pm0.8$  & $13.59\pm0.06$  & $14.20\pm0.05$ \\
HD~201345     &  $53.3\pm0.6$  &       & \ldots        & $13.54\pm0.03$  & $16.2\pm1.7$  & $14.12\pm0.04$  & $14.22\pm0.03$ \\
HD~202347     &  $62.5\pm0.7$  & 1379  &  $3.3\pm2.2$  & $13.88\pm0.04$  & \ldots        & \ldots          & \ldots         \\
HD~203374     &  $89.1\pm0.6$  & 1379  & $10.4\pm2.0$\phn  & $14.44\pm0.05$  & $23.8\pm2.2$  & $14.29\pm0.04$  & $14.67\pm0.04$ \\
HD~203938     &  $68.6\pm0.6$  & 1379  & $20.1\pm1.3$\phn  & $14.80\pm0.03$  & $<38.3$        & $<14.42$          & \ldots         \\
HD~206267     &  $76.2\pm0.6$  & 1097  &  $9.4\pm1.2$  & $14.04\pm0.04$  & $12.9\pm3.2$  & $14.00\pm0.10$  & $14.32\pm0.05$ \\
HD~206773     &  $73.1\pm0.5$  & 1097  & $12.4\pm1.5$\phn  & $14.19\pm0.04$  & $12.0\pm2.3$  & $13.97\pm0.08$  & $14.40\pm0.04$ \\
HD~207198     &  $77.1\pm0.4$  & 1097  & $12.3\pm3.7$\phn  & $14.17\pm0.04$  & $13.8\pm7.3$  & $14.04\pm0.18$  & $14.41\pm0.09$ \\
HD~207308     &  $67.5\pm0.7$  & 1379  & $15.6\pm1.1$\phn  & $14.64\pm0.03$  & $18.3\pm7.1$  & $14.23\pm0.14$  & $14.78\pm0.05$ \\
HD~207538     &  $74.9\pm0.9$  & 1379  & $14.4\pm0.6$\phn  & $14.67\pm0.03$  & $13.9\pm5.6$  & $14.05\pm0.15$  & $14.77\pm0.04$ \\
HD~208440     &  $90.5\pm0.7$  & 1097  &  $9.5\pm1.8$  & $14.04\pm0.03$  & $17.1\pm2.8$  & $14.14\pm0.07$  & $14.39\pm0.04$ \\
HD~209339     &  $73.5\pm0.5$  & 1097  &  $7.7\pm1.3$  & $13.94\pm0.03$  & $13.5\pm1.6$  & $14.03\pm0.05$  & $14.29\pm0.03$ \\
HD~210809     & $109.6\pm0.7$\phn  & 1379  &  $4.1\pm2.4$  & $13.97\pm0.02$  & \ldots        & \ldots          & \ldots         \\
HD~210839     & $106.2\pm0.7$\phn  & 1097  & $12.4\pm2.3$\phn  & $14.16\pm0.04$  & $11.8\pm3.4$  & $14.01\pm0.11$\tablenotemark{b}  & $14.39\pm0.05$ \\
HD~212791     &  $62.2\pm1.0$  &       & \ldots        & $13.65\pm0.03$  & $11.4\pm5.5$  & $13.93\pm0.17$  & $14.11\pm0.12$ \\
HD~218915     &  $77.3\pm0.4$  & 1379  &  $6.2\pm1.0$  & $14.16\pm0.04$  & \ldots        & \ldots          & \ldots         \\
HD~219188     &  $40.0\pm0.3$  &       & \ldots        & $13.45\pm0.03$  & $11.1\pm2.2$  & $13.95\pm0.08$  & $14.07\pm0.06$ \\
HD~220057     &  $40.1\pm0.7$  & 1097  &  $6.4\pm1.6$  & $13.89\pm0.05$  &  \phn$6.1\pm2.6$  & $13.68\pm0.16$  & $14.10\pm0.07$ \\
HDE~232522    &  $82.2\pm0.2$  & 1379  &  $5.8\pm0.5$  & $14.13\pm0.03$  & \ldots        & \ldots          & \ldots         \\
HDE~303308    &  $78.8\pm0.7$  & 1379  &  $8.5\pm1.9$  & $14.32\pm0.05$  & \ldots        & \ldots          & \ldots         \\
HDE~308813    &  $76.0\pm2.1$  & 1379  &  $7.8\pm2.0$  & $14.30\pm0.05$  & \ldots        & \ldots          & \ldots         \\
BD$+$35~4258  &  $66.5\pm2.0$  & 1379  &  $4.4\pm2.0$  & $14.02\pm0.06$  & $26.3\pm4.1$  & $14.35\pm0.06$  & $14.52\pm0.05$ \\
BD$+$53~2820  & $104.1\pm2.7$\phn  & 1379  &  $3.6\pm3.3$  & $13.91\pm0.04$  & \ldots        & \ldots          & \ldots         \\
CPD$-$59~2603 &  $58.9\pm0.7$  & 1379  &  $8.5\pm1.3$  & $14.33\pm0.05$  & \ldots        & \ldots          & \ldots         \\
CPD$-$59~4552 &  $82.9\pm1.4$  &       & \ldots        & $13.95\pm0.04$  & \ldots        & \ldots          & \ldots         \\
CPD$-$69~1743 &  $40.1\pm2.1$  & 1379  & $<7.1$\tablenotemark{c} & $13.53\pm0.04$  & \ldots        & \ldots          & \ldots         \\
\enddata
\tablenotetext{a}{Identification and equivalent width of the weak Cl~{\sc i} line used to constrain a simultaneous profile synthesis fit with the $\lambda1347$ line.}
\tablenotetext{b}{A small upward correction has been made to the Cl~{\sc ii} column density in this direction to account for blending between the Cl~{\sc ii} profile and the nearby H$_2$ feature.}
\tablenotetext{c}{A $3\sigma$ upper limit is provided because the equivalent width determined in the fit is less than the associated uncertainty.}
\tablenotetext{d}{Data on Cl~{\sc i}~$\lambda1097$ and Cl~{\sc ii}~$\lambda1071$ toward HD~147933 ($\rho$~Oph~A) are obtained from Copernicus observations.}
\end{deluxetable*}

\begin{figure*}
\centering
\includegraphics[width=0.44\textwidth]{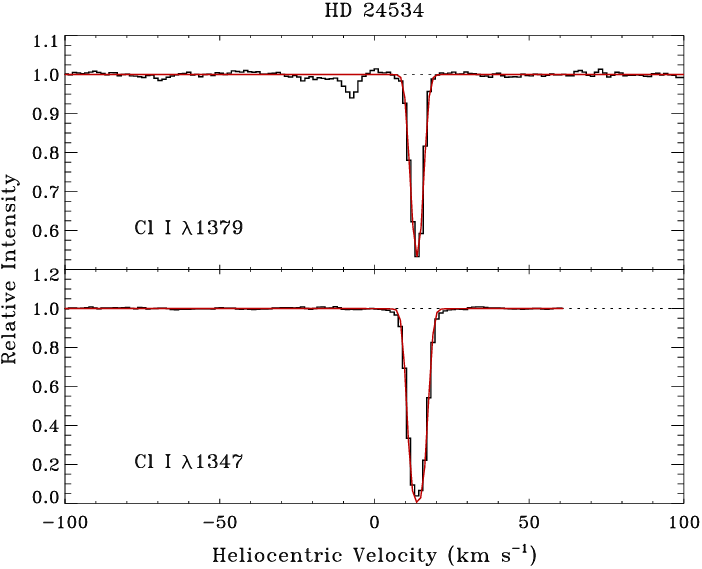}
\includegraphics[width=0.44\textwidth]{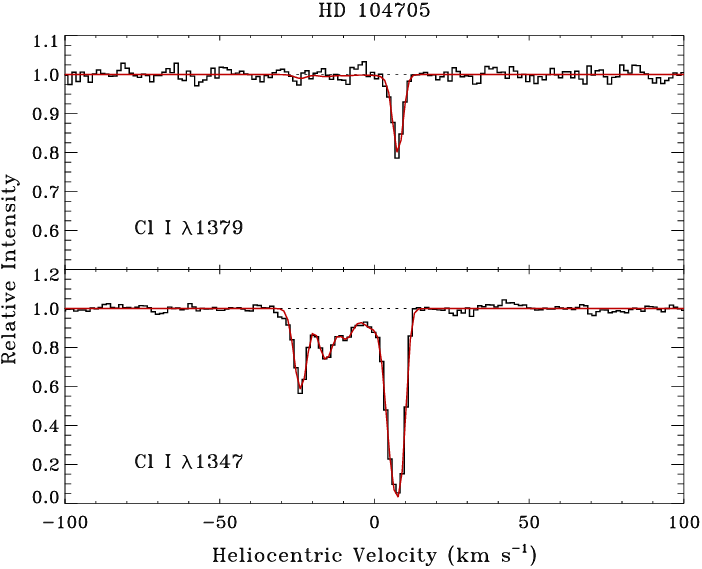}
\caption{High-resolution STIS spectra in the vicinity of the Cl~{\sc i}~$\lambda1379$ and $\lambda1347$ lines toward HD~24534 (left panels) and HD~104705 (right panels). Solid red lines represent simultaneous profile synthesis fits to the two Cl~{\sc i} lines in each direction.\label{fig:chlor_fit1}}
\end{figure*}

\begin{figure*}
\centering
\includegraphics[width=0.44\textwidth]{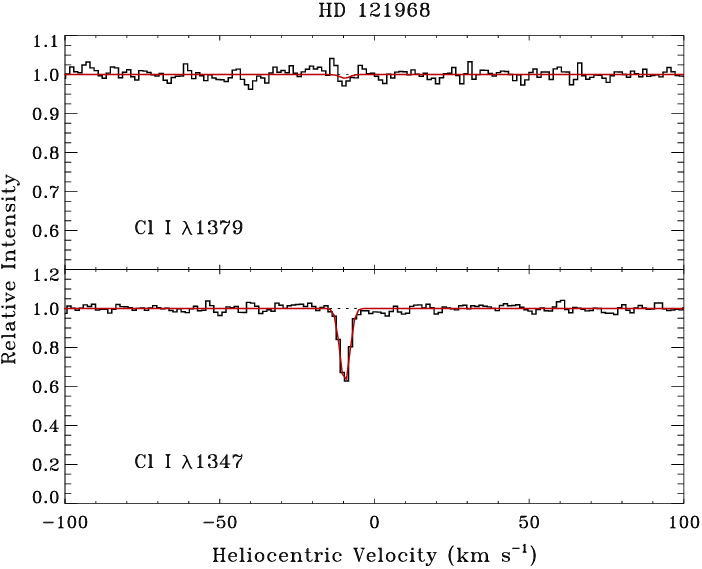}
\includegraphics[width=0.44\textwidth]{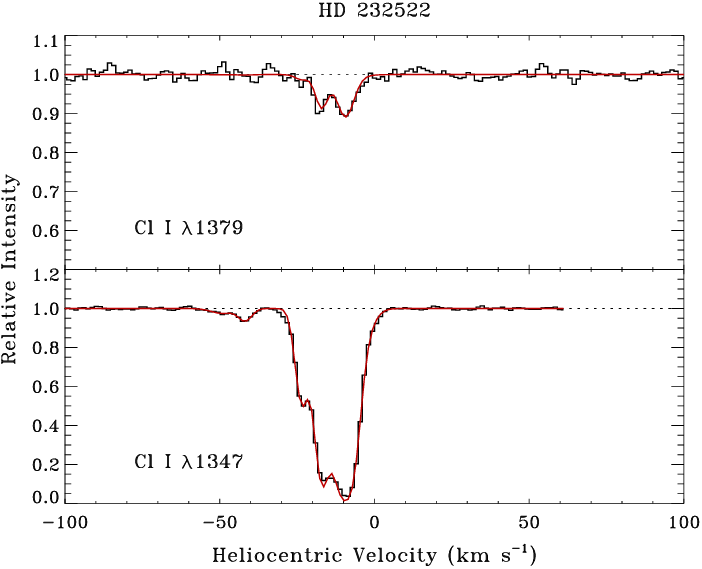}
\caption{Same as Figure~4 except toward HD~121968 (left panels) and HD~232522 (right panels).\label{fig:chlor_fit2}}
\end{figure*}

Our Cl~{\sc i} fits employed the same profile fitting routine as was used to fit the P~{\sc ii} lines, except that the two Cl~{\sc i} absorption profiles were fitted simultaneously. This means that the column densities, $b$-values, and velocities of the individual components included in the fits to the two profiles are necessarily identical, but the best fitting values of the parameters are still determined iteratively through an rms-minimizing approach. The equivalent widths and total Cl~{\sc i} column densities derived through our profile synthesis fits are provided in Table~\ref{tab:chlorine} for a total of 98 sight lines. Examples of our simultaneous fits to the Cl~{\sc i} $\lambda1379$ and $\lambda1347$ lines are presented in Figures~\ref{fig:chlor_fit1} and \ref{fig:chlor_fit2}. In a few cases, the $\lambda1379$ line, although included in the simultaneous fit, is not significantly detected (i.e., the derived equivalent width is smaller than the associated uncertainty). An example of this is provided by the line of sight to HD~121968 (Figure~\ref{fig:chlor_fit2}). In these situations, we list $3\sigma$ upper limits to the equivalent width of the $\lambda1379$ line in Table~\ref{tab:chlorine}.

\begin{figure*}
\centering
\includegraphics[width=0.44\textwidth]{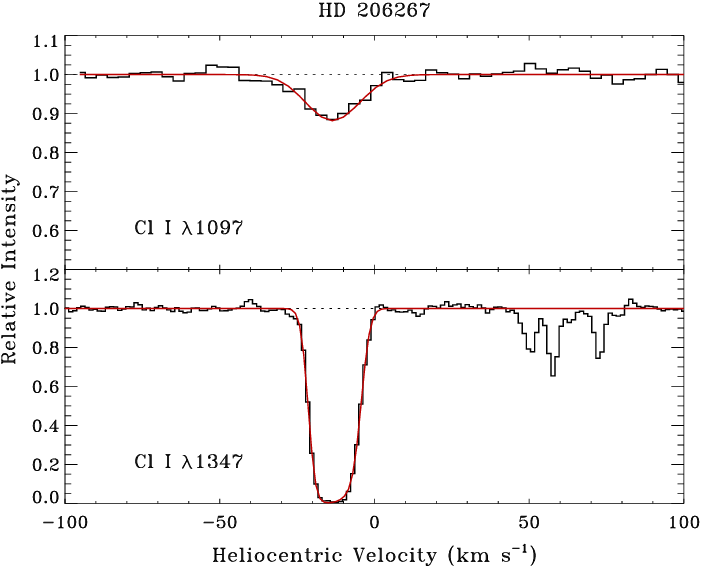}
\includegraphics[width=0.44\textwidth]{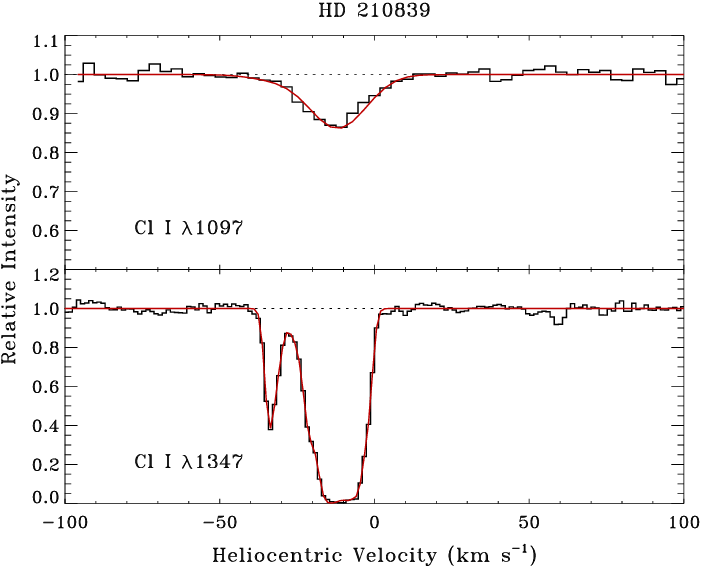}
\caption{Simultaneous profile synthesis fits to the Cl~{\sc i}~$\lambda1347$ and $\lambda1097$ lines toward HD~206267 (left panels) and HD~210839 (right panels). The $\lambda1347$ data are from high-resolution STIS spectra, while the $\lambda1097$ spectra are from FUSE observations.\label{fig:chlor_fit3}}
\end{figure*}

There are 19 sight lines included in Table~\ref{tab:chlorine} for which the Cl~{\sc i} $\lambda1379$ line is not available in the archival STIS data, but the $\lambda1347$ feature is weak enough that optical depth effects should not be a major concern. In each of these cases, the relative intensity of the $\lambda1347$ line does not drop below $\sim$0.05 and the strongest absorption component retains a Gaussian shape. In other words, the shape of the profile does not appear to be distorted by saturation in the line core. For these 19 sight lines, a profile synthesis fit to the $\lambda1347$ line alone yielded the total Cl~{\sc i} column density.

There are also a significant number of sight lines where there are no observations covering the Cl~{\sc i} $\lambda1379$ line and where the $\lambda1347$ line is much too strong to fit on its own. For many of these sight lines, we use observations of the Cl~{\sc i} $\lambda1004$, $\lambda1094$, or $\lambda1097$ transition, available from FUSE spectra, to help constrain the Cl~{\sc i} column density. Experimentally determined $f$-values for the Cl~{\sc i} transitions at 1004.7~\AA{} and 1094.8~\AA{} were recently reported by \citet{a19}. These $f$-values, from beam-foil experiments, are in good agreement with the empirical $f$-values derived for these transitions by \citet{s06}. A secure experimental $f$-value is also available for the Cl~{\sc i} $\lambda1097$ transition \citep{s93}.

The Cl~{\sc i} $\lambda1097$ transition is the weakest of the three FUSE transitions mentioned above. We therefore prefer to use this transition to constrain the Cl~{\sc i} column density provided that the line can be reliably measured. In cases where the $\lambda1097$ line is too weak to detect or is affected by noise in the spectrum, we chose one of the other two lines instead, whichever appeared to provide the most reliable measurement. We then performed a simultaneous profile synthesis fit using the $\lambda1347$ line from STIS data and whichever FUSE transition was chosen to constrain the total column density. Since the FUSE wavelength scale is known to be rather poorly calibrated, we first corrected the velocity scale of the FUSE spectrum so that the centroid velocity of the Cl~{\sc i} absorption feature was consistent with the (weighted mean) velocity of the $\lambda1347$ line before proceeding with the simultaneous fit. The velocity shift applied to the FUSE spectrum was calculated either directly from the Cl~{\sc i} $\lambda1347$ line or from the S~{\sc i} $\lambda1295$ transition if the $\lambda1347$ line was too heavily saturated. The S~{\sc i} $\lambda1295$ transition (from high-resolution STIS spectra) also provided initial values for the component parameters (i.e., the relative velocities, $b$-values, and component fractions) in cases where the Cl~{\sc i} $\lambda 1347$ line was extremely optically thick. Examples of our simultaneous profile synthesis fits to the Cl~{\sc i} $\lambda1347$ line from STIS spectra and the $\lambda1097$ feature from FUSE data are provided in Figure~\ref{fig:chlor_fit3}. The resulting total Cl~{\sc i} column densities are included in Table~\ref{tab:chlorine}.

\subsubsection{Singly Ionized Chlorine\label{subsubsec:ionized}}
Analysis of the Cl~{\sc ii} $\lambda1071$ feature, obtained from FUSE observations, is complicated by the presence of a nearby H$_2$ absorption line (as described in Section~\ref{subsec:fuse}). The velocity separation between the Cl~{\sc ii} $\lambda1071$ line and the H$_2$ (3$-$0) $R$(4) line is only 38~km~s$^{-1}$, with the H$_2$ feature positioned to the blue of the Cl~{\sc ii} line. Thus, if the Cl~{\sc ii} absorption profile includes components that are blueshifted by $\sim$40~km~s$^{-1}$ (relative to the main absorption component), then the Cl~{\sc ii} profile will be severely blended with H$_2$ absorption from the main component. Unfortunately, many of the sight lines in our Cl~{\sc i} sample show components that are blueshifted with respect to the main absorption complex (both in the Cl~{\sc i} $\lambda1347$ line and in lines from dominant ions, such as P~{\sc ii} $\lambda1301$). Typically, sight lines that exhibit multiple absorption complexes at different velocities are those that probe material in a distant spiral arm (such as the Sagittarius-Carina spiral arm or the Perseus spiral arm). In these cases, the absorption complexes at relatively low velocity correspond to material in the ``local arm'', whereas the absorption complexes displaced to negative velocities correspond to material in the more distant arm. (One example of this is provided by the Cl~{\sc i} $\lambda1347$ line toward HD~104705 shown in Figure~\ref{fig:chlor_fit1}. The absorption near $v_{\sun}\approx-25$~km~s$^{-1}$ is associated with diffuse molecular gas in the Sagittarius-Carina spiral arm.)

We do not attempt to derive Cl~{\sc ii} column densities along any sight lines where it is likely that the Cl~{\sc ii} absorption profile is severely blended with H$_2$ absorption. In order to determine whether a given sight line exhibits blended absorption, we examined both the Cl~{\sc i} $\lambda1347$ profile and the P~{\sc ii} $\lambda1301$ and/or $\lambda1532$ profile. (The H$_2$ absorption probably closely follows the Cl~{\sc i} absorption, while the Cl~{\sc ii} profile should be more similar to that of P~{\sc ii}.) We looked for sight lines where most of the absorption was positioned at velocities less than $\sim$20~km~s$^{-1}$ relative to the deepest part of the absorption profile. Sight lines with prominent absorption complexes located more than $\sim$20~km~s$^{-1}$ from the main component were rejected.

For sight lines that passed this initial inspection, we still needed to ``deblend'' the Cl~{\sc ii} and H$_2$ absorption lines. Due to the low resolution of the FUSE spectrograph ($\sim$18~km~s$^{-1}$), the wings of the Cl~{\sc ii} and H$_2$ lines will be blended even if the intrinsic spread in absorption is rather narrow. To deblend the two features, we started with a simple Voigt profile fit consisting of two components, one component for the H$_2$ line and one component for the Cl~{\sc ii} feature. We then used the parameters derived for the H$_2$ component to create a synthetic spectrum that was divided into the observed spectrum, thereby removing the absorption associated with H$_2$. Any remaining absorption was then attributed to Cl~{\sc ii} $\lambda1071$.

We then proceeded to fit the Cl~{\sc ii} $\lambda1071$ feature with our usual profile fitting routine. However, because the low resolution of the FUSE spectra makes it impossible to discern individual components, we adopted a profile template for the component structure of the Cl~{\sc ii} $\lambda1071$ line based on either the P~{\sc ii} $\lambda1532$ line or the Ge~{\sc ii} $\lambda1237$ feature. Both of these lines (which are available from STIS observations) are expected to have an equivalent width that is comparable to (but somewhat larger than) that of Cl~{\sc ii} $\lambda1071$. Given the difference in the cosmic abundances of P and Cl, and taking into account the different wavelengths and $f$-values of the two transitions, the expected equivalent width ratio between the $\lambda1071$ and $\lambda1532$ lines is $W_{1071}/W_{1532}\approx0.6$ (assuming optically thin absorption and no differences in depletion). Similarly, the expected ratio between the Cl~{\sc ii} $\lambda1071$ and Ge~{\sc ii} $\lambda1237$ lines is $W_{1071}/W_{1237}\approx0.5$. By using P~{\sc ii} $\lambda1532$ or Ge~{\sc ii} $\lambda1237$ as a template, we can therefore be reasonably well assured that our fits to the Cl~{\sc ii} $\lambda1071$ feature include all of the absorption components that are likely to be detectable.

\begin{figure*}
\centering
\includegraphics[width=0.44\textwidth]{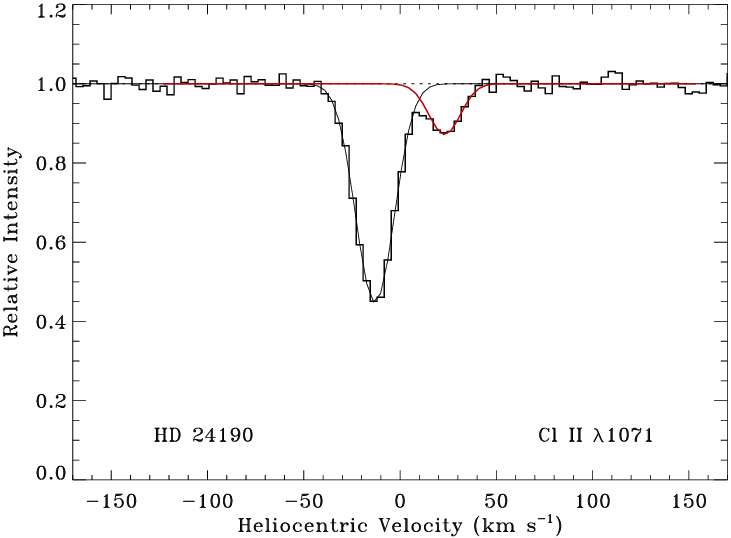}
\includegraphics[width=0.44\textwidth]{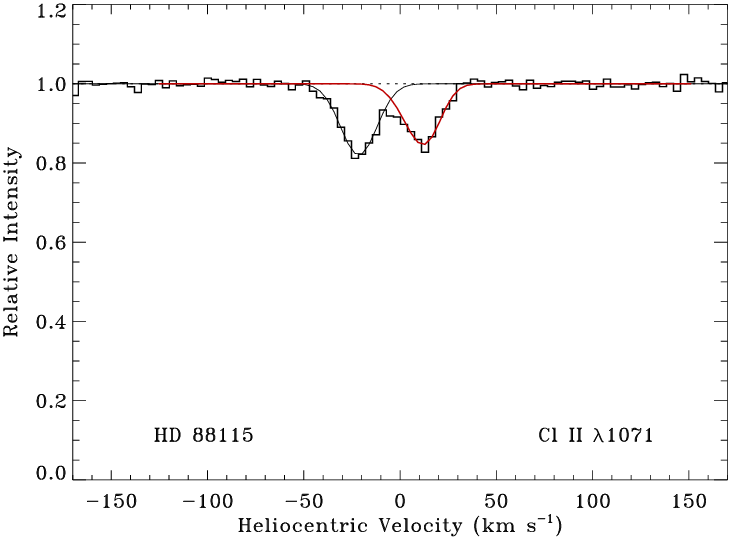}
\caption{FUSE spectra in the vicinity of the Cl~{\sc ii}~$\lambda1071$ line toward HD~24190 (left panel) and HD~88115 (right panel). Solid red lines represent profile synthesis fits to the Cl~{\sc ii} feature. The smooth black curves show the Voigt profile fits used to deblend and remove the H$_2$ lines.\label{fig:cl2_fit1}}
\end{figure*}

\begin{figure*}
\centering
\includegraphics[width=0.44\textwidth]{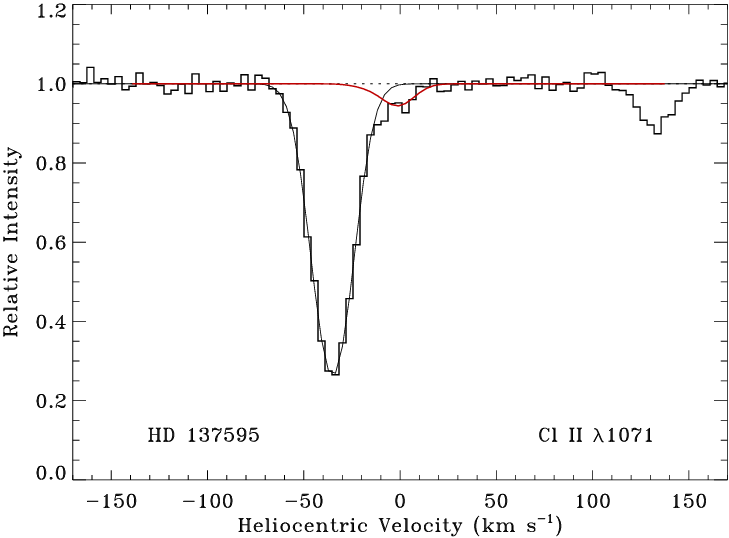}
\includegraphics[width=0.44\textwidth]{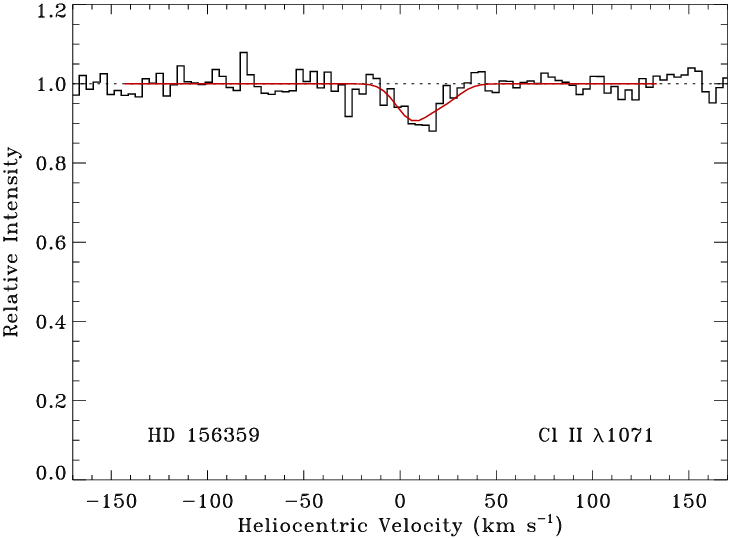}
\caption{Same as Figure~7 except toward HD~137595 (left panel) and HD~156359 (right panel). Note that the H$_2$ (3$-$0) $R$(4) line is not detected toward HD~156359.\label{fig:cl2_fit2}}
\end{figure*}

\begin{figure*}
\centering
\includegraphics[width=0.44\textwidth]{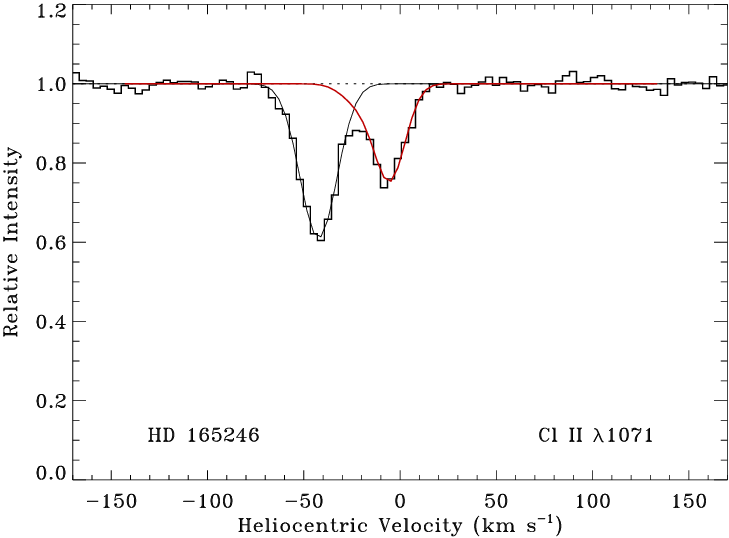}
\includegraphics[width=0.44\textwidth]{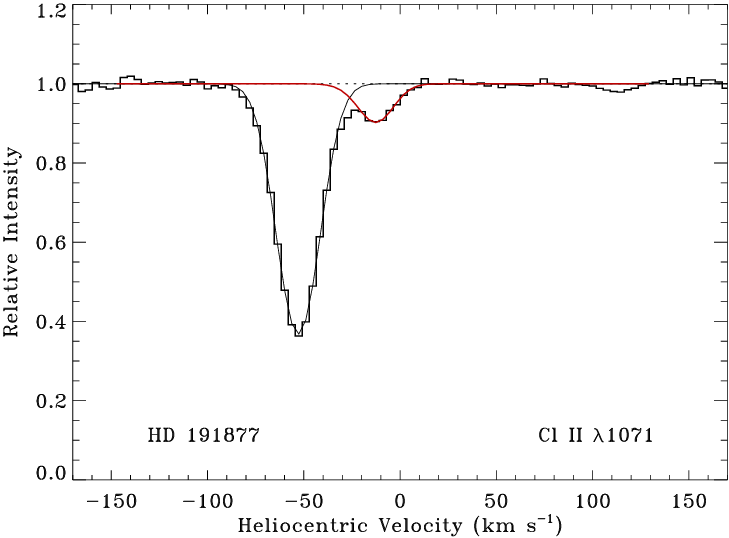}
\caption{Same as Figure~7 except toward HD~165246 (left panel) and HD~191877 (right panel).\label{fig:cl2_fit3}}
\end{figure*}

For many of our sight lines, profile fitting results were already available for P~{\sc ii} $\lambda1532$. Thus, these served as the basis for fitting the Cl~{\sc ii} $\lambda1071$ feature. In cases where observations of the P~{\sc ii} $\lambda1532$ line were not available, the Ge~{\sc ii} $\lambda1237$ line was used instead. Component structures for Ge~{\sc ii} were obtained either from new analyses of the Ge~{\sc ii} $\lambda1237$ line or from the component results published in \citet{r18}. The relative velocities, $b$-values, and component fractions (obtained from P~{\sc ii} or Ge~{\sc ii}) were held fixed in the fit to the Cl~{\sc ii} $\lambda1071$ line. The only free parameters were the total Cl~{\sc ii} column density and an overall velocity offset. The equivalent widths and total Cl~{\sc ii} column densities derived from these fits are provided in Table~\ref{tab:chlorine}. Examples of our fits to the H$_2$ and Cl~{\sc ii} features near 1071~\AA{} are presented in Figures~\ref{fig:cl2_fit1}--\ref{fig:cl2_fit3}. In these figures, the smooth black curve represents the Voigt profile fit used to deblend and remove the H$_2$ line, while the red curve indicates the final profile synthesis fit to the Cl~{\sc ii} feature. Total chlorine column densities, where $N({\rm Cl}_{\rm tot})=N({\rm Cl~\textsc{i}})+N({\rm Cl~\textsc{ii}})$, are also provided in Table~\ref{tab:chlorine}.

There are five sight lines where a relatively weak negative velocity absorption component (or group of components) appears in the P~{\sc ii} or Ge~{\sc ii} absorption profile (at a velocity more than $\sim$20~km~s$^{-1}$ from the main component), but the column density associated with the blueshifted absorption is no more than 20\% of the total line-of-sight column density. These cases were initially rejected for exhibiting blended H$_2$ and Cl~{\sc ii} absorption. Ultimately, however, we decided to fit the unblended portion of the Cl~{\sc ii} absorption and apply a correction to the resulting column density to account for the ``missing'' absorption. The magnitude of the correction factor was determined from the fractional column density of the P~{\sc ii} or Ge~{\sc ii} components not included in the profile fit. These correction factors amounted to upward revisions in the total Cl~{\sc ii} column density of 0.08 dex for HD~88115, 0.10 dex for HD~99857, 0.06 dex for HD~116852, 0.02 dex for HD~165246, and 0.05 dex for HD~210839. In most of these cases, the applied correction is comparable to the uncertainty in the Cl~{\sc ii} column density obtained from the profile fit.

For six sight lines (HD~23478, HD~144965, HD~178487, HD~179407, HD~185418, and HD~203938), we are unable to derive a Cl~{\sc ii} column density, not because the Cl~{\sc ii}~$\lambda1071$ feature is blended with H$_2$ absorption, but because the Cl~{\sc ii} line is not significantly detected. These may be cases where nearly all of the chlorine is in neutral form. We therefore calculated 3$\sigma$ upper limits to the equivalent width and column density of the Cl~{\sc ii}~$\lambda1071$ feature based on an assumed component structure (from the observed P~{\sc ii} line). These limits are provided in Table~\ref{tab:chlorine}. Unfortunately, none of the derived Cl~{\sc ii} upper limits are low enough that the Cl~{\sc ii} column density can be neglected in deriving a total chlorine abundance. We therefore do not provide total Cl column densities for these sight lines.

\subsection{Atomic and Molecular Hydrogen Column Densities\label{subsec:hydrogen}}
\begin{figure}
\centering
\includegraphics[width=0.44\textwidth]{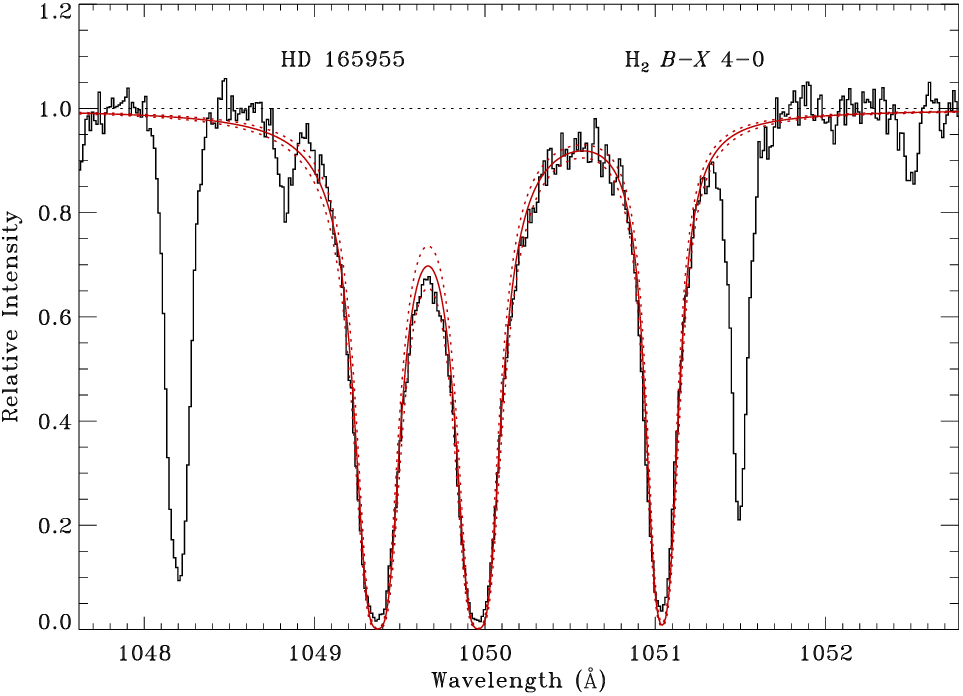}
\caption{Determination of $N({\rm H}_2)$ toward HD~165955. A synthetic spectrum (red curve) is compared to observations of the $R$(0), $R$(1), and $P$(1) lines of the $B$$-$$X$ (4$-$0) band. The solid red curve corresponds to a total H$_2$ column density of $\log N({\rm H}_2)=19.09$, while the dashed red curves indicate variations in the column density of $\pm0.07$ dex.\label{fig:h2_fit}}
\end{figure}

Column densities of atomic and molecular hydrogen for the sight lines in our combined phosphorus and chlorine sample were obtained either from \citet{j19} or from the values tabulated in \citet{j09}. These values are provided in Table~\ref{tab:hydrogen} along with the total hydrogen column densities, $N({\rm H}_{\rm tot})=N({\rm H~\textsc{i}})+2N({\rm H}_2)$, and molecular hydrogen fractions, $f({\rm H}_2)=2N({\rm H}_2)/N({\rm H}_{\rm tot})$. For the line of sight to HD~165955, \citet{j09} reports a molecular hydrogen column density of $\log N({\rm H}_2)=16.53\pm0.04$. This value was originally derived by \citet{c04}. However, based on the Cl~{\sc i} column density we obtain for this sight line, $\log N({\rm Cl~\textsc{i}})=13.23\pm0.03$, and the correlations discussed in the next section, the reported value of $N({\rm H}_2)$ toward HD~165955 appears to be much too low. We therefore undertook our own analysis of the FUSE data in this direction to determine a more accurate value for the molecular hydrogen column density. From an analysis of the $J=0$ and 1 lines of the H$_2$ (1$-$0), (2$-$0), (3$-$0), and (4$-$0) bands\footnote{For the analysis of H$_2$ absorption toward HD~165955, we used the IDL-based program H2GUI, which was originally created and used by \citet{t02}.}, we find $\log N({\rm H}_2)=19.09\pm0.07$. A representative fit to the (4$-$0) band is presented in Figure~\ref{fig:h2_fit}.

\startlongtable
\begin{deluxetable*}{lcccccc}
\tablecolumns{7}
\tabletypesize{\small}
\tablecaption{Atomic and Molecular Hydrogen Column Densities\label{tab:hydrogen}}
\tablehead{ \colhead{Star} & \colhead{log~$N$(H~{\sc i})} & \colhead{log~$N$(H$_2$)} & \colhead{log~$N$(H$_{\rm tot}$)} & \colhead{log~$f$(H$_2$)} & \colhead{$F_*$\tablenotemark{a}} & \colhead{Ref.} }
\startdata
HD~108          & $21.38\pm0.05$   & $20.45\pm0.08$   & $21.47^{+0.04}_{-0.05}$   & $-0.72\pm0.09$   & $0.53\pm0.04$   & 1 \\
HD~1383         & $21.46\pm0.05$   & $20.49\pm0.07$   & $21.54^{+0.04}_{-0.05}$   & $-0.75\pm0.08$   & $0.48\pm0.05$   & 1 \\
HD~3827         & $20.55\pm0.07$   & $18.35\pm0.25$   & $20.56^{+0.07}_{-0.08}$   & $-1.90\pm0.25$   & $0.41\pm0.07$   & 1 \\
HD~12323        & $21.19\pm0.04$   & $20.26\pm0.08$   & $21.28^{+0.04}_{-0.04}$   & $-0.72\pm0.09$   & $0.51\pm0.05$   & 1 \\
HD~13268        & $21.34\pm0.07$   & $20.46\pm0.08$   & $21.44^{+0.06}_{-0.07}$   & $-0.68\pm0.10$   & $0.45\pm0.07$   & 1 \\
HD~13745        & $21.34\pm0.05$   & $20.53\pm0.07$   & $21.46^{+0.04}_{-0.05}$   & $-0.63\pm0.08$   & $0.55\pm0.05$   & 1 \\
HD~14434        & $21.37\pm0.09$   & $20.47\pm0.07$   & $21.47^{+0.07}_{-0.09}$   & $-0.70\pm0.10$   & $0.52\pm0.04$   & 2 \\
HD~15137        & $21.24\pm0.08$   & $20.23\pm0.08$   & $21.32^{+0.07}_{-0.08}$   & $-0.79\pm0.10$   & $0.36\pm0.06$   & 1 \\
HD~23478        & $20.71\pm0.17$   & $20.48\pm0.07$   & $21.05^{+0.09}_{-0.12}$   & $-0.27\pm0.11$   & $0.00\pm0.50$   & 2 \\
HD~24190        & $21.18\pm0.06$   & $20.38\pm0.07$   & $21.30^{+0.05}_{-0.06}$   & $-0.62\pm0.08$   & $0.63\pm0.24$   & 2 \\
HD~24534        & $20.73\pm0.06$   & $20.92\pm0.04$   & $21.34^{+0.03}_{-0.04}$   & $-0.12\pm0.05$   & $0.90\pm0.06$   & 2 \\
HD~25443        & $21.29\pm0.06$   & $20.92\pm0.08$   & $21.56^{+0.05}_{-0.06}$   & $-0.34\pm0.09$   & $0.72\pm0.05$   & 1 \\
HD~37903        & $21.12\pm0.10$   & $20.85\pm0.07$   & $21.44^{+0.06}_{-0.07}$   & $-0.29\pm0.09$   & $1.15\pm0.03$   & 2 \\
HD~41161        & $21.09\pm0.05$   & $19.99\pm0.09$   & $21.15^{+0.05}_{-0.05}$   & $-0.86\pm0.10$   & $0.35\pm0.04$   & 1 \\
HD~46223        & $21.46\pm0.04$   & $20.67\pm0.06$   & $21.58^{+0.03}_{-0.04}$   & $-0.61\pm0.07$   & $0.86\pm0.03$   & 1 \\
HD~52266        & $21.22\pm0.04$   & $20.00\pm0.08$   & $21.27^{+0.04}_{-0.04}$   & $-0.97\pm0.09$   & $0.56\pm0.05$   & 1 \\
HD~53975        & $21.08\pm0.04$   & $19.15\pm0.09$   & $21.09^{+0.04}_{-0.04}$   & $-1.64\pm0.10$   & $0.38\pm0.03$   & 1 \\
HD~63005        & $21.24\pm0.03$   & $20.17\pm0.09$   & $21.31^{+0.03}_{-0.03}$   & $-0.84\pm0.09$   & $0.68\pm0.04$   & 1 \\
HD~66788        & $21.23\pm0.04$   & $19.72\pm0.14$   & $21.26^{+0.04}_{-0.04}$   & $-1.24\pm0.14$   & $0.55\pm0.04$   & 1 \\
HD~72754        & $21.17\pm0.12$   & $20.35\pm0.10$   & $21.28^{+0.10}_{-0.13}$   & $-0.63\pm0.13$   & $0.76\pm0.10$   & 2 \\
HD~73882        & $21.11\pm0.11$   & $21.08\pm0.10$   & $21.57^{+0.08}_{-0.09}$   & $-0.19\pm0.12$   & $0.68\pm0.07$   & 2 \\
HD~75309        & $21.10\pm0.03$   & $20.16\pm0.06$   & $21.19^{+0.03}_{-0.03}$   & $-0.73\pm0.07$   & $0.58\pm0.04$   & 1 \\
HD~79186        & $21.18\pm0.09$   & $20.72\pm0.09$   & $21.41^{+0.07}_{-0.08}$   & $-0.39\pm0.11$   & $0.69\pm0.03$   & 2 \\
HD~88115        & $21.03\pm0.06$   & $19.25\pm0.14$   & $21.04^{+0.06}_{-0.07}$   & $-1.49\pm0.15$   & $0.39\pm0.08$   & 1 \\
HD~89137        & $21.03\pm0.07$   & $20.02\pm0.09$   & $21.11^{+0.06}_{-0.07}$   & $-0.79\pm0.11$   & $0.45\pm0.04$   & 1 \\
HD~90087        & $21.19\pm0.05$   & $19.88\pm0.07$   & $21.23^{+0.05}_{-0.05}$   & $-1.05\pm0.08$   & $0.36\pm0.07$   & 1 \\
HD~91597        & $21.40\pm0.06$   & $19.70\pm0.05$   & $21.42^{+0.06}_{-0.07}$   & $-1.42\pm0.07$   & $0.44\pm0.05$   & 2 \\
HD~91651        & $21.15\pm0.06$   & $19.07\pm0.03$   & $21.16^{+0.06}_{-0.07}$   & $-1.79\pm0.07$   & $0.27\pm0.04$   & 2 \\
HD~91824        & $21.12\pm0.04$   & $19.81\pm0.09$   & $21.16^{+0.04}_{-0.04}$   & $-1.05\pm0.10$   & $0.33\pm0.04$   & 1 \\
HD~91983        & $21.15\pm0.06$   & $20.10\pm0.07$   & $21.22^{+0.05}_{-0.06}$   & $-0.82\pm0.09$   & $0.31\pm0.05$   & 1 \\
HD~92554        & $21.34\pm0.09$   & $19.24\pm0.07$   & $21.35^{+0.09}_{-0.11}$   & $-1.81\pm0.11$   & $0.15\pm0.10$   & 1 \\
HD~93129        & $21.47\pm0.07$   & $20.21\pm0.07$   & $21.52^{+0.06}_{-0.07}$   & $-1.00\pm0.09$   & $0.49\pm0.05$   & 1 \\
HD~93205        & $21.36\pm0.05$   & $19.73\pm0.12$   & $21.38^{+0.05}_{-0.05}$   & $-1.35\pm0.13$   & $0.33\pm0.05$   & 1 \\
HD~93222        & $21.47\pm0.03$   & $19.77\pm0.09$   & $21.49^{+0.03}_{-0.03}$   & $-1.42\pm0.09$   & $0.33\pm0.04$   & 1 \\
HD~93843        & $21.30\pm0.05$   & $19.62\pm0.10$   & $21.32^{+0.05}_{-0.05}$   & $-1.40\pm0.11$   & $0.50\pm0.04$   & 1 \\
HD~94493        & $21.10\pm0.06$   & $20.09\pm0.06$   & $21.18^{+0.05}_{-0.06}$   & $-0.79\pm0.08$   & $0.32\pm0.05$   & 1 \\
HD~97175        & $20.96\pm0.06$   & $20.10\pm0.09$   & $21.07^{+0.05}_{-0.06}$   & $-0.66\pm0.10$   & $0.49\pm0.05$   & 1 \\
HD~99857        & $21.27\pm0.07$   & $20.30\pm0.05$   & $21.35^{+0.06}_{-0.07}$   & $-0.75\pm0.08$   & $0.45\pm0.05$   & 1 \\
HD~99890        & $21.12\pm0.05$   & $19.56\pm0.09$   & $21.14^{+0.05}_{-0.05}$   & $-1.28\pm0.10$   & $0.16\pm0.05$   & 1 \\
HD~100199       & $21.18\pm0.06$   & $20.15\pm0.08$   & $21.25^{+0.05}_{-0.06}$   & $-0.80\pm0.09$   & $0.41\pm0.09$   & 1 \\
HD~101190       & $21.24\pm0.03$   & $20.43\pm0.04$   & $21.36^{+0.02}_{-0.03}$   & $-0.63\pm0.05$   & $0.51\pm0.05$   & 1 \\
HD~103779       & $21.17\pm0.05$   & $19.90\pm0.08$   & $21.21^{+0.05}_{-0.05}$   & $-1.01\pm0.09$   & $0.29\pm0.05$   & 1 \\
HD~104705       & $21.15\pm0.06$   & $20.02\pm0.07$   & $21.21^{+0.05}_{-0.06}$   & $-0.89\pm0.09$   & $0.40\pm0.05$   & 1 \\
HD~108639       & $21.36\pm0.04$   & $20.01\pm0.10$   & $21.40^{+0.04}_{-0.04}$   & $-1.09\pm0.11$   & $0.34\pm0.04$   & 1 \\
HD~109399       & $21.11\pm0.05$   & $20.01\pm0.09$   & $21.17^{+0.05}_{-0.05}$   & $-0.86\pm0.10$   & $0.44\pm0.05$   & 1 \\
HD~114886       & $21.34\pm0.06$   & $20.30\pm0.07$   & $21.41^{+0.05}_{-0.06}$   & $-0.81\pm0.09$   & $0.47\pm0.05$   & 1 \\
HD~115455       & $21.38\pm0.05$   & $20.55\pm0.07$   & $21.49^{+0.04}_{-0.05}$   & $-0.64\pm0.08$   & $0.41\pm0.06$   & 1 \\
HD~116852       & $20.96\pm0.04$   & $19.75\pm0.09$   & $21.01^{+0.04}_{-0.04}$   & $-0.96\pm0.10$   & $0.48\pm0.04$   & 1 \\
HD~121968       & $20.58\pm0.12$   & $18.70\pm0.10$   & $20.59^{+0.12}_{-0.16}$   & $-1.59\pm0.15$   & $0.26\pm0.06$   & 2 \\
HD~122879       & $21.31\pm0.06$   & $20.33\pm0.07$   & $21.39^{+0.05}_{-0.06}$   & $-0.76\pm0.08$   & $0.51\pm0.04$   & 1 \\
HD~124314       & $21.41\pm0.06$   & $20.41\pm0.09$   & $21.49^{+0.05}_{-0.06}$   & $-0.78\pm0.10$   & $0.47\pm0.05$   & 1 \\
HD~137595       & $20.97\pm0.06$   & $20.59\pm0.06$   & $21.23^{+0.04}_{-0.05}$   & $-0.34\pm0.07$   & $0.85\pm0.05$   & 1 \\
HD~144965       & $20.97\pm0.09$   & $20.74\pm0.07$   & $21.31^{+0.06}_{-0.07}$   & $-0.27\pm0.09$   & $1.07\pm0.09$   & 1 \\
HD~147683       & $21.20\pm0.15$   & $20.68\pm0.12$   & $21.41^{+0.11}_{-0.14}$   & $-0.42\pm0.15$   & $0.56\pm0.47$   & 2 \\
HD~147888       & $21.68\pm0.08$   & $20.45\pm0.05$   & $21.73^{+0.07}_{-0.09}$   & $-0.98\pm0.09$   & $1.15\pm0.12$   & 1 \\
HD~147933       & $21.63\pm0.09$   & $20.57\pm0.15$   & $21.70^{+0.08}_{-0.10}$   & $-0.83\pm0.16$   & $1.09\pm0.08$   & 2 \\
HD~148422       & $21.24\pm0.09$   & $20.15\pm0.11$   & $21.31^{+0.08}_{-0.10}$   & $-0.85\pm0.13$   & $0.21\pm0.09$   & 1 \\
HD~148937       & $21.48\pm0.06$   & $20.68\pm0.06$   & $21.60^{+0.05}_{-0.05}$   & $-0.62\pm0.08$   & $0.56\pm0.07$   & 1 \\
HD~151805       & $21.33\pm0.05$   & $20.35\pm0.07$   & $21.41^{+0.04}_{-0.05}$   & $-0.76\pm0.08$   & $0.54\pm0.04$   & 1 \\
HD~152590       & $21.37\pm0.07$   & $20.47\pm0.10$   & $21.47^{+0.06}_{-0.07}$   & $-0.70\pm0.11$   & $0.52\pm0.05$   & 1 \\
HD~156359       & $20.80\pm0.10$   & $18.07\pm0.21$   & $20.80^{+0.10}_{-0.13}$   & $-2.43\pm0.22$   & $0.24\pm0.09$   & 1 \\
HD~157857       & $21.26\pm0.09$   & $20.68\pm0.10$   & $21.44^{+0.07}_{-0.08}$   & $-0.46\pm0.12$   & $0.62\pm0.04$   & 2 \\
HD~163522       & $21.14\pm0.08$   & $19.66\pm0.19$   & $21.17^{+0.08}_{-0.09}$   & $-1.21\pm0.20$   & $0.34\pm0.08$   & 1 \\
HD~165246       & $21.41\pm0.03$   & $20.14\pm0.07$   & $21.45^{+0.03}_{-0.03}$   & $-1.01\pm0.07$   & $0.78\pm0.04$   & 1 \\
HD~165955       & $21.10\pm0.06$   & $19.09\pm0.07$\tablenotemark{b}   & $21.11^{+0.06}_{-0.07}$   & $-1.72\pm0.09$   & $0.42\pm0.04$   & 2 \\
HD~167402       & $21.13\pm0.05$   & $20.06\pm0.10$   & $21.20^{+0.05}_{-0.05}$   & $-0.84\pm0.11$   & $0.25\pm0.07$   & 1 \\
HD~168076       & $21.65\pm0.23$   & $20.68\pm0.08$   & $21.73^{+0.20}_{-0.37}$   & $-0.75\pm0.21$   & $0.68\pm0.17$   & 2 \\
HD~168941       & $21.18\pm0.05$   & $20.11\pm0.08$   & $21.25^{+0.04}_{-0.05}$   & $-0.84\pm0.09$   & $0.68\pm0.06$   & 1 \\
HD~170740       & $21.09\pm0.06$   & $20.86\pm0.08$   & $21.43^{+0.05}_{-0.06}$   & $-0.27\pm0.09$   & $0.87\pm0.09$   & 1 \\
HD~177989       & $20.99\pm0.05$   & $20.16\pm0.12$   & $21.10^{+0.05}_{-0.05}$   & $-0.64\pm0.13$   & $0.66\pm0.05$   & 1 \\
HD~178487       & $21.22\pm0.04$   & $20.51\pm0.09$   & $21.36^{+0.04}_{-0.04}$   & $-0.55\pm0.10$   & $0.74\pm0.07$   & 1 \\
HD~179407       & $21.20\pm0.06$   & $20.23\pm0.09$   & $21.28^{+0.05}_{-0.06}$   & $-0.75\pm0.10$   & $0.63\pm0.09$   & 1 \\
HD~185418       & $21.19\pm0.05$   & $20.72\pm0.07$   & $21.41^{+0.04}_{-0.05}$   & $-0.39\pm0.08$   & $0.74\pm0.04$   & 1 \\
HD~190918       & $21.38\pm0.06$   & $19.84\pm0.08$   & $21.40^{+0.06}_{-0.07}$   & $-1.26\pm0.10$   & $0.46\pm0.03$   & 2 \\
HD~191877       & $21.03\pm0.05$   & $20.00\pm0.17$   & $21.10^{+0.05}_{-0.06}$   & $-0.80\pm0.17$   & $0.53\pm0.05$   & 1 \\
HD~192035       & $21.20\pm0.04$   & $20.63\pm0.09$   & $21.39^{+0.04}_{-0.05}$   & $-0.46\pm0.10$   & $0.74\pm0.07$   & 1 \\
HD~192639       & $21.29\pm0.09$   & $20.73\pm0.10$   & $21.48^{+0.07}_{-0.08}$   & $-0.45\pm0.12$   & $0.64\pm0.04$   & 2 \\
HD~195455       & $20.61\pm0.04$   & $18.42\pm0.19$   & $20.62^{+0.04}_{-0.04}$   & $-1.89\pm0.19$   & $0.39\pm0.07$   & 1 \\
HD~195965       & $20.92\pm0.05$   & $20.28\pm0.08$   & $21.08^{+0.04}_{-0.05}$   & $-0.50\pm0.09$   & $0.45\pm0.05$   & 1 \\
HD~198478       & $21.32\pm0.12$   & $20.76\pm0.33$   & $21.51^{+0.16}_{-0.26}$   & $-0.45\pm0.35$   & $0.75\pm0.14$   & 1 \\
HD~198781       & $20.93\pm0.07$   & $20.48\pm0.08$   & $21.16^{+0.05}_{-0.06}$   & $-0.38\pm0.09$   & $0.79\pm0.05$   & 1 \\
HD~201345       & $21.00\pm0.05$   & $19.36\pm0.12$   & $21.02^{+0.05}_{-0.05}$   & $-1.36\pm0.13$   & $0.34\pm0.06$   & 1 \\
HD~202347       & $20.83\pm0.08$   & $19.98\pm0.10$   & $20.94^{+0.07}_{-0.08}$   & $-0.66\pm0.12$   & $0.66\pm0.08$   & 1 \\
HD~203374       & $21.20\pm0.05$   & $20.67\pm0.07$   & $21.40^{+0.04}_{-0.05}$   & $-0.43\pm0.08$   & $0.66\pm0.04$   & 1 \\
HD~203938       & $21.48\pm0.15$   & $21.00\pm0.06$   & $21.70^{+0.10}_{-0.13}$   & $-0.40\pm0.11$   & $0.99\pm0.64$   & 2 \\
HD~206267       & $21.22\pm0.06$   & $20.85\pm0.08$   & $21.49^{+0.05}_{-0.06}$   & $-0.34\pm0.09$   & $0.80\pm0.05$   & 1 \\
HD~206773       & $21.09\pm0.07$   & $20.41\pm0.06$   & $21.24^{+0.05}_{-0.06}$   & $-0.53\pm0.08$   & $0.60\pm0.04$   & 1 \\
HD~207198       & $21.28\pm0.07$   & $20.79\pm0.05$   & $21.50^{+0.05}_{-0.05}$   & $-0.41\pm0.07$   & $0.80\pm0.05$   & 1 \\
HD~207308       & $21.20\pm0.06$   & $20.80\pm0.06$   & $21.45^{+0.04}_{-0.05}$   & $-0.35\pm0.07$   & $0.82\pm0.05$   & 1 \\
HD~207538       & $21.27\pm0.06$   & $20.85\pm0.06$   & $21.52^{+0.04}_{-0.05}$   & $-0.36\pm0.07$   & $0.85\pm0.05$   & 1 \\
HD~208440       & $21.24\pm0.06$   & $20.28\pm0.11$   & $21.33^{+0.05}_{-0.06}$   & $-0.75\pm0.12$   & $0.65\pm0.05$   & 1 \\
HD~209339       & $21.20\pm0.04$   & $20.13\pm0.06$   & $21.27^{+0.04}_{-0.04}$   & $-0.84\pm0.07$   & $0.50\pm0.05$   & 1 \\
HD~210809       & $21.31\pm0.06$   & $19.97\pm0.12$   & $21.35^{+0.06}_{-0.06}$   & $-1.08\pm0.13$   & $0.31\pm0.06$   & 1 \\
HD~210839       & $21.24\pm0.05$   & $20.80\pm0.05$   & $21.48^{+0.04}_{-0.04}$   & $-0.38\pm0.06$   & $0.78\pm0.04$   & 1 \\
HD~212791       & $21.11\pm0.12$   & $19.42\pm0.11$   & $21.13^{+0.12}_{-0.16}$   & $-1.41\pm0.15$   & $0.57\pm0.08$   & 2 \\
HD~218915       & $21.20\pm0.07$   & $20.16\pm0.06$   & $21.27^{+0.06}_{-0.07}$   & $-0.81\pm0.08$   & $0.39\pm0.06$   & 1 \\
HD~219188       & $20.72\pm0.07$   & $19.38\pm0.11$   & $20.76^{+0.07}_{-0.08}$   & $-1.08\pm0.12$   & $0.37\pm0.06$   & 1 \\
HD~220057       & $20.95\pm0.14$   & $20.27\pm0.09$   & $21.10^{+0.11}_{-0.14}$   & $-0.53\pm0.13$   & $0.69\pm0.13$   & 1 \\
HD~224151       & $21.35\pm0.05$   & $20.55\pm0.06$   & $21.47^{+0.04}_{-0.05}$   & $-0.62\pm0.07$   & $0.49\pm0.05$   & 1 \\
HDE~232522      & $21.12\pm0.04$   & $20.18\pm0.08$   & $21.21^{+0.04}_{-0.04}$   & $-0.73\pm0.09$   & $0.41\pm0.04$   & 1 \\
HDE~303308      & $21.41\pm0.03$   & $20.20\pm0.05$   & $21.46^{+0.03}_{-0.03}$   & $-0.96\pm0.06$   & $0.42\pm0.05$   & 1 \\
HDE~308813      & $21.20\pm0.06$   & $20.23\pm0.07$   & $21.28^{+0.05}_{-0.06}$   & $-0.75\pm0.08$   & $0.52\pm0.06$   & 1 \\
BD$+$35~4258    & $21.24\pm0.03$   & $19.61\pm0.11$   & $21.26^{+0.03}_{-0.03}$   & $-1.35\pm0.11$   & $0.38\pm0.06$   & 1 \\
BD$+$53~2820    & $21.35\pm0.05$   & $20.10\pm0.11$   & $21.40^{+0.05}_{-0.05}$   & $-1.00\pm0.12$   & $0.44\pm0.07$   & 1 \\
CPD$-$59~2603   & $21.43\pm0.04$   & $20.02\pm0.12$   & $21.46^{+0.04}_{-0.04}$   & $-1.14\pm0.12$   & $0.51\pm0.05$   & 1 \\
CPD$-$59~4552   & $21.28\pm0.05$   & $20.45\pm0.08$   & $21.39^{+0.04}_{-0.05}$   & $-0.64\pm0.09$   & $0.40\pm0.05$   & 1 \\
CPD$-$69~1743   & $21.16\pm0.04$   & $19.85\pm0.15$   & $21.20^{+0.04}_{-0.04}$   & $-1.05\pm0.15$   & $0.40\pm0.07$   & 1 \\
\enddata
\tablenotetext{a}{Sight line depletion factor.}
\tablenotetext{b}{The H$_2$ column density derived by \citet{c04}, and adopted in \citet{j09}, for the line of sight to HD~165955 is too low. The value listed here is from our own analysis of the H$_2$ data in this direction.}
\tablerefs{(1) \citet{j19}, (2) \citet{j09}.}
\end{deluxetable*}

\section{ANALYSIS\label{sec:analysis}}
\subsection{Column Density Correlations\label{subsec:correlations}}
Owing to the unique chemistry that links neutral chlorine to molecular hydrogen, the column densities of Cl~{\sc i} and H$_2$ have long been expected to be well correlated \citep[e.g.,][]{jy78,hb84,m12}. Most recently, \citet{m12}, using Cl measurements from the Copernicus satellite, found a correlation between $N({\rm Cl~\textsc{i}})$ and $N({\rm H}_2)$ that had a slope of $0.92\pm0.19$ (in $\log$-$\log$ space) and a correlation coefficient greater than 0.6. \citet{b15} found that a similar relationship holds for high-redshift damped Lyman-$\alpha$ (DLA) systems where both H$_2$ and Cl~{\sc i} are detected. Other correlations, such as that between $N({\rm Cl}_{\rm tot})$ and $N({\rm H}_{\rm tot})$, have also been noted in the literature \citep{hb84,m12}. In light of our greatly expanded sample of Cl~{\sc i} and Cl~{\sc ii} column densities (Table~\ref{tab:chlorine}), it is worthwhile to revisit these correlations.

\begin{figure*}
\centering
\includegraphics[width=0.75\textwidth]{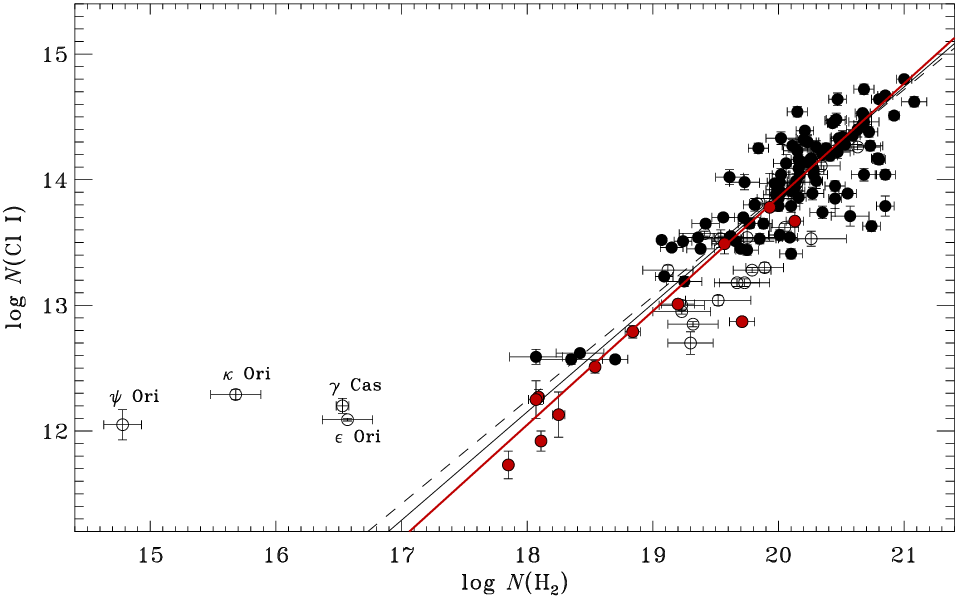}
\caption{Correlation between $\log N$(Cl~{\sc i}) and $\log N$(H$_2$). Solid black circles represent chlorine measurements obtained in this investigation from STIS and FUSE observations; open circles represent measurements obtained in previous studies from Copernicus observations \citep{m12,br15}. Solid red circles represent Cl~{\sc i} and H$_2$ measurements obtained for high-redshift absorption systems toward background quasars \citep{b15}. The dashed black line represents a least-squares linear fit to the STIS+FUSE sample, while the solid black line indicates a linear fit to the entire Galactic sample (excluding the four discrepant sight lines with very low H$_2$ column densities). The solid red line shows a least-squares linear fit to the combined Galactic and extragalactic sample (excluding the four outliers).\label{fig:cl_corr1}}
\end{figure*}

In Figure~\ref{fig:cl_corr1}, we compare our new determinations of neutral chlorine column densities (from STIS and FUSE observations) with the corresponding molecular hydrogen column densities from Table~\ref{tab:hydrogen}. Also included in Figure~\ref{fig:cl_corr1} are the Cl~{\sc i} measurements obtained in previous studies based on Copernicus observations \citep{m12,br15}\footnote{\citet{br15} derived Cl~{\sc i} and Cl~{\sc ii} column densities from Copernicus observations for a sample of sight lines with very low H$_2$ column densities as part of a Masters Thesis at the University of Toledo. For convenience, the Copernicus measurements of \citet{m12} and \citet{br15} are compiled in Appendix~\ref{app:copernicus}.} and the Cl~{\sc i} and H$_2$ column densities derived for high-redshift absorption systems toward background quasars \citep{b15}. A strong correlation between $N({\rm Cl~\textsc{i}})$ and $N({\rm H}_2)$ is found for Galactic sight lines with $\log N({\rm H}_2)\gtrsim18$. Least-squares linear fits\footnote{The least-squares linear fits described in this paper were performed using the IDL procedure FITEXY, which accounts for uncertainties in both the $x$ and $y$ coordinates \citep{p07}.} to the ``STIS+FUSE'' sample and the sample that also includes Copernicus measurements (``all Galactic'') yield linear correlation coefficients of $\sim$0.8 and slopes of $\sim$0.8--0.9 (see Figure~\ref{fig:cl_corr1} and Table~\ref{tab:correlations}), consistent with the results of \citet{m12}. Four sight lines with $\log N({\rm H}_2)<17$ do not follow the trend exhibited by the other Galactic sight lines and are not included in the linear fits.

\begin{deluxetable*}{lcccccc}
\tablecolumns{7}
\tabletypesize{\small}
\tablecaption{Column Density Correlations\label{tab:correlations}}
\tablehead{ \colhead{$Y$} & \colhead{$X$} & \colhead{$A$} & \colhead{$B$} & \colhead{$r$\tablenotemark{a}} & \colhead{$N$\tablenotemark{b}} & \colhead{Sample} }
\startdata
Cl~{\sc i} & H$_2$ & $0.824\pm0.016$ & $-2.59\pm0.31$ & 0.840 & 98 & STIS+FUSE \\
 & & $0.862\pm0.015$ & $-3.37\pm0.31$ & 0.841 & 118 & all Galactic\tablenotemark{c} \\
 & & $0.906\pm0.012$ & $-4.26\pm0.23$ & 0.897 & 130 & Galactic+DLAs\tablenotemark{c} \\
Cl~{\sc i} & H$_{\rm tot}$ & $2.862\pm0.106$ & $-46.96\pm2.25$\phn & 0.731 & 98 & STIS+FUSE \\
 & & $2.478\pm0.061$ & $-38.74\pm1.29$\phn & 0.828 & 119 & all Galactic \\
Cl$_{\rm tot}$ & H$_{\rm tot}$ & $1.065\pm0.057$ & $-8.26\pm1.21$ & 0.845 & 62 & STIS+FUSE \\
  & & $1.030\pm0.022$ & $-7.52\pm0.45$ & 0.931 & 88 & all Galactic\\
\enddata
\tablecomments{$\log N(Y)=B+A\times\log N(X)$}
\tablenotetext{a}{Linear correlation coefficient.}
\tablenotetext{b}{Number of sight lines in the sample.}
\tablenotetext{c}{Excludes outliers: $\gamma$~Cas, $\psi$~Ori, $\epsilon$~Ori, and $\kappa$~Ori.}
\end{deluxetable*}

As already discussed by \citet{b15}, the relationship between Cl~{\sc i} and H$_2$ exhibited by high-redshift DLAs is very similar to that found for Galactic sight lines. A least-squares linear fit to the combined Galactic and extragalactic sample (``Galactic+DLAs'') yields a slope of $0.906\pm0.012$ and a linear correlation coefficient of 0.897 (Table~\ref{tab:correlations}). The extragalactic measurements help to extend the trend of Cl~{\sc i} versus H$_2$ to lower neutral chlorine column densities (i.e., below $\log N({\rm Cl~\textsc{i}})\sim12.5$). The only Galactic sight lines with such low values of $N({\rm Cl~\textsc{i}})$ (i.e., $\gamma$~Cas, $\psi$~Ori, $\epsilon$~Ori, and $\kappa$~Ori) have H$_2$ column densities that are much lower than the values that would be predicted by the linear trends. These sight lines probe molecular gas in the transition region where H$_2$ is not yet fully self-shielded. A similar pattern is seen when the column densities of other trace neutral species (e.g., K~{\sc i} and Na~{\sc i}) are plotted versus $N({\rm H}_2)$ \citep{wh01}. The K~{\sc i} and Na~{\sc i} column densities examined by \citet{wh01} exhibit a plateau for molecular hydrogen column densities in the range $15\lesssim\log N({\rm H}_2)\lesssim19$ (see their Figures 26 and 27).

\citet{wh01} find nearly linear relationships for $N({\rm K~\textsc{i}})$ versus $N({\rm H}_2)$ and $N({\rm Na~\textsc{i}})$ versus $N({\rm H}_2)$ for $\log N({\rm H}_2)>18.5$, very similar to our findings for Cl~{\sc i}. The fact that the slope of the relationship between $N({\rm Cl~\textsc{i}})$ and $N({\rm H}_2)$ in the high column density regime is slightly less than one is most likely related to the increase in the fraction of hydrogen in molecular form in the portion of the cloud where the H$_2$ resides. As discussed below (see Section~\ref{subsec:depletions2}), the gas-phase abundance (i.e., depletion) of chlorine shows very little dependence on the molecular hydrogen fraction. Thus, while the ratio of neutral chlorine to total hydrogen remains roughly constant in the molecular portion of the cloud, the fraction of molecular hydrogen increases as a function of $N({\rm H}_2)$.

\begin{figure}
\centering
\includegraphics[width=0.44\textwidth]{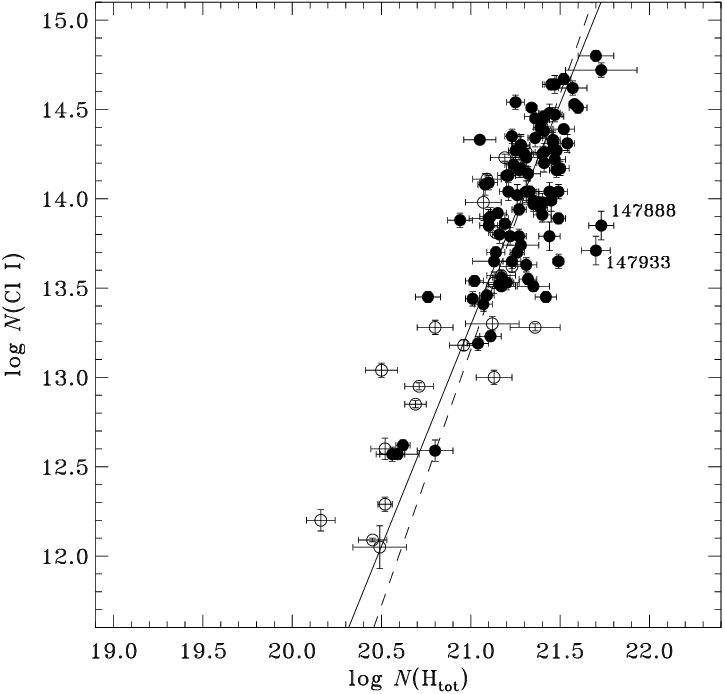}
\caption{Correlation between $\log N$(Cl~{\sc i}) and $\log N$(H$_{\rm tot}$). The plotting symbols have the same meaning as in Figure~\ref{fig:cl_corr1}. Two sight lines with anomalously low Cl~{\sc i} column densities are identified. (These sight lines were not excluded from the linear fits.)\label{fig:cl_corr2}}
\end{figure}

\citet{wh01} find nearly quadratic relationships between the column densities of K~{\sc i} and Na~{\sc i} and the total hydrogen column density. A quadratic dependence of $N({\rm K~\textsc{i}})$ and $N({\rm Na~\textsc{i}})$ on $N({\rm H}_{\rm tot})$ is expected if the ionization equilibrium is dominated by photoionization and radiative recombination, if the electron fraction is roughly constant, and if the individual clouds have a roughly uniform thickness \citep{h74,wh01}. In Figure~\ref{fig:cl_corr2}, we plot the observed trend of $N({\rm Cl~\textsc{i}})$ versus $N({\rm H}_{\rm tot})$. Least-squares linear fits to the ``STIS+FUSE'' and ``all Galactic'' samples yield slopes in the range 2.5--2.9 (see Table~\ref{tab:correlations}), much steeper than the corresponding trends involving K~{\sc i} and Na~{\sc i}. The steeper relationship for Cl~{\sc i} is due to the conversion of Cl$^+$ to Cl$^0$ in molecule-rich sight lines, where Cl~{\sc i} becomes the dominant ion. Indeed, the fraction of chlorine in neutral form, $f({\rm Cl~\textsc{i}})=N({\rm Cl~\textsc{i}})/N({\rm Cl}_{\rm tot})$, increases steeply as a function of $N({\rm H}_{\rm tot})$.

Two sight lines are identified in Figure~\ref{fig:cl_corr2} as having anomalously low Cl~{\sc i} column densities relative to the total amount of hydrogen along the lines of sight. The two sight lines, HD~147933 ($\rho$~Oph~A) and HD~147888 ($\rho$~Oph~D), are part of the same stellar system and probe the same complex of diffuse molecular clouds \citep[e.g.,][]{sn08}. \citet{wh01} describe these sight lines, and several others in the Sco-Oph region, as being ``discrepant'' because they exhibit several anomalous column density ratios. For example, both $\rho$~Oph~A and $\rho$~Oph~D have anomalously low K~{\sc i} and Na~{\sc i} column densities relative to $N({\rm H}_{\rm tot})$ \citep{wh01,sn08}. Furthermore, while both sight lines exhibit severe depletions of many different elements from the gas-phase \citep[e.g.,][]{r23}, the molecular hydrogen fractions are unusually low. An enhanced UV radiation field may be responsible for the anomalously low column densities of Cl~{\sc i}, K~{\sc i}, Na~{\sc i}, and H$_2$ in these directions. Indeed, from an analysis of carbon ionization balance, \citet{jt11} find that the UV radiation field toward HD~147888 is enhanced by a factor of $\sim$16 over the average interstellar field.

\begin{figure}
\centering
\includegraphics[width=0.44\textwidth]{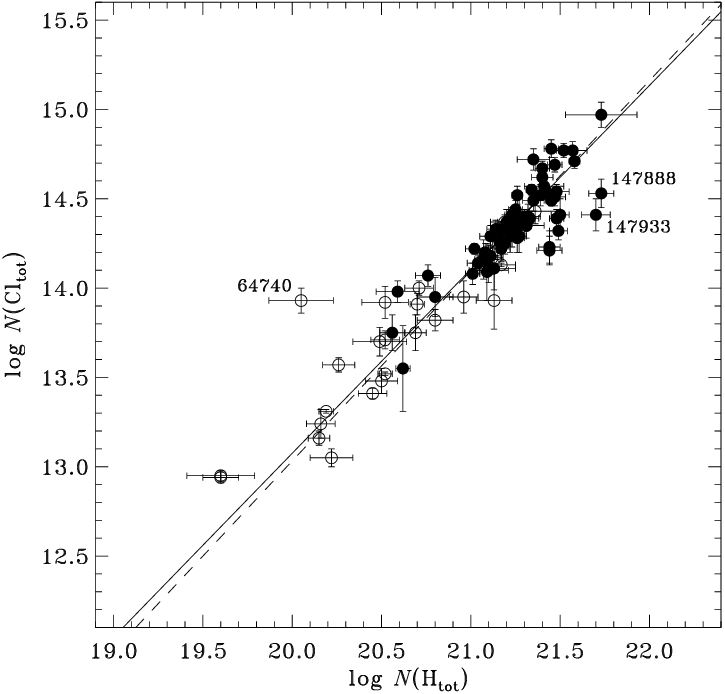}
\caption{Correlation between $\log N$(Cl$_{\rm tot}$) and $\log N$(H$_{\rm tot}$). The plotting symbols have the same meaning as in Figure~\ref{fig:cl_corr1}. Several outliers are indicated. (These sight lines were not excluded from the linear fits.)\label{fig:cl_corr3}}
\end{figure}

In Figure~\ref{fig:cl_corr3}, we plot the observed relationship between $N({\rm Cl}_{\rm tot})$ and $N({\rm H}_{\rm tot})$. Least-squares linear fits to the ``STIS+FUSE'' and ``all Galactic'' samples yield slopes of $\sim$1.0--1.1 and linear correlation coefficients of $\sim$0.8--0.9. The tight linear correlation between the total chlorine column density and the total hydrogen column density is an indication that the abundance and/or depletion of chlorine does not vary appreciably with $N({\rm H}_{\rm tot})$. The weighted mean value of $\log [N({\rm Cl}_{\rm tot})/N({\rm H}_{\rm tot})]$ for the ``all Galactic'' sample is $-6.893\pm0.008$, which is consistent with (but somewhat higher than) the value of $-6.99\pm0.04$ reported by \citet{m12}. The sight lines to HD~147888 and HD~147933 do not appear to be significant outliers in Figure~\ref{fig:cl_corr3} like they are in Figure~\ref{fig:cl_corr2}. Still, the total Cl column densities toward the $\rho$~Oph stars are $\sim$0.4--0.6 dex below that determined for the line of sight to HD~168076, which has a similar total hydrogen column density. The sight line to HD 64740 does appear to be an outlier in Figure~\ref{fig:cl_corr3}, although the hydrogen column density is not very well determined in this direction \citep{b83,j09}.

\subsection{Phosphorus and Chlorine Depletion Parameters\label{subsec:depletions}}
The main objective of this investigation is the redetermination of depletion parameters for the elements phosphorus and chlorine. Previous studies of P abundances \citep[e.g.,][]{l05,j09} used now outdated oscillator strengths for the important P~{\sc ii} transitions, leading to an overestimation of P~{\sc ii} column densities, and, consequently, an underestimation of the depletion of P onto interstellar grains. In his analysis of Cl depletion, \citet{j09} considered only Cl~{\sc ii} column densities, neglecting any contribution from Cl~{\sc i} to the derived total Cl abundances. However, of the 68 sight lines in Table~\ref{tab:chlorine} with measurements (or upper limits) for both Cl~{\sc i} and Cl~{\sc ii}, nearly half (32) have $N({\rm Cl~\textsc{i}})>N({\rm Cl~\textsc{ii}})$. Moreover, the neutral chlorine fraction, $f({\rm Cl~\textsc{i}})$, increases systematically with $f({\rm H}_2)$ (see Section~\ref{subsec:depletions2}). The neglect of Cl~{\sc i} will therefore significantly impact the interpretation of any perceived trends involving the depletion of Cl onto interstellar grains.

As in \citet{r18}, we derive element depletion parameters following the methodology of \citet{j09}. In this formalism, the logarithmic depletion of element $X$, defined as $[X/{\rm H}]=\log(X/{\rm H})-\log(X/{\rm H})_{\sun}$, depends on the sight-line depletion strength factor $F_*$ according to:

\begin{equation}
[X/{\rm H}]=B_X+A_X(F_*-z_X),
\end{equation}

\noindent
where the depletion parameters $A_X$, $B_X$, and $z_X$ are unique to each element. The slope parameter $A_X$ indicates how quickly a given element becomes depleted as the growth of dust grains progresses in interstellar clouds. The intercept parameter $B_X$ indicates the expected depletion of element $X$ at $F_*=z_X$, where $z_X$ represents a weighted mean value of $F_*$ for the particular set of sight lines with depletion measurements available for the element. Values of the coefficients $A_X$ and $B_X$ are obtained for each element through the evaluation of a least-squares linear fit, with $[X/{\rm H}]$ as the dependent variable and $F_*$ the independent variable. \citep[The reason for the additional term involving $z_X$ in Equation (1) is that for a particular choice of $z_X$ there is a near zero covariance between the formal fitting errors for the solutions of $A_X$ and $B_X$; see][]{j09}.

\begin{figure*}
\centering
\includegraphics[width=0.44\textwidth]{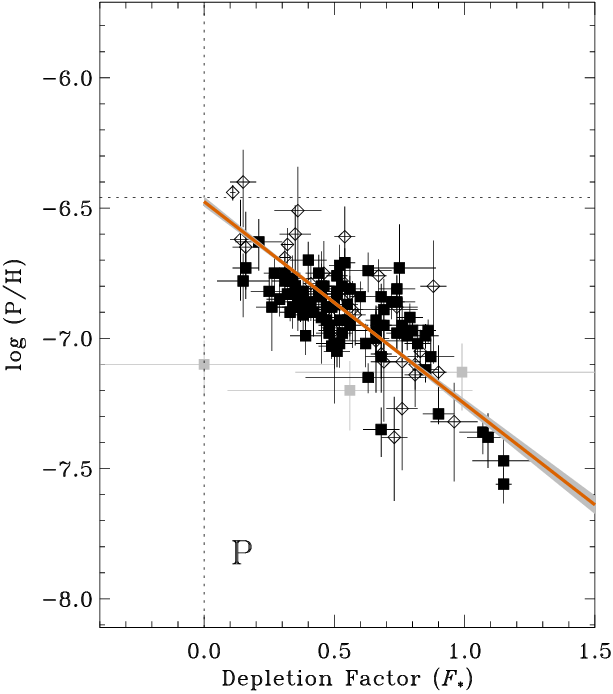}
\includegraphics[width=0.44\textwidth]{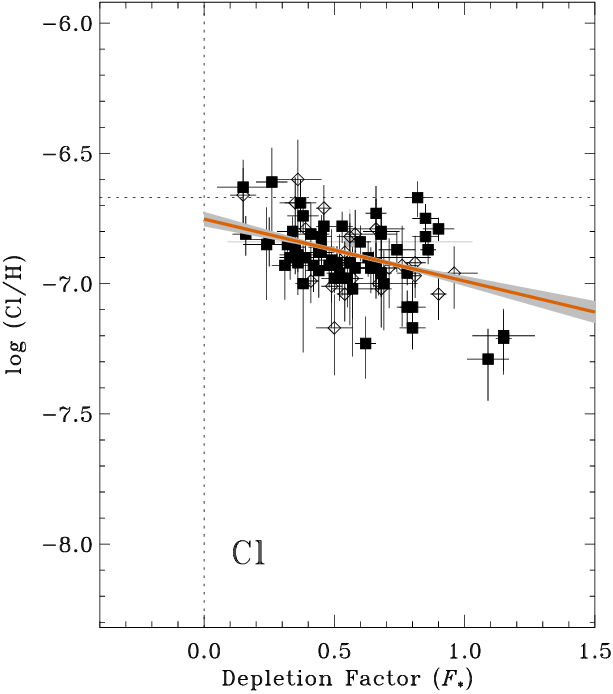}
\caption{Gas-phase P and Cl abundances as a function of the sight line depletion factor ($F_*$). Solid squares represent abundances derived in this work from STIS and FUSE observations; open diamonds represent measurements obtained in previous studies from Copernicus and GHRS observations \citep{j09,m12,br15}. Grey symbols denote sight lines where $\sigma(F_*)\ge0.30$. The solid orange lines represent linear fits to the data derived according to the methodology of \citet{j09}. The shaded gray regions indicate the 1$\sigma$ errors on the fit parameters. The horizontal dotted line in each panel gives the adopted solar system abundance from \citet{l03}.\label{fig:depletion}}
\end{figure*}

Values of the depletion strength factor $F_*$ for the sight lines in our combined P and Cl sample are obtained primarily from \citet{j19}, although, in some cases, we use the $F_*$ values provided by \citet{j09}. The adopted values are listed in Table~\ref{tab:hydrogen}. The $F_*$ values from \citet{j19} are preferred because they were derived solely from depletion measurements for Mg and Mn and thus are entirely independent from the elements considered in this investigation. Regardless of our preferrence, however, the two sets of $F_*$ values are comparable. For the 70 sight lines in our sample that have depletion strength factors listed in both \citet{j19} and \citet{j09}, the differences in the values are typically at the 1$\sigma$ or 2$\sigma$ level. (The mean absolute difference is 1.1$\sigma$.)

In Figure~\ref{fig:depletion}, we plot the gas-phase P and Cl abundances\footnote{Note that $\log({\rm P}/{\rm H})=\log N({\rm P~\textsc{ii}})-\log N({\rm H}_{\rm tot})$, while $\log({\rm Cl}/{\rm H})=\log N({\rm Cl}_{\rm tot})-\log N({\rm H}_{\rm tot})$.} as a function of $F_*$. Measurements obtained in this investigation (from the data presented in Tables~\ref{tab:phosphorus}, \ref{tab:chlorine}, and \ref{tab:hydrogen}) are represented by solid squares, while measurements for additional sight lines examined in previous investigations are represented by open diamonds. In the case of phosphorus, we include the P~{\sc ii} measurements tabulated by \citet{j09} for sight lines not included in our STIS sample. Most of these additional P~{\sc ii} measurements were derived from Copernicus or Goddard High Resolution Spectrograph (GHRS) observations \citep[see][and references therein]{j09}. All of the P~{\sc ii} column densities obtained from \citet{j09} were corrected so as to be consistent with the set of $f$-values adopted in this investigation (Table~\ref{tab:lines}). However, many of these previously reported P abundances have large associated uncertainties. We therefore retained only those abundance measurements with logarithmic uncertainties less than or equal to 0.176 dex (corresponding to a relative uncertainty of 50\% or better). For our analysis of chlorine depletions, we include, in addition to our own measurements, the total Cl column densities reported by \citet{m12} and \citet[][see also Appendix~\ref{app:copernicus}]{br15}, which were based on Copernicus observations.

Depletion parameters for P and Cl were determined through least-squares linear fits to the trends of $[{\rm P}/{\rm H}]$ and $[{\rm Cl}/{\rm H}]$ versus $F_*$, adopting the functional form expressed in Equation (1). The resulting values of $A_X$, $B_X$, and $z_X$ for the two elements are given in Table~\ref{tab:elem_depl_par}. The fits themselves are depicted by solid orange lines in Figure~\ref{fig:depletion} with shaded gray regions indicating the 1$\sigma$ errors on the fit parameters. The horizontal dashed lines in Figure~\ref{fig:depletion} represent the adopted solar system abundances of P and Cl \citep[][see Table~\ref{tab:elem_depl_par}]{l03}. Since the values obtained for the $B_X$ parameters depend on the specific choice of reference abundances, the uncertainties in the solar system abundances were added in quadrature to the formal fitting errors for $B_{\rm P}$ and $B_{\rm Cl}$ to arrive at the uncertainties listed for these parameters in Table~\ref{tab:elem_depl_par}. (Note that the slope parameters $A_X$ are entirely independent of the adopted solar reference abundances.)

\begin{deluxetable*}{lcccccccc}
\tablecolumns{9}
\tabletypesize{\small}
\tablecaption{Element Depletion Parameters\label{tab:elem_depl_par}}
\tablehead{ \colhead{Elem.} & \colhead{log~($X$/H)$_{\sun}$\tablenotemark{a}} & \colhead{$A_X$} & \colhead{$B_X$} & \colhead{$z_X$} & \colhead{[$X$/H]$_0$} & \colhead{[$X$/H]$_1$} & \colhead{$\chi^2$} & \colhead{$\nu$} }
\startdata
P & $-6.46\pm0.04$ & $-0.776\pm0.035$ & $-0.418\pm0.041$ & 0.520 & $-0.015\pm0.045$ & $-0.791\pm0.044$ & 297.4 & 126 \\
Cl & $-6.67\pm0.06$ & $-0.238\pm0.046$ & $-0.223\pm0.061$ & 0.593 & $-0.082\pm0.067$ & $-0.320\pm0.063$ & 182.7 & 81 \\
\enddata
\tablenotetext{a}{Recommended solar system abundance from \citet{l03}.}
\end{deluxetable*}

The intercept parameters $B_X$ are sensitive to the specific set of depletion measurements available for a given element (because they depend on $z_X$, which will vary from one element to the next). Thus, in order to facilitate a more straightforward comparison of the depletion results for different elements, it is important to evaluate two additional depletion parameters. The two parameters:

\begin{equation}
[X/{\rm H}]_0=B_X-A_Xz_X
\end{equation}

\noindent
and

\begin{equation}
[X/{\rm H}]_1=B_X+A_X(1-z_X),
\end{equation}

\noindent
indicate the expected depletions at $F_*=0$ and $F_*=1$, respectively. We interpret the $[X/{\rm H}]_0$ values as representing the initial amounts of depletion present in the diffuse ISM before significant grain growth has occurred (or after the outer portions of the grains have been destroyed by the passage of an interstellar shock). The $[X/{\rm H}]_1$ values represent the depletions associated with a prototypical diffuse molecular cloud. (\citet{j09} used the $v_{\sun}=-15$~km~s$^{-1}$ component toward $\zeta$~Oph as the standard for a cloud with $F_*=1$.) The values we obtain for $[{\rm P}/{\rm H}]_0$, $[{\rm P}/{\rm H}]_1$, $[{\rm Cl}/{\rm H}]_0$, and $[{\rm Cl}/{\rm H}]_1$ are provided in Table~\ref{tab:elem_depl_par}. (The errors in these quantities were determined according to the relations given in \citet{j09}).

The last two columns in Table~\ref{tab:elem_depl_par} give the chi-squared value ($\chi^2$) associated with each of the linear fits along with the number of degrees of freedom $\nu$ (i.e., the number of observations minus two). In each case, the reduced chi-squared value ($\chi^2/\nu$) is relatively high (i.e., 2.4 and 2.3 for the P and Cl fits, respectively). This could indicate that there is real intrinsic scatter in the gas-phase P and Cl abundances at a given value of $F_*$. Alternatively, the poor goodness-of-fit statistics could mean that the errors associated with the P and Cl abundances have been underestimated. The depletion trend for Cl, in particular, seems irregular. Some sight lines with moderate-to-large depletion strengths (such as HD~157857 and HD~206267) have relatively low gas-phase Cl abundances, while others with similar or even larger depletion strengths (such as X~Per, HD~207308, and HD~207538) have much higher total Cl abundances. The sight lines with the largest values of $F_*$ (HD~37903, HD~147888, and HD~147933) all have Cl abundances that are lower than expected based on the linear trend shown in Figure~\ref{fig:depletion}.

An important fundamental result of our depletion analysis is that the ``initial'' depletions of P and Cl (i.e., the $[X/{\rm H}]_0$ values) no longer indicate that the P and Cl abundances are ``supersolar'' at $F_*=0$, in contrast to the results presented in \citet{j09}. Both $[{\rm P}/{\rm H}]_0$ and $[{\rm Cl}/{\rm H}]_0$ are slightly less than zero (but consistent with zero at approximately the 1$\sigma$ level), indicating very little (if any) P and Cl depletion in the low-density ISM. The depletion slope for P is relatively steep, however ($A_{\rm P}\approx-0.8$), while the slope for Cl is rather shallow ($A_{\rm Cl}\approx-0.2$). These results will be discussed further, and compared with the results for many other elements, in Section~\ref{subsec:dust}.

\begin{figure*}
\centering
\includegraphics[width=0.44\textwidth]{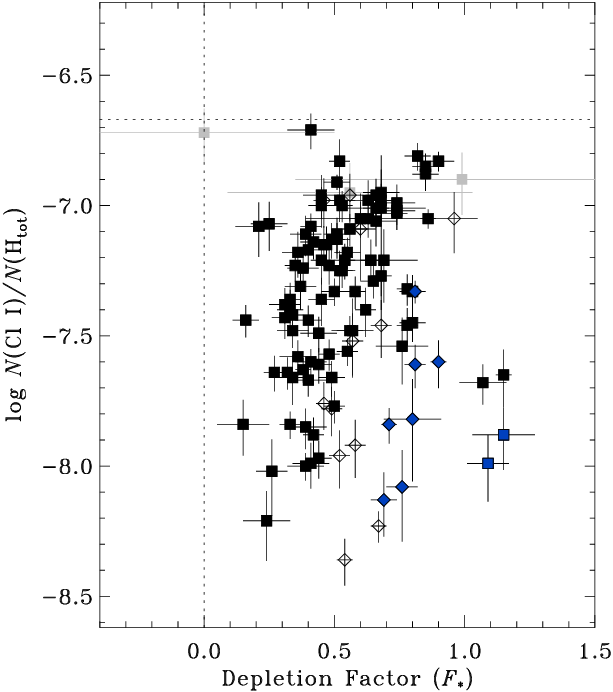}
\includegraphics[width=0.44\textwidth]{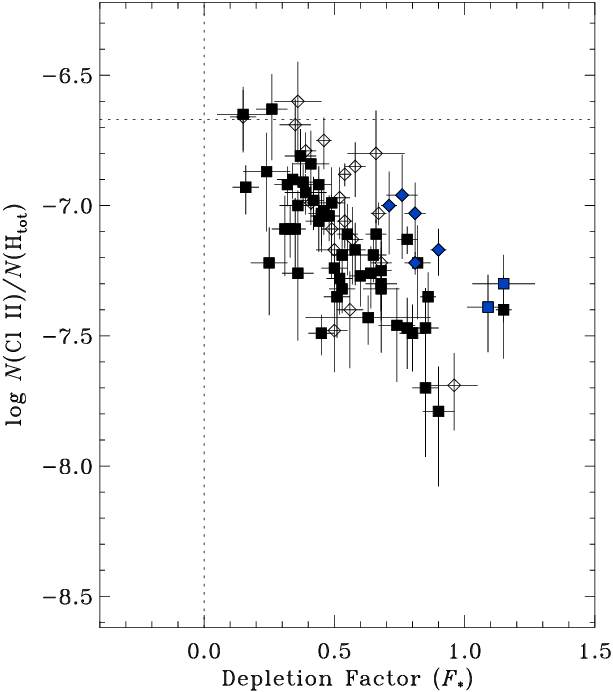}
\caption{Abundances of Cl~{\sc i} (left panel) and Cl~{\sc ii} (right panel) plotted as a function of the sight line depletion factor ($F_*$). Squares represent abundances derived in this work from STIS and FUSE observations; diamonds represent measurements obtained in previous studies from Copernicus and GHRS observations \citep{j09,m12,br15}. Grey symbols denote sight lines where $\sigma(F_*)\ge0.30$. Blue symbols are used to identify the discrepant Sco-Oph sight lines \citep[see][]{wh01,wc10}. The horizontal dotted line in each panel gives the adopted solar system abundance from \citet{l03}.\label{fig:depletion2}}
\end{figure*}

\subsection{Additional Insights on Depletion\label{subsec:depletions2}}
We can gain additional insight into the possible cause of the irregular depletion trend for Cl by examining separately the abundance trends for neutral and singly-ionized Cl. In Figure~\ref{fig:depletion2}, we plot the abundances of Cl~{\sc i} and Cl~{\sc ii} against the sight line depletion strength factors. In these plots, we have explicitly identified (with blue symbols) the discrepant Sco-Oph sight lines discussed in \citet{wh01} \citep[see also][]{wc10,w20}. These sight lines include: 1~Sco, $\pi$~Sco, $\delta$~Sco, $\beta^1$~Sco, $\omega^1$~Sco, $\nu$~Sco, $\sigma$~Sco, $\rho$~Oph~A, and $\rho$~Oph~D. For most of the sight lines in our sample, we find a very steep increase in the Cl~{\sc i} abundance with increasing $F_*$ (presumably due to the neutralization of Cl$^+$ in increasingly depleted H$_2$-rich gas). However, the discrepant Sco-Oph sight lines (and several other sight lines, including $\kappa$~Ori, HD~37903, HD~72754, HD~144965, HD~165246, HD~206267, HD~207198, and HD~210839) do not follow this trend, exhibiting Cl~{\sc i} abundances that are much lower than expected for sight lines having relatively large values of $F_*$.\footnote{It should be noted that the sight-line depletion strength factor $F_*$ increases systematically with increasing $f({\rm H}_2)$ for $\log f({\rm H}_2)>-1.0$ if the discrepant Sco-Oph sight lines are excluded \citep[see][]{w20}.}

As shown in the righthand panel of Figure~\ref{fig:depletion2}, the Cl~{\sc ii} abundance drops precipitously with increasing $F_*$ (again, owing to the rapid neutralization of Cl$^+$ ions as the H$_2$ concentration increases). The discrepant Sco-Oph sight lines (plus several others, such as $\psi$~Ori, $\kappa$~Ori, HD~37903, and HD~165246) show enhanced Cl~{\sc ii} abundances relative to other sight lines with similar values of $F_*$. The Sco-Oph sight lines, and the other discrepant sight lines, likely probe regions with enhanced UV radiation fields. This is certainly true of HD~37903, which probes the photodissociation region (PDR) associated with the reflection nebula NGC~2023. Likewise, HD~206267 is the exciting star of the H~{\sc ii} region IC~1396 and HD~165246 probes gas in the outskirts of the Lagoon Nebula (M8). An enhanced UV radiation field would significantly reduce the neutral chlorine abundance (both through the photoionization of Cl$^0$ and the photodissociation of H$_2$), and would enhance the abundance of Cl$^+$. These effects could then account for the anomalous Cl~{\sc i} and Cl~{\sc ii} abundances seen along the discrepant sight lines noted above.

If the Sco-Oph sight lines, and the other discrepant sight lines, are excluded from consideration, and the Cl depletion trend shown in Figure~\ref{fig:depletion} is re-evaluated, the slope of the relation between the gas-phase Cl abundance and the sight line depletion factor becomes completely flat (i.e., $A_{\rm Cl}=+0.00\pm0.05$). Conversely, if only the discrepant sight lines are considered, the slope of the relation is relatively steep  (i.e., $A_{\rm Cl}=-0.59\pm0.18$). If the same discrepant sight lines are excluded from the depletion analysis for P, the slope of the relation between the gas-phase P abundance and $F_*$ is relatively unchanged (i.e., $A_{\rm P}=-0.74\pm0.04$). However, if only the discrepant sight lines are included, the depletion slope for P becomes significantly steeper (i.e., $A_{\rm P}=-1.49\pm0.18$). For both P and Cl, the gas-phase abundances are much better correlated with $F_*$ when only the discrepant sight lines are considered. (For these fits, the $\chi^2/\nu$ values are 0.7 for P and 1.1 for Cl.)

\begin{figure*}
\centering
\includegraphics[width=0.44\textwidth]{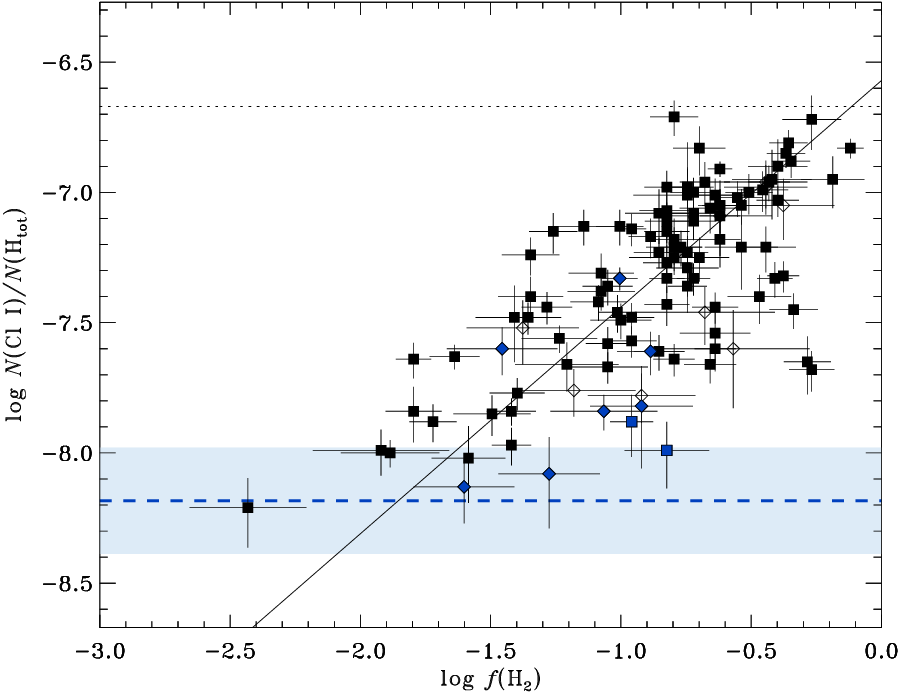}
\includegraphics[width=0.44\textwidth]{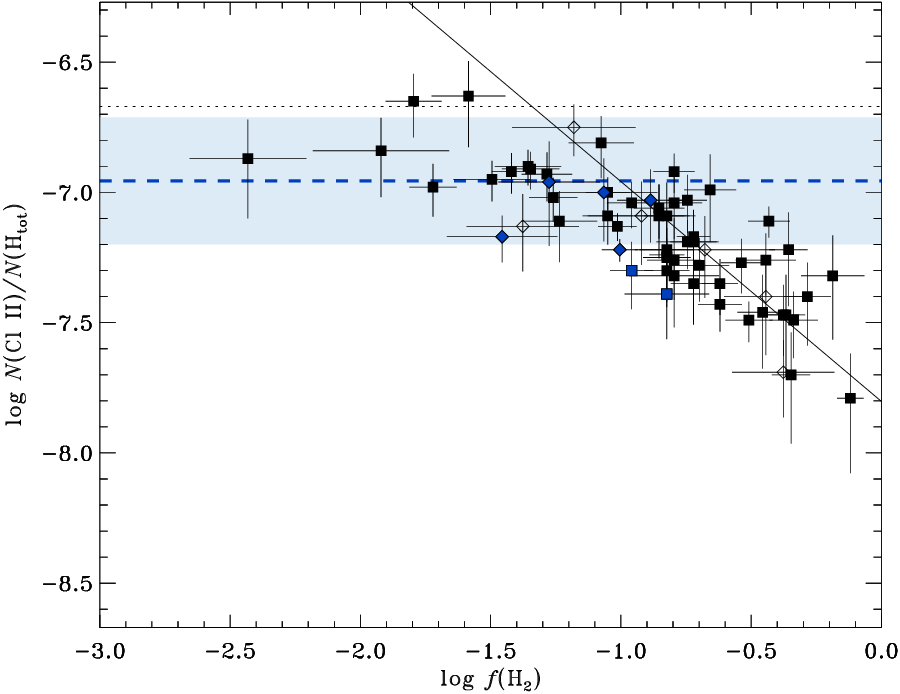}
\includegraphics[width=0.44\textwidth]{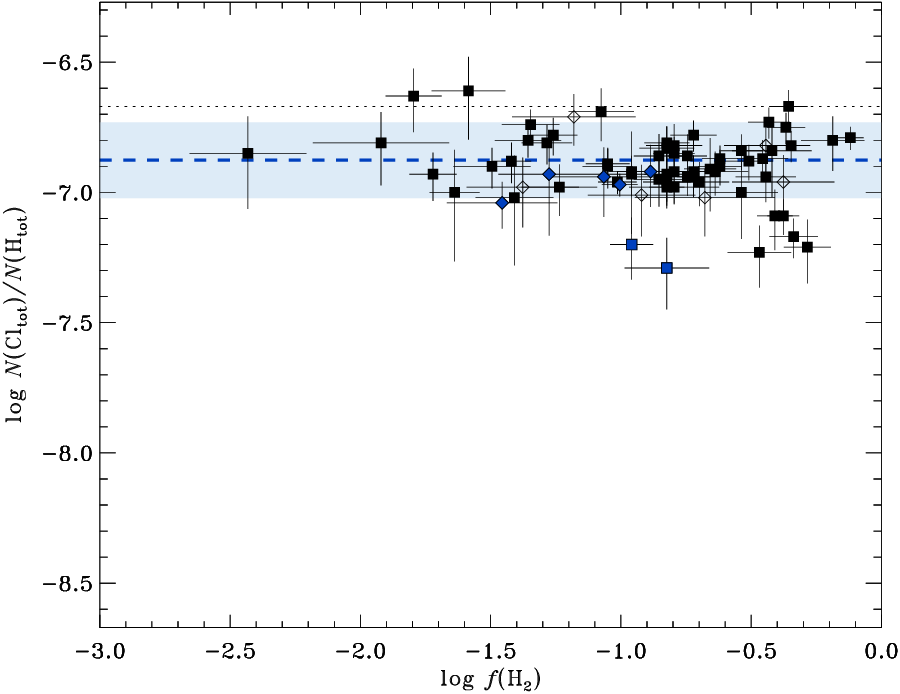}
\includegraphics[width=0.44\textwidth]{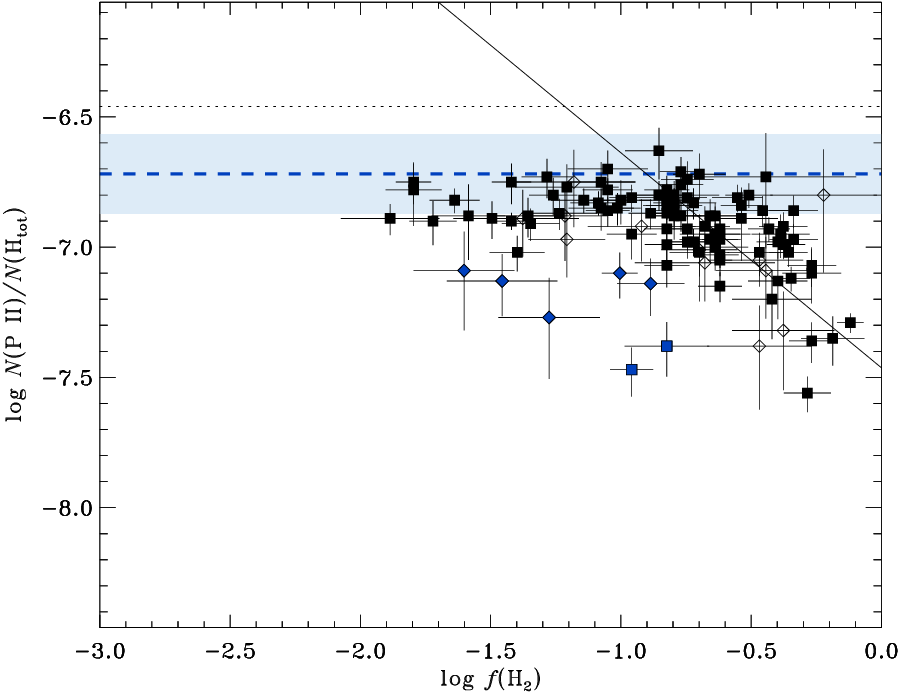}
\caption{Gas-phase abundances of Cl~{\sc i} (upper left), Cl~{\sc ii} (upper right), total Cl (lower left), and P~{\sc ii} (lower right) plotted as a function of the line-of-sight molecular hydrogen fraction. Squares represent abundances derived in this work from STIS and FUSE observations; diamonds represent measurements obtained in previous studies from Copernicus and GHRS observations \citep{j09,m12,br15}. Blue symbols are used to identify the discrepant Sco-Oph sight lines \citep[see][]{wh01,wc10}. The black dotted line in each panel indicates the adopted solar system abundance from \citet{l03}. The blue dashed line and light blue shaded region represent the mean and standard deviation of the abundances for sight lines with $\log f({\rm H}_2)<-3.0$. Solid lines indicate linear fits with approximate slopes of $+0.9$ for Cl~{\sc i} (for sight lines with $\log f({\rm H}_2)>-3.0$) and $-0.8$ for Cl~{\sc ii} and P~{\sc ii} (for sight lines with $\log f({\rm H}_2)>-1.0$).\label{fig:abun_fh2}}
\end{figure*}

The implication is that the depletion rates are enhanced in regions with enhanced ionization. This would then point to the importance of ion-grain reactions in driving the depletion of elements from the gas phase. The accretion of atoms onto grain surfaces may be enhanced by the presence of very small negatively-charged dust grains \citep{wd99}. Larger grains tend to be positively charged, yet most of the abundant refractory elements in the ISM are singly ionized. The cross sections for collisions between ions and small grains are therefore enhanced, while those for larger grains are diminished \citep[see][]{wd99}. Our results indicate that there is no enhanced depletion of Cl in relatively dense (heavily depleted) regions where Cl is predominantly neutral (e.g., toward X~Per, HD~207308, and HD~207538). However, there is enhanced Cl depletion in heavily depleted regions where Cl is mostly ionized (e.g., toward HD~37903, HD~147888, and HD~147933).

We can also examine the relationships that exist between the gas-phase P and Cl abundances and the (line-of-sight) molecular hydrogen fractions. In Figure~\ref{fig:abun_fh2}, we plot the gas-phase abundances of Cl~{\sc i}, Cl~{\sc ii}, total Cl, and P~{\sc ii} as a function of $f({\rm H}_2)$.\footnote{Note that in all four panels of Figure~\ref{fig:abun_fh2}, there are sight lines that are not shown because they have $\log f({\rm H}_2)<-3.0$. The gas-phase abundances in these directions are generally consistent with a single value for a given species such that the abundances exhibit a ``plateau'' at low molecular fraction. The blue dashed line and light blue shaded region in each panel represent the mean and standard deviation of the abundances for sight lines that are not shown in the figure.} As expected, the neutral chlorine abundance (upper left panel) increases systematically with the molecular hydrogen fraction. A linear fit indicates a slope of $\sim$0.9, similar to results for other trace neutral species, such as Na~{\sc i} and C~{\sc i} \citep[e.g.,][]{w20}. There is considerable scatter in this relation, however, presumably because the total hydrogen column densities include varying amounts of atomic hydrogen not associated with the molecular portions of the interstellar clouds (see Section~\ref{subsec:fractions}). Conversely, the abundance of singly-ionized chlorine (upper right panel of Figure~\ref{fig:abun_fh2}) shows a steady decrease with increasing $f({\rm H}_2)$, especially for sight lines with $\log f({\rm H}_2)\gtrsim-1.0$. The observed trends of Cl~{\sc i} and Cl~{\sc ii} (relative to total H) versus $f({\rm H}_2)$ are a direct consequence of the rapid conversion of Cl$^+$ to Cl$^0$ in molecule-rich gas \citep{j74,nw09}.

While the Cl~{\sc i} abundance increases with increasing $f({\rm H}_2)$, and the Cl~{\sc ii} abundance decreases, the total Cl abundance (lower left panel of Figure~\ref{fig:abun_fh2}) shows virtually no dependence on the line-of-sight molecular fraction As already mentioned, there are some sight lines with unusually low total Cl abundances (such as HD~37903, HD~147888, HD~147933, HD~157857, and HD~206267), but the overall trend is exceptionally flat. In contrast, the gas-phase P abundances (lower right panel) show a relatively constant value for $\log f({\rm H}_2)\lesssim-1.0$ (with the exception of the Sco-Oph sight lines) and a sharp decline for $\log f({\rm H}_2)\gtrsim-1.0$. This behavior is typical of elements that exhibit moderate-to-strong depletions in cold interstellar clouds. For example, similar trends versus $f({\rm H}_2)$ are seen for the elements B \citep{r11}, Mg \citep{c06,js07b,w20}, Ti \citep{wc10,w20}, and Fe \citep{js07a,w20}. Such trends are clear indications of the increase in gas-phase element depletions in increasingly dense, molecule-rich gas.

The discrepant Sco-Oph sight lines are prominent outliers in the plot showing P/H versus $f({\rm H}_2)$ (lower right panel of Figure~\ref{fig:abun_fh2}). These sight lines are known to exhibit strong depletions of (for example) Mg, P, Ti, and Fe \citep[e.g.,][this work]{wc10,w20}, yet contain relatively little H$_2$. An enhanced UV radiation field may be the source of these discrepancies. An enhanced flux of UV radiation (for example, if the absorbing gas is situated in close proximity to the background star) would contribute to H$_2$ photodissociation and would raise the ionization level of atomic species. An increase in the concentration of ionized species in the denser parts of the clouds could then lead to enhanced depletion through ion-grain reactions \citep[e.g.,][]{wd99}. These results tend to corroborate our findings on Cl depletion, which indicate that enhanced depletion of Cl generally only occurs along sight lines with elevated levels of Cl ionization.

\subsection{Neutral Chlorine Fractions\label{subsec:fractions}}
Because the first ionization potential of chlorine is less than that of hydrogen, chlorine is predominantly ionized in low-density, diffuse atomic gas. However, in molecular gas (i.e., in regions where the H$_2$ molecules are optically thick), Cl$^+$ reacts rapidly with H$_2$, yielding HCl$^+$ and H. The molecular ion HCl$^+$ then undergoes dissociative recombination or reacts further with H$_2$, yielding H$_2$Cl$^+$ and H. Finally, the H$_2$Cl$^+$ is destroyed through dissociative recombination, producing HCl, Cl, and H \citep[see][]{j74,nw09}. Since the rate coefficient for the initial reaction ($k=1.0\times10^{-9}$ cm$^3$~s$^{-1}$), when combined with a typical density for neutral diffuse gas ($n_{\rm H}\gtrsim10$ cm$^{-3}$), is considerably larger than the Cl photoionization rate for the average interstellar radiation field ($\Gamma\lesssim4.8\times10^{-11}$ s$^{-1}$), chlorine will be predominantly neutral wherever H$_2$ is abundant.

With these considerations, \citet{jy78} proposed a simple model to account for observations of Cl~{\sc i} and Cl~{\sc ii} in the ISM. Their model consists of two zones: (1) a region where the H$_2$ molecules are optically thick, chlorine is entirely neutral, and there is some associated atomic hydrogen, and (2) a region where the H$_2$ molecules are optically thin, chlorine is ionized, and the hydrogen is predominantly (neutral) atomic. Following \citet{s02}, Equation (A7) in \citet{jy78} may be written

\begin{equation}
N_1({\rm H~\textsc{i}})=N({\rm H}_{\rm tot})[f({\rm Cl~\textsc{i}})-f({\rm H}_2)],
\end{equation}

\noindent
where $N_1({\rm H~\textsc{i}})$ is the column density of H~{\sc i} that is associated with the optically-thick H$_2$ region, $N({\rm H}_{\rm tot})$ is the total column density of hydrogen along the line of sight, $f({\rm Cl~\textsc{i}})$ is the neutral chlorine fraction for the line of sight, and $f({\rm H}_2)$ is the line-of-sight molecular hydrogen fraction. With this simple model, it is possible to use integrated line-of-sight column densities of Cl~{\sc i}, Cl~{\sc ii}, H~{\sc i}, and H$_2$ to derive estimates for the total hydrogen column density and molecular hydrogen fraction associated with just the optically-thick H$_2$ region.

We apply Equation (4) to all of the sight lines in our chlorine sample with measurements of both Cl~{\sc i} and Cl~{\sc ii} (Table~\ref{tab:chlorine}), using values of $N({\rm H}_{\rm tot})$ and $f({\rm H}_2)$ from Table~\ref{tab:hydrogen}. The results are presented in Table~\ref{tab:fractions}. Note, in particular, the last two columns of Table~\ref{tab:fractions}, which give $N_1({\rm H}_{\rm tot})=N_1({\rm H~\textsc{i}})+2N({\rm H}_2)$ and $f_1({\rm H}_2)=2N({\rm H}_2)/N_1({\rm H}_{\rm tot})$, the total hydrogen column density and molecular hydrogen fraction, respectively, in the optically-thick H$_2$ region. We also include in Table~\ref{tab:fractions} those sight lines with Cl~{\sc i} and Cl~{\sc ii} measurements from Copernicus observations, provided that they also have measured values of $N({\rm H~\textsc{i}})$ and $N({\rm H}_2)$ \citep[][Appendix~\ref{app:copernicus}]{m12,br15}.

\startlongtable
\begin{deluxetable*}{lccccccc}
\tablecolumns{8}
\tabletypesize{\small}
\tablecaption{Neutral Chlorine Fractions and Derived Parameters\label{tab:fractions}}
\tablehead{ \colhead{Star} & \colhead{log~$N$(H$_{\rm tot}$)} & \colhead{log~$f$(Cl~{\sc i})} & \colhead{log~$f$(H$_2$)} & \colhead{log~$N_1$(H~{\sc i})\tablenotemark{a}} & \colhead{log~$N$(H$_2$)} & \colhead{log~$N_1$(H$_{\rm tot}$)\tablenotemark{b}} & \colhead{log~$f_1$(H$_2$)\tablenotemark{c}} }
\startdata
HD~108         & $21.47^{+0.04}_{-0.05}$   & $-0.22\pm0.05$   & $-0.72\pm0.09$   & $21.09\pm0.09$   & $20.45\pm0.08$   & $21.26\pm0.07$   & $-0.50\pm0.10$ \\
HD~3827        & $20.56^{+0.07}_{-0.08}$   & $-1.17\pm0.11$   & $-1.90\pm0.25$   & $19.29\pm0.15$   & $18.35\pm0.25$   & $19.38\pm0.14$   & $-0.73\pm0.27$ \\
HD~12323       & $21.28^{+0.04}_{-0.04}$   & $-0.20\pm0.07$   & $-0.72\pm0.09$   & $20.93\pm0.11$   & $20.26\pm0.08$   & $21.08\pm0.09$   & $-0.52\pm0.11$ \\
HD~15137       & $21.32^{+0.07}_{-0.08}$   & $-0.26\pm0.08$   & $-0.79\pm0.10$   & $20.90\pm0.13$   & $20.23\pm0.08$   & $21.06\pm0.10$   & $-0.53\pm0.12$ \\
HD~24190       & $21.30^{+0.05}_{-0.06}$   & $-0.15\pm0.05$   & $-0.62\pm0.08$   & $20.96\pm0.10$   & $20.38\pm0.07$   & $21.15\pm0.07$   & $-0.46\pm0.10$ \\
HD~24534       & $21.34^{+0.03}_{-0.04}$   & $-0.05\pm0.03$   & $-0.12\pm0.05$   & $20.51\pm0.26$   & $20.92\pm0.04$   & $21.30\pm0.06$   & $-0.08\pm0.07$ \\
HD~37903       & $21.44^{+0.06}_{-0.07}$   & $-0.45\pm0.11$   & $-0.29\pm0.09$   & \ldots\tablenotemark{d}   & $20.85\pm0.07$   & $20.99\pm0.18$   & $+0.16\pm0.19$ \\
HD~41161       & $21.15^{+0.05}_{-0.05}$   & $-0.38\pm0.07$   & $-0.86\pm0.10$   & $20.60\pm0.12$   & $19.99\pm0.09$   & $20.78\pm0.09$   & $-0.49\pm0.12$ \\
HD~46223       & $21.58^{+0.03}_{-0.04}$   & $-0.18\pm0.04$   & $-0.61\pm0.07$   & $21.21\pm0.08$   & $20.67\pm0.06$   & $21.41\pm0.06$   & $-0.43\pm0.08$ \\
HD~52266       & $21.27^{+0.04}_{-0.04}$   & $-0.55\pm0.15$   & $-0.97\pm0.09$   & $20.51\pm0.23$   & $20.00\pm0.08$   & $20.72\pm0.16$   & $-0.42\pm0.17$ \\
HD~53975       & $21.09^{+0.04}_{-0.04}$   & $-0.63\pm0.16$   & $-1.64\pm0.10$   & $20.41\pm0.18$   & $19.15\pm0.09$   & $20.46\pm0.16$   & $-1.01\pm0.18$ \\
HD~63005       & $21.31^{+0.03}_{-0.03}$   & $-0.31\pm0.09$   & $-0.84\pm0.09$   & $20.84\pm0.13$   & $20.17\pm0.09$   & $21.00\pm0.10$   & $-0.52\pm0.13$ \\
HD~66788       & $21.26^{+0.04}_{-0.04}$   & $-0.58\pm0.09$   & $-1.24\pm0.14$   & $20.56\pm0.12$   & $19.72\pm0.14$   & $20.67\pm0.10$   & $-0.65\pm0.17$ \\
HD~73882       & $21.57^{+0.08}_{-0.09}$   & $-0.15\pm0.07$   & $-0.19\pm0.12$   & $20.28\pm0.76$   & $21.08\pm0.10$   & $21.41\pm0.15$   & $-0.03\pm0.17$ \\
HD~75309       & $21.19^{+0.03}_{-0.03}$   & $-0.39\pm0.07$   & $-0.73\pm0.07$   & $20.53\pm0.14$   & $20.16\pm0.06$   & $20.80\pm0.09$   & $-0.34\pm0.10$ \\
HD~88115       & $21.04^{+0.06}_{-0.07}$   & $-0.96\pm0.05$   & $-1.49\pm0.15$   & $19.94\pm0.11$   & $19.25\pm0.14$   & $20.09\pm0.09$   & $-0.54\pm0.16$ \\
HD~89137       & $21.11^{+0.06}_{-0.07}$   & $-0.39\pm0.08$   & $-0.79\pm0.11$   & $20.49\pm0.15$   & $20.02\pm0.09$   & $20.72\pm0.10$   & $-0.40\pm0.13$ \\
HD~90087       & $21.23^{+0.05}_{-0.05}$   & $-0.68\pm0.05$   & $-1.05\pm0.08$   & $20.31\pm0.10$   & $19.88\pm0.07$   & $20.55\pm0.07$   & $-0.37\pm0.10$ \\
HD~91597       & $21.42^{+0.06}_{-0.07}$   & $-1.09\pm0.05$   & $-1.42\pm0.07$   & $20.05\pm0.12$   & $19.70\pm0.05$   & $20.33\pm0.07$   & $-0.33\pm0.09$ \\
HD~91824       & $21.16^{+0.04}_{-0.04}$   & $-0.46\pm0.07$   & $-1.05\pm0.10$   & $20.57\pm0.10$   & $19.81\pm0.09$   & $20.70\pm0.08$   & $-0.59\pm0.12$ \\
HD~91983       & $21.22^{+0.05}_{-0.06}$   & $-0.50\pm0.10$   & $-0.82\pm0.09$   & $20.44\pm0.19$   & $20.10\pm0.07$   & $20.72\pm0.12$   & $-0.32\pm0.13$ \\
HD~92554       & $21.35^{+0.09}_{-0.11}$   & $-1.21\pm0.07$   & $-1.81\pm0.11$   & $20.01\pm0.12$   & $19.24\pm0.07$   & $20.14\pm0.10$   & $-0.59\pm0.12$ \\
HD~94493       & $21.18^{+0.05}_{-0.06}$   & $-0.79\pm0.05$   & $-0.79\pm0.08$   & \ldots\tablenotemark{d}   & $20.09\pm0.06$   & $20.38\pm0.11$   & $+0.01\pm0.12$ \\
HD~97175       & $21.07^{+0.05}_{-0.06}$   & $-0.75\pm0.12$   & $-0.66\pm0.10$   & \ldots\tablenotemark{d}   & $20.10\pm0.09$   & $20.31\pm0.18$   & $+0.09\pm0.20$ \\
HD~99857       & $21.35^{+0.06}_{-0.07}$   & $-0.50\pm0.08$   & $-0.75\pm0.08$   & $20.51\pm0.19$   & $20.30\pm0.05$   & $20.86\pm0.10$   & $-0.26\pm0.11$ \\
HD~99890       & $21.14^{+0.05}_{-0.05}$   & $-0.63\pm0.06$   & $-1.28\pm0.10$   & $20.41\pm0.09$   & $19.56\pm0.09$   & $20.52\pm0.07$   & $-0.66\pm0.11$ \\
HD~109399      & $21.17^{+0.05}_{-0.05}$   & $-0.66\pm0.07$   & $-0.86\pm0.10$   & $20.09\pm0.22$   & $20.01\pm0.09$   & $20.52\pm0.11$   & $-0.21\pm0.14$ \\
HD~116852      & $21.01^{+0.04}_{-0.04}$   & $-0.64\pm0.07$   & $-0.96\pm0.10$   & $20.09\pm0.15$   & $19.75\pm0.09$   & $20.37\pm0.09$   & $-0.32\pm0.12$ \\
HD~121968      & $20.59^{+0.12}_{-0.16}$   & $-1.41\pm0.07$   & $-1.59\pm0.15$   & $18.72\pm0.29$   & $18.70\pm0.10$   & $19.19\pm0.14$   & $-0.18\pm0.16$ \\
HD~137595      & $21.23^{+0.04}_{-0.05}$   & $-0.06\pm0.06$   & $-0.34\pm0.07$   & $20.85\pm0.14$   & $20.59\pm0.06$   & $21.17\pm0.08$   & $-0.28\pm0.10$ \\
HD~147683      & $21.41^{+0.11}_{-0.14}$   & $-0.11\pm0.08$   & $-0.42\pm0.15$   & $21.00\pm0.21$   & $20.68\pm0.12$   & $21.29\pm0.13$   & $-0.31\pm0.17$ \\
HD~147888      & $21.73^{+0.07}_{-0.09}$   & $-0.68\pm0.11$   & $-0.98\pm0.09$   & $20.74\pm0.21$   & $20.45\pm0.05$   & $21.05\pm0.12$   & $-0.30\pm0.13$ \\
HD~147933      & $21.70^{+0.08}_{-0.10}$   & $-0.70\pm0.11$   & $-0.83\pm0.16$   & $20.42\pm0.44$   & $20.57\pm0.15$   & $21.00\pm0.19$   & $-0.13\pm0.23$ \\
HD~152590      & $21.47^{+0.06}_{-0.07}$   & $-0.29\pm0.08$   & $-0.70\pm0.11$   & $20.96\pm0.14$   & $20.47\pm0.10$   & $21.18\pm0.10$   & $-0.41\pm0.14$ \\
HD~156359      & $20.80^{+0.10}_{-0.13}$   & $-1.36\pm0.13$   & $-2.43\pm0.22$   & $19.40\pm0.16$   & $18.07\pm0.21$   & $19.44\pm0.15$   & $-1.07\pm0.24$ \\
HD~157857      & $21.44^{+0.07}_{-0.08}$   & $-0.17\pm0.09$   & $-0.46\pm0.12$   & $20.96\pm0.20$   & $20.68\pm0.10$   & $21.27\pm0.12$   & $-0.29\pm0.15$ \\
HD~165246      & $21.45^{+0.03}_{-0.03}$   & $-0.49\pm0.07$   & $-1.01\pm0.07$   & $20.80\pm0.10$   & $20.14\pm0.07$   & $20.96\pm0.08$   & $-0.52\pm0.10$ \\
HD~165955      & $21.11^{+0.06}_{-0.07}$   & $-0.95\pm0.07$   & $-1.72\pm0.09$   & $20.08\pm0.10$   & $19.09\pm0.07$   & $20.16\pm0.09$   & $-0.77\pm0.11$ \\
HD~167402      & $21.20^{+0.05}_{-0.05}$   & $-0.23\pm0.10$   & $-0.84\pm0.11$   & $20.84\pm0.13$   & $20.06\pm0.10$   & $20.97\pm0.11$   & $-0.60\pm0.14$ \\
HD~168076      & $21.73^{+0.20}_{-0.37}$   & $-0.24\pm0.08$   & $-0.75\pm0.21$   & $21.33\pm0.23$   & $20.68\pm0.08$   & $21.49\pm0.17$   & $-0.51\pm0.19$ \\
HD~168941      & $21.25^{+0.04}_{-0.05}$   & $-0.17\pm0.07$   & $-0.84\pm0.09$   & $20.97\pm0.10$   & $20.11\pm0.08$   & $21.08\pm0.08$   & $-0.66\pm0.11$ \\
HD~177989      & $21.10^{+0.05}_{-0.05}$   & $-0.08\pm0.07$   & $-0.64\pm0.13$   & $20.88\pm0.11$   & $20.16\pm0.12$   & $21.02\pm0.09$   & $-0.56\pm0.14$ \\
HD~190918      & $21.40^{+0.06}_{-0.07}$   & $-0.37\pm0.05$   & $-1.26\pm0.10$   & $20.97\pm0.08$   & $19.84\pm0.08$   & $21.03\pm0.07$   & $-0.89\pm0.10$ \\
HD~191877      & $21.10^{+0.05}_{-0.06}$   & $-0.27\pm0.05$   & $-0.80\pm0.17$   & $20.69\pm0.11$   & $20.00\pm0.17$   & $20.84\pm0.10$   & $-0.54\pm0.19$ \\
HD~192035      & $21.39^{+0.04}_{-0.05}$   & $-0.13\pm0.08$   & $-0.46\pm0.10$   & $20.98\pm0.16$   & $20.63\pm0.09$   & $21.26\pm0.10$   & $-0.33\pm0.13$ \\
HD~192639      & $21.48^{+0.07}_{-0.08}$   & $-0.28\pm0.06$   & $-0.45\pm0.12$   & $20.72\pm0.25$   & $20.73\pm0.10$   & $21.20\pm0.12$   & $-0.17\pm0.15$ \\
HD~195455      & $20.62^{+0.04}_{-0.04}$   & $-0.92\pm0.24$   & $-1.89\pm0.19$   & $19.64\pm0.27$   & $18.42\pm0.19$   & $19.69\pm0.25$   & $-0.97\pm0.29$ \\
HD~195965      & $21.08^{+0.04}_{-0.05}$   & $-0.12\pm0.07$   & $-0.50\pm0.09$   & $20.73\pm0.14$   & $20.28\pm0.08$   & $20.96\pm0.09$   & $-0.38\pm0.12$ \\
HD~201345      & $21.02^{+0.05}_{-0.05}$   & $-0.69\pm0.04$   & $-1.36\pm0.13$   & $20.23\pm0.08$   & $19.36\pm0.12$   & $20.33\pm0.07$   & $-0.67\pm0.13$ \\
HD~203374      & $21.40^{+0.04}_{-0.05}$   & $-0.23\pm0.06$   & $-0.43\pm0.08$   & $20.73\pm0.20$   & $20.67\pm0.07$   & $21.17\pm0.09$   & $-0.20\pm0.11$ \\
HD~206267      & $21.49^{+0.05}_{-0.06}$   & $-0.28\pm0.06$   & $-0.34\pm0.09$   & $20.31\pm0.49$   & $20.85\pm0.08$   & $21.21\pm0.12$   & $-0.06\pm0.14$ \\
HD~206773      & $21.24^{+0.05}_{-0.06}$   & $-0.21\pm0.05$   & $-0.53\pm0.08$   & $20.76\pm0.12$   & $20.41\pm0.06$   & $21.04\pm0.07$   & $-0.32\pm0.09$ \\
HD~207198      & $21.50^{+0.05}_{-0.05}$   & $-0.24\pm0.10$   & $-0.41\pm0.07$   & $20.75\pm0.28$   & $20.79\pm0.05$   & $21.26\pm0.11$   & $-0.16\pm0.12$ \\
HD~207308      & $21.45^{+0.04}_{-0.05}$   & $-0.14\pm0.06$   & $-0.35\pm0.07$   & $20.90\pm0.17$   & $20.80\pm0.06$   & $21.31\pm0.08$   & $-0.21\pm0.10$ \\
HD~207538      & $21.52^{+0.04}_{-0.05}$   & $-0.09\pm0.05$   & $-0.36\pm0.07$   & $21.09\pm0.14$   & $20.85\pm0.06$   & $21.42\pm0.07$   & $-0.27\pm0.09$ \\
HD~208440      & $21.33^{+0.05}_{-0.06}$   & $-0.35\pm0.05$   & $-0.75\pm0.12$   & $20.75\pm0.12$   & $20.28\pm0.11$   & $20.97\pm0.09$   & $-0.39\pm0.14$ \\
HD~209339      & $21.27^{+0.04}_{-0.04}$   & $-0.35\pm0.04$   & $-0.84\pm0.07$   & $20.75\pm0.07$   & $20.13\pm0.06$   & $20.92\pm0.05$   & $-0.49\pm0.08$ \\
HD~210839      & $21.48^{+0.04}_{-0.04}$   & $-0.23\pm0.07$   & $-0.38\pm0.06$   & $20.69\pm0.23$   & $20.80\pm0.05$   & $21.24\pm0.08$   & $-0.14\pm0.10$ \\
HD~212791      & $21.13^{+0.12}_{-0.16}$   & $-0.46\pm0.12$   & $-1.41\pm0.15$   & $20.62\pm0.17$   & $19.42\pm0.11$   & $20.67\pm0.15$   & $-0.95\pm0.18$ \\
HD~219188      & $20.76^{+0.07}_{-0.08}$   & $-0.62\pm0.07$   & $-1.08\pm0.12$   & $19.96\pm0.13$   & $19.38\pm0.11$   & $20.14\pm0.10$   & $-0.46\pm0.14$ \\
HD~220057      & $21.10^{+0.11}_{-0.14}$   & $-0.21\pm0.09$   & $-0.53\pm0.13$   & $20.61\pm0.21$   & $20.27\pm0.09$   & $20.89\pm0.13$   & $-0.32\pm0.15$ \\
BD$+$35~4258   & $21.26^{+0.03}_{-0.03}$   & $-0.50\pm0.08$   & $-1.35\pm0.11$   & $20.69\pm0.09$   & $19.61\pm0.11$   & $20.76\pm0.08$   & $-0.85\pm0.13$ \\
$\gamma$~Cas   & $20.16^{+0.08}_{-0.10}$   & $-1.03\pm0.10$   & $-3.33\pm0.09$   & $19.13\pm0.12$   & $16.53\pm0.05$   & $19.13\pm0.12$   & $-2.30\pm0.13$ \\
$\epsilon$~Per & $20.50^{+0.09}_{-0.12}$   & $-0.44\pm0.08$   & $-0.68\pm0.27$   & $19.70\pm0.35$   & $19.52\pm0.26$   & $20.07\pm0.23$   & $-0.24\pm0.32$ \\
$\psi$~Ori     & $20.49^{+0.15}_{-0.23}$   & $-1.65\pm0.14$   & $-5.41\pm0.20$   & $18.84\pm0.19$   & $14.78\pm0.15$   & $18.84\pm0.19$   & $-3.76\pm0.23$ \\
$\lambda$~Ori  & $20.80^{+0.10}_{-0.12}$   & $-0.54\pm0.07$   & $-1.38\pm0.21$   & $20.19\pm0.13$   & $19.12\pm0.20$   & $20.26\pm0.12$   & $-0.84\pm0.22$ \\
$\epsilon$~Ori & $20.45^{+0.08}_{-0.10}$   & $-1.32\pm0.03$   & $-3.58\pm0.21$   & $19.12\pm0.09$   & $16.57\pm0.20$   & $19.13\pm0.08$   & $-2.26\pm0.21$ \\
$\kappa$~Ori   & $20.52^{+0.04}_{-0.04}$   & $-1.23\pm0.04$   & $-4.54\pm0.20$   & $19.29\pm0.06$   & $15.68\pm0.20$   & $19.29\pm0.06$   & $-3.31\pm0.20$ \\
139~Tau        & $20.96^{+0.08}_{-0.09}$   & $-0.77\pm0.10$   & $-0.92\pm0.21$   & $19.65\pm0.43$   & $19.73\pm0.20$   & $20.18\pm0.22$   & $-0.15\pm0.27$ \\
1~Sco          & $21.13^{+0.10}_{-0.13}$   & $-0.93\pm0.16$   & $-1.60\pm0.20$   & $20.10\pm0.22$   & $19.23\pm0.18$   & $20.20\pm0.18$   & $-0.67\pm0.24$ \\
$\pi$~Sco      & $20.69^{+0.06}_{-0.07}$   & $-0.90\pm0.10$   & $-1.07\pm0.20$   & $19.28\pm0.41$   & $19.32\pm0.20$   & $19.79\pm0.21$   & $-0.16\pm0.27$ \\
$\delta$~Sco   & $21.17^{+0.08}_{-0.09}$   & $-0.57\pm0.02$   & $-1.45\pm0.21$   & $20.54\pm0.09$   & $19.41\pm0.20$   & $20.60\pm0.08$   & $-0.89\pm0.21$ \\
$\beta^1$~Sco  & $21.14^{+0.04}_{-0.04}$   & $-0.36\pm0.01$   & $-1.00\pm0.07$   & $20.66\pm0.04$   & $19.83\pm0.06$   & $20.78\pm0.04$   & $-0.64\pm0.07$ \\
$\omega^1$~Sco & $21.23^{+0.07}_{-0.09}$   & $-0.68\pm0.08$   & $-0.88\pm0.13$   & $20.11\pm0.27$   & $20.05\pm0.11$   & $20.55\pm0.13$   & $-0.20\pm0.16$ \\
$\sigma$~Sco   & $21.36^{+0.14}_{-0.22}$   & $-1.15\pm0.08$   & $-1.27\pm0.20$   & $19.60\pm0.47$   & $19.79\pm0.15$   & $20.21\pm0.20$   & $-0.12\pm0.23$ \\
$\chi$~Oph     & $21.31^{+0.10}_{-0.13}$   & $-0.09\pm0.03$   & $-0.38\pm0.20$   & $20.90\pm0.23$   & $20.63\pm0.18$   & $21.22\pm0.15$   & $-0.28\pm0.22$ \\
$\mu$~Nor      & $21.19^{+0.08}_{-0.10}$   & $-0.14\pm0.05$   & $-0.44\pm0.16$   & $20.77\pm0.19$   & $20.45\pm0.15$   & $21.06\pm0.13$   & $-0.31\pm0.19$ \\
$\gamma$~Ara   & $20.71^{+0.08}_{-0.09}$   & $-1.05\pm0.04$   & $-1.18\pm0.24$   & $19.07\pm0.50$   & $19.23\pm0.23$   & $19.66\pm0.24$   & $-0.13\pm0.31$ \\
\enddata
\tablenotetext{a}{Column density of atomic hydrogen associated with the optically-thick H$_2$ region.}
\tablenotetext{b}{Total hydrogen column density associated with the optically-thick H$_2$ region.}
\tablenotetext{c}{Fraction of hydrogen in molecular form in the optically-thick H$_2$ region.}
\tablenotetext{d}{The derived value of $N_1$(H~{\sc i}) for this sight line is negative. However, the data are consistent with $f_1$(H$_2$) = 1.}
\end{deluxetable*}

\begin{figure}
\centering
\includegraphics[width=0.44\textwidth]{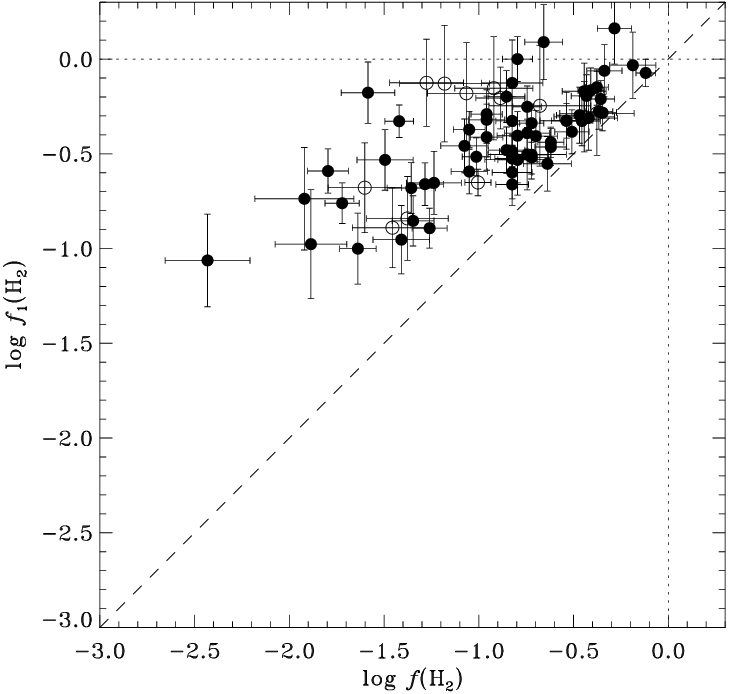}
\caption{Molecular fraction associated with the optically-thick H$_2$ region, denoted $f_1$(H$_2$), plotted against the line-of-sight molecular fraction, $f$(H$_2$). The plotting symbols have the same meaning as in Figure~\ref{fig:cl_corr1}. The diagonal dashed line indicates the line of equality between $f_1$(H$_2$) and $f$(H$_2$). Four sight lines with very low values of $f$(H$_2$) (i.e., $\gamma$~Cas, $\psi$~Ori, $\epsilon$~Ori, and $\kappa$~Ori) are not shown in this plot.\label{fig:f1h2_fh2}}
\end{figure}

From the results presented in Table~\ref{tab:fractions}, we find that, to varying degrees, the molecular hydrogen fraction in the optically-thick H$_2$ region is always larger than the line-of-sight integrated value. This is illustrated in Figure~\ref{fig:f1h2_fh2}, where the derived values of $f_1({\rm H}_2)$ are plotted against the observed line-of-sight values $f({\rm H}_2)$. All of the data points lie above the line of equality represented by the diagonal dashed line in the figure. For the median sight line in our sample, the molecular fraction in the optically-thick H$_2$ region is a factor of 2.8 larger than the line-of-sight value. Note that, according to Equation (4), the ratio $f_1({\rm H}_2)/f({\rm H}_2)=1/f({\rm Cl~\textsc{i}})$. Thus, the larger the neutral chlorine fraction, the smaller the contrast between $f_1({\rm H}_2)$ and $f({\rm H}_2)$. Another way to state this is that the neutral chlorine fraction is equivalent to the fraction of total hydrogen associated with the optically-thick H$_2$ region: $f({\rm Cl~\textsc{i}})=N({\rm Cl~\textsc{i}})/N({\rm Cl}_{\rm tot})=N_1({\rm H}_{\rm tot})/N({\rm H}_{\rm tot})$.

The largest neutral chlorine fractions (i.e., $f({\rm Cl~\textsc{i}})\gtrsim0.8$) are seen toward HD~24534 (X~Per), HD~137595, HD~177989, HD~207538, and $\chi$~Oph. For these sight lines, more than 80\% of the interstellar material is associated with the H$_2$-bearing gas. The smallest fractions ($f({\rm Cl~\textsc{i}})\lesssim0.05$) are found toward HD~121968, HD~156359, $\psi$~Ori, and $\epsilon$~Ori. The sight line to HD~121968 is interesting because, while the molecular material constitutes only $\sim$4\% of the total hydrogen column, the molecular fraction in that material is high (i.e., $f_1({\rm H}_2)\approx0.65$). Similar results are found toward $\sigma$~Sco and $\gamma$~Ara, where $f({\rm Cl~\textsc{i}})\approx0.07$ and 0.09, respectively, but $f_1({\rm H}_2)\approx0.76$ and 0.74. (The uncertainties associated with these values are relatively large, however; see Table~\ref{tab:fractions}) For the median sight line in our sample, $\sim$36\% of the total hydrogen column is associated with the optically-thick H$_2$ region.

An interesting pattern emerges from the data plotted in Figure~\ref{fig:f1h2_fh2}. For sight lines with $\log f({\rm H}_2)\gtrsim-0.85$, the $f_1({\rm H}_2)$ values adhere more closely to the line of equality (the diagonal dashed line in Figure~\ref{fig:f1h2_fh2}). If a data point lies along this line of equality, then all of the hydrogen along the line of sight is associated with the H$_2$-bearing gas. (Note this does not mean that all of the hydrogen is in the form of H$_2$.) Values of  $\log f({\rm H}_2)\gtrsim-0.85$ roughly correspond to molecular hydrogen column densities of  $\log N({\rm H}_2)\gtrsim20.0$. This threshold value in $f({\rm H}_2)$ also approximately corresponds to the point at which the gas-phase abundances (of P and other depleted elements, such as Mg, Ti, and Fe) begin to fall sharply with increasing $f({\rm H}_2)$ \citep[Figure~\ref{fig:abun_fh2}; see also Figure~21 in][]{w20}.

One of the motivations for the analysis of neutral chlorine fractions described here is to determine whether any of the sight lines in our sample contain ``translucent'' interstellar material. A translucent cloud is generally defined as having a total visual extinction $A_V\gtrsim1$~mag, a molecular hydrogen fraction $f({\rm H}_2)\gtrsim0.9$, and a kinetic temperature $T_k\lesssim40$~K \citep{vb88,r02}. Previous H$_2$ surveys with FUSE \citep[e.g.,][]{r02,r09} have generally failed to detect true translucent clouds along any of the sight lines surveyed. This failure may (at least partly) be due to the fact that the line-of-sight molecular fractions that are examined are lower limits to the true molecular fractions in the portions of the clouds where the H$_2$ molecules reside (Figure~\ref{fig:f1h2_fh2}). \citet{s02} recognized that an analysis of neutral chlorine fractions could help to identify any translucent material along an otherwise complex line of sight.

There are several sight lines in Table~\ref{tab:fractions} where the results for $f_1({\rm H}_2)$ indicate that the molecular fraction in the H$_2$-bearing gas could be as high as $\sim$1. Two intriguing cases are presented by the sight lines to HD~94493 and HD~97175. In both cases, the outcome for $N_1({\rm H~\textsc{i}})$ is negative because $f({\rm Cl~\textsc{i}})$ is slightly smaller than $f({\rm H}_2)$. (This is an unphysical result, based on our simple model, because there cannot be more molecular hydrogen than total hydrogen in the H$_2$-bearing gas.) Nevertheless, within the uncertainties, both sight lines could have a value of $f_1({\rm H}_2)\approx1$. Neither of these sight lines would traditionally be characterized as a translucent sight line, however. The total visual extinction toward HD~94493 is $A_V=0.82\pm0.18$ \citep{v04}. The extinction may be somewhat less toward HD~97175 based on a comparison of the $E(\bv)$ values (Table~\ref{tab:sample}). The H$_2$ rotational temperatures in the two directions are $T_{01}=79$~K and 74~K, respectively \citep{j19}, consistent with the average interstellar value of $\sim$80~K.

The outcome for $N_1({\rm H~\textsc{i}})$ toward HD~37903 is also negative. However, again, the result for $f_1({\rm H}_2)$ is consistent with 1 (within the uncertainties). This is an interesting sight line, which has been included in many previous surveys of ``translucent clouds'' \citep[e.g.,][]{c01,r09}. The total visual extinction in this direction is $A_V=1.28\pm0.15$ and the H$_2$ rotational temperature is $T_{01}=68\pm7$~K \citep{r09}. However, this sight line shows enhanced depletion of Cl, which tends to complicate the analysis of the neutral chlorine fraction (see below). Nevertheless, the ``true'' molecular fraction seems to be quite high in this direction, despite the fact that only $\sim$36\% of the total H is associated with the H$_2$-bearing gas.

Three other sight lines are worth mentioning in this context. The derived values of $f_1({\rm H}_2)$ toward HD~24534 (X~Per), HD~73882, and HD~206267 are consistent (within the quoted 1$\sigma$ uncertainties) with $f_1({\rm H}_2)\approx1$. The values are $f_1({\rm H}_2)=0.84\pm0.15$ toward X~Per, $0.93\pm0.46$ toward HD~73882, and $0.88\pm0.33$ toward HD~206267. In all three cases, the $f_1({\rm H}_2)$ values are larger than the line-of-sight molecular fractions, which are $f({\rm H}_2)=0.76\pm0.10$ toward X~Per, $0.65\pm0.21$ toward HD~73882, and $0.46\pm0.11$ toward HD~206267. The values of total visual extinction are also fairly high in these directions: $A_V=2.05\pm0.18$ (X~Per), $2.36\pm0.23$ (HD~73882), and $1.41\pm0.18$ (HD~206267) \citep{r09}. The H$_2$ rotational temperatures, however, are somewhat larger than the values expected for translucent clouds: $T_{01}=57\pm4$~K (X~Per), $51\pm6$~K (HD~73882), and $65\pm5$~K (HD~206267) \citep{r02}. There may be stratification in the molecular gas, however, since the kinetic temperatures derived from C$_2$ excitation are lower than the H$_2$ rotational temperatures: $T_k=45\pm10$~K toward X~Per, $20\pm5$~K toward HD~73882, and $35\pm5$~K toward HD~206267 \citep{s07}.

\begin{figure}
\centering
\includegraphics[width=0.44\textwidth]{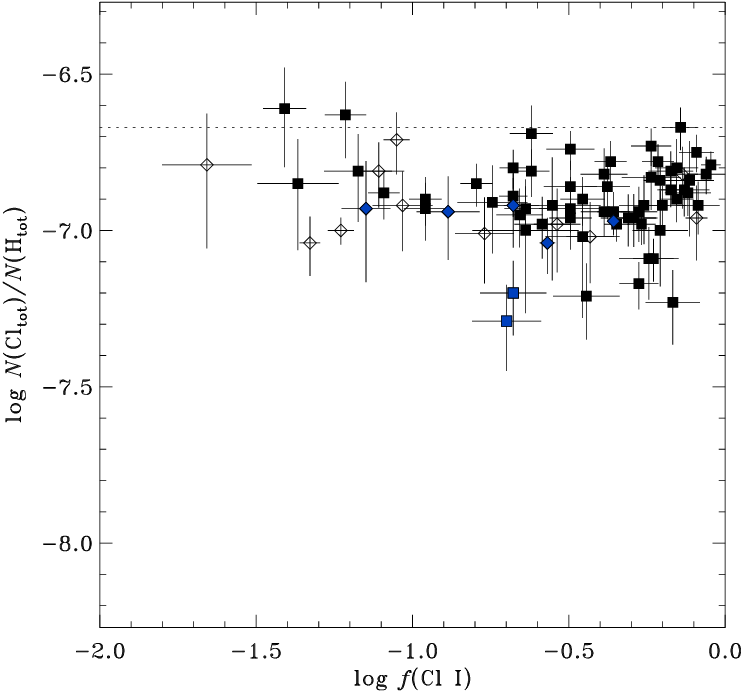}
\caption{Total gas-phase Cl abundance plotted as a function of the fraction of neutral chlorine along the line of sight. Plotting symbols have the same meaning as in Figure~\ref{fig:abun_fh2}. (Note the differences in the scales of the axes between this figure and Figure~\ref{fig:abun_fh2}.)\label{fig:cl_fcl1}}
\end{figure}

An explicit assumption in the derivation of Equation (4) is that the depletion of Cl does not vary between the zone where the H$_2$ is optically thick and the zone where the hydrogen is predominantly atomic \citep{jy78}. Justification for this assumption is provided by Figure~\ref{fig:abun_fh2}, where it is shown that the gas-phase Cl abundance does not vary significantly as a function of the line-of-sight molecular hydrogen fraction. As demonstrated in Figure~\ref{fig:cl_fcl1}, the gas-phase Cl abundance also does not vary as a function of the neutral chlorine fraction. (Recall that the neutral chlorine fraction is equivalent to the fraction of the total hydrogen column associated with the optically-thick H$_2$ region.) To illustrate this point, consider the line of sight toward X~Per, where $\sim$90\% of the interstellar material is associated with the H$_2$-bearing gas. This sight line has a total gas-phase Cl abundance of $\log ({\rm Cl}/{\rm H})=-6.79^{+0.04}_{-0.05}$, indicating a logarithmic depletion of $[{\rm Cl}/{\rm H}]=-0.12$. In contrast, $\sim$2\% of the interstellar material toward $\psi$~Ori is associated with the H$_2$ component. However, the total gas-phase Cl abundance is identical to that toward X~Per: $\log ({\rm Cl}/{\rm H})=-6.79^{+0.16}_{-0.27}$.

If the depletion of Cl does vary between the two zones, then the analysis of the neutral chlorine fractions becomes more complicated. In general, if the depletion level is higher in the zone where Cl is predominantly neutral, then the derived values of $f_1({\rm H}_2)$ will be lower \citep[see the discussion in][]{s02}. However, while there are a handful of sight lines that exhibit enhanced Cl depletion (see Section~\ref{subsec:depletions2}), these appear to be special cases. Depletion results for the vast majority of sight lines in our Cl sample indicate that Equation~(4) should be generally valid.

\section{DISCUSSION\label{sec:discussion}}
\subsection{Implications for Dust Formation and Evolution\label{subsec:dust}}
The primary purpose of our reexamination of phosphorus and chlorine abundances in the diffuse ISM is to clarify the depletion behaviors of these elements. In his landmark study of gas-phase element depletions, \citet{j09} found that the depletion trends for P and Cl indicated that the abundances of these elements were supersolar at $F_*=0$. With our new evaluation of the depletion properties of P and Cl (Figure~\ref{fig:depletion} and Table~\ref{tab:elem_depl_par}), we have demonstrated that this is not the case. We find values of $[{\rm P}/{\rm H}]_0$ and $[{\rm Cl}/{\rm H}]_0$ that are slightly less than zero (but consistent with zero within the uncertainties), indicating very little (if any) depletion of P or Cl in the low-density diffuse ISM.

\begin{figure*}
\centering
\includegraphics[width=0.65\textwidth]{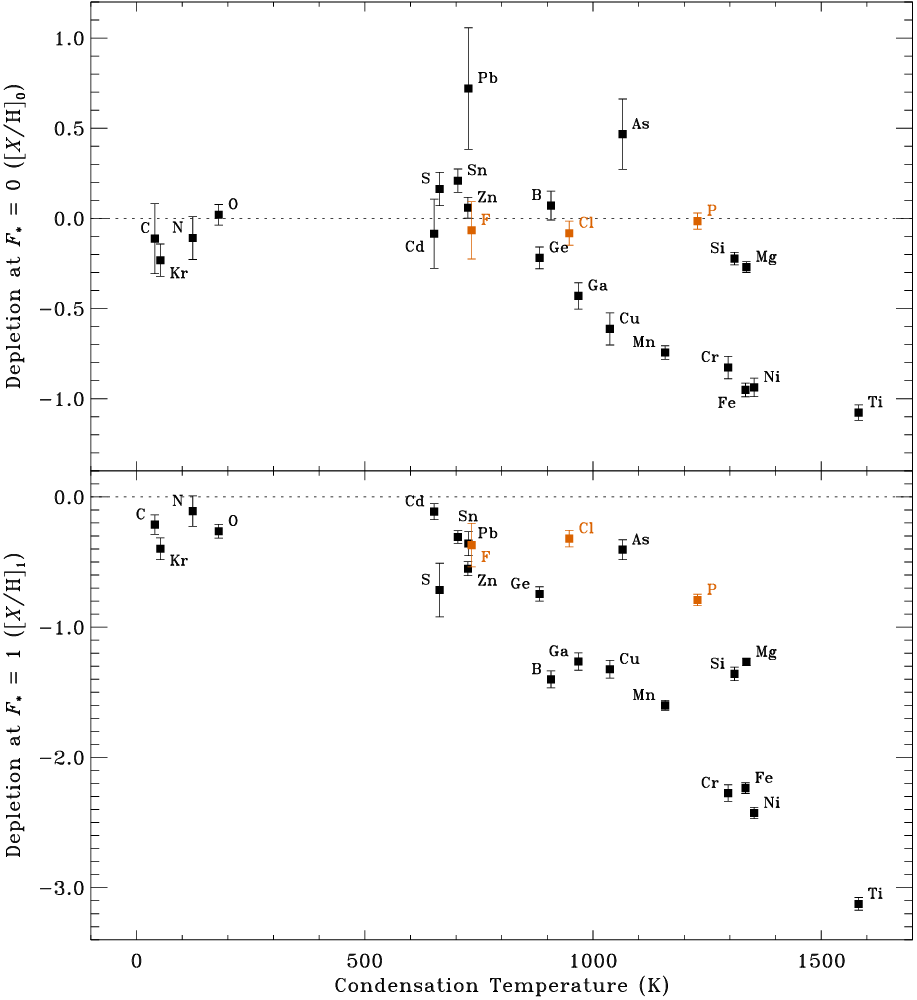}
\caption{Element depletions (at $F_*=0$ and $F_*=1$) as a function of the condensation temperature ($T_C$) from \citet{l03}. Orange symbols are used for results obtained in this work; black symbols show results for additional elements examined in previous investigations \citep{j09,r18}. Note the depletions for Mn and Cu \citep[from][]{j09} have been adjusted here by +0.165 dex and $-$0.016 dex, respectively, to account for updated $f$-values \citep{th05,b09}.\label{fig:depl_tc}}
\end{figure*}

With our improved constraints on P and Cl depletions, it is worthwhile reexamining the depletion results for all of the elements that have been analyzed via the \citet{j09} formalism. In Figure~\ref{fig:depl_tc}, we plot the depletions at $F_*=0$ and $F_*=1$ for 24 different elements as a function of the 50\% condensation temperatures ($T_C$) calculated by \citet{l03}. (This is an updated version of a similar figure presented in \citet{r18}. New results obtained in this investigation are indicated by orange symbols in Figure~\ref{fig:depl_tc}.) Note that, in addition to the new depletion parameters obtained for P and Cl, we also include in Figure~\ref{fig:depl_tc} depletion results for the element F. These values are obtained from an analysis of F depletions reported in the literature \citep[][see Appendix~\ref{app:fluorine}]{f05,sn07}.

Our new results on P, Cl, and F depletions help to clarify the situation regarding gas-phase element depletions at $F_*=0$. A fairly coherent trend now emerges, where the depletions at $F_*=0$ are consistent with zero for elements with $T_C\lesssim800$~K. For $T_C\gtrsim800$~K, the elements Ge, Ga, Cu, Mn, Cr, Fe, Ni, and Ti exhibit a clear trend of increasing depletion with increasing $T_C$, although the trend appears to flatten for $T_C\gtrsim1200$~K. The elements P, Si, and Mg do not participate in this trend, however, as they exhibit very little depletion at $F_*=0$ despite having large condensation temperatures. \citep[Note that the initial depletions for the elements As and Pb are highly uncertain because these elements are detected only along sight lines with relatively high values of $F_*$; see][]{r18}.

At $F_*=1$, another fairly regular pattern emerges. The volatile elements (i.e., C, N, O, and Kr) exhibit somewhat more depletion at $F_*=1$ than at $F_*=0$ (except for N, which has a depletion slope of $A_{\rm N}=0$). Meanwhile, a nearly linear trend of increasing depletion with increasing $T_C$ is seen for most of the more refractory elements from Cd at $T_C=650$~K to Ti at $T_C=1580$~K. \citep[Note that the depletion results for S are also uncertain; see][]{j09}. The elements Cl, As, P, Si, and Mg, however, do not follow this linear trend. These elements seem to form a second branch of the trend displaced toward higher gas-phase abundance.\footnote{Recent work by \citet{lf23} indicates that the 50\% condensation temperature for Cl may be significantly lower than that adopted here. \citet{lf23} find a value of $T_C=427$~K for Cl, which would place it among the low-depletion elements with $T_C\lesssim800$~K.}

In the context of dust grain formation, the depletions at $F_*=0$ are probably a good representation of the dust-phase abundances that pertain to the cores of the grains that emerge from stellar outflows and supernova ejecta. After being released into the ISM, the grains produced by stellar sources are subject to interstellar shocks, which may strip the outer portions of the grains leaving only the resilient cores. There are therefore two possibilities that might explain the elevated gas-phase abundances of P, Si, and Mg at $F_*=0$. Either these elements are inhibited from condensing into the grains produced in some stellar sources, or these elements are primarily incorporated into grain mantles that are more easily disrupted or destroyed through sputtering in interstellar shocks.

Observations of dust shells around O-rich AGB stars \citep[e.g.,][]{k13} are consistent with two distinct populations of dust grains condensing at different distances from the stellar photosphere. Alumina grains (Al$_2$O$_3$) condense close to the stellar surface (at $\sim$2 $R_*$), while silicate grains condense at larger distances ($\sim$4--5 $R_*$) where the gas temperature is lower. \citet{k13} find that some O-rich AGB stars have both an alumina shell and a silicate shell, while others have only an alumina dust shell. Furthermore, the stars with silicate shells tend to have much higher mass-loss rates. For the less-evolved stars with low mass-loss rates, the density in the cooler regions of the outflow (farther from the stellar surface) may be too low for significant SiO condensation. \citep[SiO is thought to be the seed particle for silicate nucleation; e.g.,][]{p11,g13}. Since the condensation temperature for Al$_2$O$_3$ is significantly larger than that for SiO, alumina condensation takes place closer to the stellar photosphere where the gas density is higher.

The fact that some AGB stars do not form silicate dust shells may help to explain why the gas-phase abundances of Si and Mg are significantly larger than those of other refractory elements (e.g., Fe) at $F_*=0$ (Figure~\ref{fig:depl_tc}). \citet{k13} argue that, owing to the relatively low cosmic abundance of Al, alumina grains alone cannot account for the observed dust-to-gas ratios in the outflows of AGB stars lacking silicate dust shells. The alumina grains that condense near the stellar surface may become seeds for further dust growth at larger radii and lower dust temperatures. Metallic Fe grains are good candidates for a dust growth species that would contribute mass but would not alter the spectral signature of Al$_2$O$_3$. It is interesting to note that many of the elements that exhibit a clear dust condensation sequence at $F_*=0$ (Ge, Ga, Cu, Cr, Fe, and Ni) condense into a host phase of Fe alloy \citep{l03}.

\citet{g16} model dust and molecule formation in the inner wind of the O-rich AGB star IK~Tau. These authors consider the effects of periodic shocks on the gas induced by stellar pulsation and follow the resulting non-equilibrium chemistry in the shocked gas layers. They find that some molecules (e.g., CO, PN, and HCl) form in the post-shock gas of the shocked photosphere and maintain almost constant abundances in the inner wind. The near constant abundances indicate that these molecules do not participate in complex chemistries or in the formation of dust clusters in the post-shock gas. The modelled abundances of PN and HCl are comparable to (and even somewhat larger than) the cosmic abundances of P and Cl, indicating that all of the P and Cl atoms are locked up in these species. The predictions of the \citet{g16} model are in agreement with our interstellar measurements, which show that there is essentially no depletion of P and very little (if any) depletion of Cl at $F_*=0$.

\citet{j17} offer an alternative explanation for why the depletions of Si and Mg are different from those of the iron group elements. They postulate that the differences in depletion are a natural consequence of the composition and structure of interstellar silicate grains. In the \citet{j17} model, the silicate group elements (O, Si, and Mg) form Mg-rich olivine and pyroxene-type amorphous silicate structures, while the iron group elements (Fe, Cr, Ni, Ti, and Mn) are present as Fe-rich nano-inclusions within the amorphous silicate matrix. \citet{j17} argue that the iron group elements show less variation in depletion because, as nano-inclusions, they are better protected by the surrounding silicate matrix and thereby better able to withstand the effects of sputtering in interstellar shocks.

\citet{d16} argues that most of the Fe missing from the gas phase in interstellar clouds must have precipitated from the ISM gas by a cold accretion onto preexisting silicate or carbon grains. He bases this conclusion on the fact that most Fe is produced in Type Ia SNe yet observational evidence suggests that these objects do not produce significant quantities of dust \citep[e.g.,][]{b07}. While some thermally-condensed Fe dust may be contributed by core-collapse SNe and AGB star winds, \citet{d16} estimates that less than 35\% of the Fe that is injected into the ISM is in solid form. The observed fraction of Fe in interstellar dust ranges from 89\% at $F_*=0$ to 99\% at $F_*=1$ \citep{j09}. Thus, if the argument advanced by \citet{d16} is correct, there must be a very efficient cycling of material between the cold and warm phases of the diffuse ISM.

In a series of dust evolution models based on a hydrodynamic simulation of the ISM, \citet{z16,z18} are able to reproduce the observed trends of increasing Si and Fe depletion with gas density by adopting a grain size distribution that includes a population of very small negatively-charged dust grains. Coulomb interactions between negatively-charged grains and ionized gas species enhance collision rates and lead to an increase in grain growth in the cold neutral medium (CNM). The rates of dust growth in the ISM in the \citet{z16,z18} models far exceed the rates of dust production by stellar sources.

\citet{p21}, however, question the efficiency of grain growth in the diffuse ISM. They argue that if accretion were as efficient as in the \citet{z16,z18} models, the small grains would grow to the point where they become positively charged, thereby halting any further accretion. \citet{p21} suggest that the strong depletions seen in the CNM (i.e., at $F_*=1$) represent the dust-phase abundances of grains produced by evolved stars and supernovae. The variation in depletions (e.g., between $F_*=1$ and $F_*=0$) would then mostly be due to the destruction of dust grains in supernova shocks, although the dust destruction efficiency would need to be reduced.

Our results on interstellar P and Cl depletions provide evidence that grain growth in fact does occur in the diffuse ISM and that the depletion rates are enhanced by ion-grain reactions (Section~\ref{subsec:depletions2}). We find enhanced Cl depletions along sight lines with higher than expected levels of Cl ionization (e.g., along sight lines probing PDRs, such as HD~37903, HD~147933, and HD~147888). Moreover, if we examine the P and Cl depletion trends for just those sight lines exhibiting elevated levels of ionization, we find much steeper depletion slopes (i.e., rates of depletion) than for the general samples. Clearly, these increased depletion rates are related to the environmental conditions in the interstellar clouds and not to the injection of depleted gas from stellar sources. The elevated levels of ionization strongly suggest that efficient ion-grain reactions are the source of the enhanced depletions.

\subsection{The Molecular Content of Diffuse Clouds}
\begin{figure}
\centering
\includegraphics[width=0.44\textwidth]{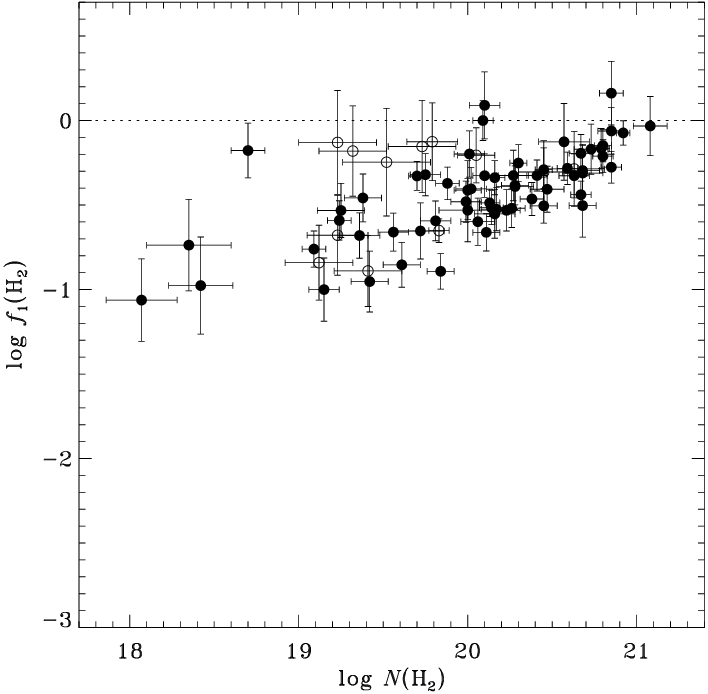}
\caption{Molecular fraction associated with the optically-thick H$_2$ region, $f_1$(H$_2$), plotted against the column density of molecular hydrogen, $N$(H$_2$). The plotting symbols have the same meaning as in Figure~\ref{fig:cl_corr1}. Four sight lines with very low values of $N$(H$_2$) (i.e., $\gamma$~Cas, $\psi$~Ori, $\epsilon$~Ori, and $\kappa$~Ori) are not shown in this plot.\label{fig:f1h2_nh2}}
\end{figure}

It is commonly assumed that the molecular hydrogen fraction in the H$_2$-bearing portion of a diffuse cloud is larger than the value derived from the integrated column densities of H~{\sc i} and H$_2$ along a line of sight. However, it is not often possible to directly assess the true local value of the molecular hydrogen fraction. The analysis described in Section~\ref{subsec:fractions} involving neutral chlorine fractions allows us to obtain estimates for the molecular fractions in the H$_2$-bearing gas, provided that our simple two-zone model for Cl chemistry is reasonably correct. We find that all sight lines with $\log N({\rm H}_2)\gtrsim18$ have molecular fractions in the optically-thick H$_2$ region of $f_1({\rm H}_2)\gtrsim0.1$ (Figure~\ref{fig:f1h2_nh2}). The only sight lines with $f_1({\rm H}_2)$ values significantly less than 0.1 are those that probe molecular gas in the transition region where H$_2$ is not yet fully self-shielded (i.e., $\gamma$~Cas, $\psi$~Ori, $\epsilon$~Ori, and $\kappa$~Ori). For $\log N({\rm H}_2)\gtrsim18$, the $f_1({\rm H}_2)$ values increase only gradually until the molecular hydrogen column reaches a value of $\log N({\rm H}_2)\approx20$, after which there is a much steeper rise. The H$_2$-bearing gas is essentially fully molecular at $\log N({\rm H}_2)\approx21$.

There are some exceptions to this general trend. For example, there is a group of sight lines (that includes HD~121968, $\gamma$~Ara, $\pi$~Sco, $\epsilon$~Per, 139~Tau, and $\sigma$~Sco) with $\log N({\rm H}_2)\lesssim20$ and $f_1({\rm H}_2)\gtrsim0.5$. The most prominent outlier among these is HD~121968, which has $\log N({\rm H}_2)\approx18.7$ and $f_1({\rm H}_2)\approx0.65$. This sight line probes gas in the lower halo at a distance of $\sim$3.3~kpc above the Galactic plane. While much of that pathlength is evidently filled with low-density neutral, atomic gas, the sight line also apparently crosses a small, moderately-dense diffuse molecular cloud. Kinematic evidence suggests that this small cloud is located close to the Galactic disk rather than near the position of the background star. The other outliers mentioned above \citep[which are from the Copernicus sample; see][Appendix~\ref{app:copernicus}]{m12} have rather large errors associated with their derived values of $f_1({\rm H}_2)$.

The median value of $f_1({\rm H}_2)$ for our sample of 78 sight lines with determinations of $f({\rm Cl~\textsc{i}})$ is 0.40, which may be compared to the median value of $f({\rm H}_2)$ for the same sight lines, which is 0.15. A value of $f({\rm H}_2)\approx0.4$ might then be adopted as a typical molecular hydrogen fraction for the H$_2$-bearing portions of diffuse clouds. This value is somewhat lower than the values that have been adopted in previous efforts to analyze the physical conditions in diffuse molecular gas. \citet{jt11}, for example, adopt a value of $f({\rm H}_2)=0.6$ for the CNM in their analysis of thermal pressures from C~{\sc i} fine-structure excitation. (Many of the same sight lines studied in that paper are included in our analysis.) From Figure~\ref{fig:f1h2_nh2}, we find that values of $f_1({\rm H}_2)$ as high as 0.6 are generally only seen along sight lines with relatively high molecular hydrogen column densities (i.e., $\log N({\rm H}_2)\gtrsim20.5$), and these sight lines represent only $\sim$13\% of the sample.

\section{SUMMARY AND CONCLUSIONS\label{sec:conclusions}}
We have presented a comprehensive examination of interstellar P and Cl abundances based on an analysis of archival HST/STIS and FUSE spectra. Column densities of P~{\sc ii}, Cl~{\sc i}, and Cl~{\sc ii} were derived through detailed profile synthesis fits for a combined sample of over 100 sight lines. P~{\sc ii} column densities were obtained from the weak P~{\sc ii}~$\lambda1532$ line wherever possible. Measurements of P~{\sc ii}~$\lambda1301$ were included only in cases where there was no significant blending with O~{\sc i}~$\lambda1302$. Cl~{\sc i} column densities were derived through simultaneous fits to the Cl~{\sc i}~$\lambda1347$ line (from STIS spectra) and either the $\lambda1379$ line or the $\lambda1097$, $\lambda1094$, or $\lambda1004$ transition (from FUSE data). Cl~{\sc ii} column densities were obtained from the Cl~{\sc ii}~$\lambda1071$ transition, after fitting and removing absorption from a nearby H$_2$ feature.

Experimentally determined oscillator strengths were used for most of the P~{\sc ii}, Cl~{\sc i}, and Cl~{\sc ii} lines included in our analysis (Table~\ref{tab:lines}). However, since no recent experimental $f$-value is available for the P~{\sc ii}~$\lambda1532$ transition, an empirical $f$-value was determined from the data for thirteen sight lines that have high-resolution STIS spectra covering both P~{\sc ii}~$\lambda1301$ and P~{\sc ii}~$\lambda1532$. Our empirical $f$-value for the $\lambda1532$ transition, $f(1532)=0.00737$, is very similar to the results of recent theoretical calculations \citep{t03,ff06}. These theoretical results and our empirical determination indicate that a downward revision of $\sim$0.4 dex is required for P~{\sc ii} column densities obtained using the $f$-value for P~{\sc ii}~$\lambda1532$ reported in \citet{m03}.

We reexamined the relationships between various Cl and H species, finding significant correlations between $N$(Cl~{\sc i}) and $N$(H$_2$) and between $N$(Cl$_{\rm tot}$) and $N$(H$_{\rm tot}$). The generally good correlation between Cl~{\sc i} and H$_2$ results from the neutralization of chlorine ions in regions where H$_2$ is abundant. The tight linear relationship between the total column densities of Cl and H indicates that the abundance and/or depletion of Cl does not vary appreciably with $N$(H$_{\rm tot}$). While the relationships we find are similar to those presented in previous investigations \citep[e.g.,][]{hb84,m12}, our results apply to a much larger, more representative sample of interstellar sight lines.

We derived depletion parameters for P and Cl in accordance with the methodology of \citet{j09}. For P, we find that the depletion slope is somewhat shallower than that reported in \citet{j09}. This difference may be a result of our much larger sample or it may be related to our use of weak, unblended P~{\sc ii} transitions (from high quality STIS observations). For Cl, we find that the depletion slope is much shallower compared to the slope obtained by \citet{j09}. In this case, the difference in slope is almost entirely due to our inclusion of Cl~{\sc i} column densities in determinations of total Cl abundances. A major improvement over the \citet{j09} results is that the initial depletions of P and Cl no longer indicate that the abundances of these elements are supersolar at $F_*=0$. In large part, this is due to improved oscillator strengths for the relevant P~{\sc ii} (and Cl~{\sc i}) transitions.

In general, we find that Cl is only lightly depleted along most interstellar sight lines, regardless of the fraction of hydrogen in molecular form (or the fraction of Cl in neutral form). Indeed, the existence of a \emph{negative} depletion slope for Cl appears to be dependent on an enhanced rate of depletion for a subset of sight lines with anomalously low Cl~{\sc i} abundances and higher-than expected Cl~{\sc ii} abundances. Many of these sight lines probe PDR-like regions where the flux of UV radiation is likely to be enhanced. The implication is that an elevated level of ionization in relatively dense gas leads to an enhanced rate of accretion of the ions onto grain surfaces. This, in turn, implies the existence of a population of small, negatively-charged grains, which accrete ionized species more efficiently through Coulomb interactions. We take this as concrete evidence of the growth of dust grains in diffuse interstellar clouds.

From an analysis of neutral chlorine fractions, we obtained estimates for the molecular hydrogen fractions that pertain to the H$_2$-bearing gas along the lines of sight in our survey. We find that for all sight lines with $\log N({\rm H}_2)\gtrsim18$ (where H$_2$ transitions to being fully self shielded), the H$_2$-bearing gas has a molecular fraction of at least 10\% and that the gas becomes essentially fully molecular at $\log N({\rm H}_2)\approx21$. A typical value of the molecular fraction in the H$_2$-bearing portion of a diffuse cloud is 0.4. This might then be adopted as a representative value for the CNM in general for studies seeking to analyze the chemistry and physical conditions in diffuse molecular clouds.

\acknowledgments
We acknowledge the efforts of undergraduate students Jennifer Hobbs and Luke Russell, who performed a preliminary analysis of Cl abundances as part of the Pre-Major in Astronomy Program (Pre-MAP) at the University of Washington. The profile fitting routine used in this work (ISMOD), which was originally created by Yaron Sheffer, was updated by Johnathan Rice to accommodate simultaneous fits to multiple absorption features. We thank Dan Welty for providing us with the H2GUI package. Our research has made use of the SIMBAD database operated at CDS, Strasbourg, France. Support for this work was provided by the Space Telescope Science Institute through grants HST-AR-15807.001-A to Eureka Scientific and HST-AR-15807.002-A to the University of Toledo. Observations were obtained from the MAST data archive at the Space Telescope Science Institute. The specific observations analyzed can be accessed via the following DOI: \dataset[10.17909/qfnj-yt76]{https://doi.org/10.17909/qfnj-yt76}. STScI is operated by the Association of Universities for Research in Astronomy, Inc., under NASA contract NAS5-26555.

\facilities{HST(STIS), FUSE, Copernicus}
\software{ISMOD \citep{s08}, STSDAS, H2GUI \citep{t02}, LTOOLS}

\appendix
\section{Chlorine Measurements from Copernicus Observations\label{app:copernicus}}
To supplement our sample of Cl measurements from STIS and FUSE observations, we have included in our analysis additional Cl abundance determinations based on observations made with the Copernicus satellite. The Copernicus sample consists of the results presented in \citet{m12}, who studied sight lines probing the transition from atomic to molecular gas, and the results obtained by \citet{br15}, who focused on sight lines with very low H$_2$ column densities as part of a Masters Thesis at the University of Toledo. \citet{br15} derived Cl~{\sc i} and Cl~{\sc ii} column densities via ISMOD fits to the Cl~{\sc i}~$\lambda1088$, Cl~{\sc i}~$\lambda1347$, and Cl~{\sc ii}~$\lambda1071$ lines, adopting component structures from high-resolution ground-based observations of K~{\sc i}, Ca~{\sc ii}~K, and Na~{\sc i}~D. A detailed comparison between the Cl~{\sc i} and Cl~{\sc ii} column densities derived by \citet{br15} and those obtained in previous Copernicus studies of the same sight lines \citep{jy78,hb84,j86} is presented in \citet{br15}. In general, the \citet{br15} results agree with those of the previous studies within the mutual uncertainties, although the uncertainties in \citet{br15} tend to be smaller.\footnote{For further details, see \citet{br15}, available at the following URL: \url{http://rave.ohiolink.edu/etdc/view?acc_num=toledo1431040914}.} For convenience, the results on $N$(Cl~{\sc i}) and $N$(Cl~{\sc ii}) from both \citet{m12} and \citet{br15} are provided in Table~\ref{tab:copernicus}. Column densities of H~{\sc i} and H$_2$ for the Copernicus sight lines, along with the sight line depletion factors, are obtained from \citet{j09}, except in the case of 67~Oph, which is not included in \citet{j09}. For this sight line, H~{\sc i} and H$_2$ column densities are obtained from \citet{s77} and \citet{b78}. Values of $N$(Cl$_{\rm tot}$) are listed in Table~\ref{tab:copernicus} even in cases where Cl~{\sc i} is not detected, provided that the upper limit on $N$(Cl~{\sc i}) is low enough that Cl~{\sc i} will not contribute meaningfully to the total Cl column density. The same is true for values of $N$(H$_{\rm tot}$) in cases where H$_2$ is not detected.

\startlongtable
\begin{deluxetable*}{lccccccccc}
\tablecolumns{10}
\tabletypesize{\small}
\tablecaption{Chlorine Column Densities from Copernicus Observations\label{tab:copernicus}}
\tablehead{ \colhead{Star} & \colhead{Name} & \colhead{log~$N$(Cl~{\sc i})} & \colhead{log~$N$(Cl~{\sc ii})} & \colhead{log~$N$(Cl$_{\rm tot}$)} & \colhead{Ref.} & \colhead{log~$N$(H~{\sc i})\tablenotemark{a}} & \colhead{log~$N$(H$_2$)\tablenotemark{a}} & \colhead{log~$N$(H$_{\rm tot}$)} & \colhead{$F_*$\tablenotemark{a}} }
\startdata
HD~5394    & $\gamma$~Cas    & $12.20\pm0.06$   & $13.19\pm0.09$  & $13.24\pm0.08$  & 1 & $20.16\pm0.08$  & $16.53\pm0.05$  & $20.16^{+0.08}_{-0.10}$  & $0.52\pm0.04$ \\
HD~23408   & 20~Tau          & $13.54\pm0.09$   & $<13.15$        & \ldots          & 2 & \ldots          & $19.75\pm0.26$  & \ldots                & \ldots        \\
HD~23630   & $\eta$~Tau      & $13.53\pm0.07$   & $<13.18$        & \ldots          & 2 & \ldots          & $19.54\pm0.18$  & \ldots                & \ldots        \\
HD~24760   & $\epsilon$~Per  & $13.04\pm0.04$   & $13.28\pm0.10$  & $13.48\pm0.07$  & 2 & $20.40\pm0.08$  & $19.52\pm0.26$  & $20.50^{+0.09}_{-0.12}$  & $0.68\pm0.04$ \\
HD~30614   & $\alpha$~Cam    & $14.11\pm0.03$   & $<13.40$        & \ldots          & 2 & $20.90\pm0.08$  & $20.34\pm0.15$  & $21.09^{+0.08}_{-0.09}$  & $0.46\pm0.04$ \\
HD~35715   & $\psi$~Ori      & $12.05\pm0.12$   & $13.69\pm0.08$  & $13.70\pm0.08$  & 1 & $20.49\pm0.15$  & $14.78\pm0.15$  & $20.49^{+0.15}_{-0.23}$  & $0.66\pm0.11$ \\
HD~36486   & $\delta$~Ori    & $<11.41$         & $13.31\pm0.01$  & $13.31\pm0.01$  & 1 & $20.19\pm0.04$  & $14.68\pm0.26$  & $20.19^{+0.04}_{-0.04}$  & $0.54\pm0.02$ \\
HD~36861   & $\lambda$~Ori   & $13.28\pm0.04$   & $13.67\pm0.08$  & $13.82\pm0.06$  & 2 & $20.78\pm0.10$  & $19.12\pm0.20$  & $20.80^{+0.10}_{-0.12}$  & $0.57\pm0.04$ \\
HD~37043   & $\iota$~Ori     & $<11.40$         & $13.16\pm0.04$  & $13.16\pm0.04$  & 1 & $20.15\pm0.06$  & $14.69\pm0.20$  & $20.15^{+0.06}_{-0.07}$  & $0.41\pm0.03$ \\
HD~37128   & $\epsilon$~Ori  & $12.09\pm0.01$   & $13.39\pm0.03$  & $13.41\pm0.03$  & 1 & $20.45\pm0.08$  & $16.57\pm0.20$  & $20.45^{+0.08}_{-0.10}$  & $0.54\pm0.03$ \\
HD~37468   & $\sigma$~Ori    & $12.60\pm0.06$   & $13.67\pm0.05$  & $13.71\pm0.05$  & 1 & $20.52\pm0.08$  & $<18.30$        & $20.52^{+0.08}_{-0.10}$  & $0.58\pm0.04$ \\
HD~38771   & $\kappa$~Ori    & $12.29\pm0.04$   & $13.49\pm0.01$  & $13.52\pm0.01$  & 1 & $20.52\pm0.04$  & $15.68\pm0.20$  & $20.52^{+0.04}_{-0.04}$  & $0.67\pm0.03$ \\
HD~40111   & 139~Tau         & $13.18\pm0.03$   & $13.87\pm0.11$  & $13.95\pm0.09$  & 2 & $20.90\pm0.08$  & $19.73\pm0.20$  & $20.96^{+0.08}_{-0.09}$  & $0.49\pm0.04$ \\
HD~57060   & 29~CMa          & $<13.00$         & $13.22\pm0.09$  & \ldots          & 1 & $20.70\pm0.08$  & $15.78\pm0.18$  & $20.70^{+0.08}_{-0.10}$  & $0.50\pm0.05$ \\
HD~57061   & $\tau$~CMa      & $<12.00$         & $13.91\pm0.06$  & $13.91\pm0.06$  & 1 & $20.70\pm0.04$  & $15.48\pm0.18$  & $20.70^{+0.04}_{-0.04}$  & $0.39\pm0.04$ \\
HD~64740   & \ldots          & $<12.30$         & $13.93\pm0.07$  & $13.93\pm0.07$  & 1 & $20.05\pm0.18$  & $<14.95$        & $20.05^{+0.18}_{-0.31}$  & $0.27\pm0.30$ \\
HD~64760   & \ldots          & $<11.89$         & $13.57\pm0.04$  & $13.57\pm0.04$  & 1 & $20.26\pm0.09$  & $<14.60$        & $20.26^{+0.09}_{-0.11}$  & $0.35\pm0.06$ \\
HD~65575   & $\chi$~Car      & $<12.28$         & $13.52\pm0.05$  & $13.52\pm0.05$  & 1 & $<20.74$        & $<14.78$        & \ldots                & \ldots        \\
HD~65818   & $\nu$~Pup       & $<12.54$         & $13.92\pm0.09$  & $13.92\pm0.09$  & 1 & $20.52\pm0.13$  & $15.08\pm0.30$  & $20.52^{+0.13}_{-0.19}$  & $0.36\pm0.09$ \\
HD~106490  & $\delta$~Cru    & $<11.95$         & $13.31\pm0.07$  & $13.31\pm0.07$  & 1 & $<19.69$        & $<14.08$        & \ldots                & \ldots        \\
HD~108248  & $\alpha^1$~Cru  & $<11.38$         & $12.94\pm0.03$  & $12.94\pm0.03$  & 1 & $19.60\pm0.10$  & $<14.18$        & $19.60^{+0.10}_{-0.13}$  & $0.15\pm0.05$ \\
HD~118716  & $\epsilon$~Cen  & $<11.54$         & $12.95\pm0.03$  & $12.95\pm0.03$  & 1 & $19.60\pm0.19$  & $<14.08$        & $19.60^{+0.19}_{-0.35}$  & $0.15\pm0.16$ \\
HD~127972  & $\eta$~Cen      & $<11.67$         & $13.21\pm0.20$  & $13.21\pm0.20$  & 1 & $<19.48$        & $<14.18$        & \ldots                & \ldots        \\
HD~136298  & $\delta$~Lup    & $<11.81$         & $13.11\pm0.07$  & $13.11\pm0.07$  & 1 & $<19.76$        & $<14.26$        & \ldots                & \ldots        \\
HD~141637  & 1~Sco           & $13.00\pm0.04$   & $13.88\pm0.17$  & $13.93\pm0.16$  & 2 & $21.12\pm0.10$  & $19.23\pm0.18$  & $21.13^{+0.10}_{-0.13}$  & $0.69\pm0.05$ \\
HD~143018  & $\pi$~Sco       & $12.85\pm0.02$   & $13.69\pm0.12$  & $13.75\pm0.10$  & 2 & $20.65\pm0.06$  & $19.32\pm0.20$  & $20.69^{+0.06}_{-0.07}$  & $0.71\pm0.03$ \\
HD~143118  & $\eta$~Lup      & $12.03\pm0.10$   & $12.85\pm0.14$  & $12.91\pm0.13$  & 1 & $<20.12$        & $<14.23$        & \ldots                & \ldots        \\
HD~143275  & $\delta$~Sco    & $13.57\pm0.02$   & $14.00\pm0.01$  & $14.13\pm0.01$  & 2 & $21.15\pm0.08$  & $19.41\pm0.20$  & $21.17^{+0.08}_{-0.09}$  & $0.90\pm0.03$ \\
HD~144217  & $\beta^1$~Sco   & $13.81\pm0.01$   & $13.92\pm0.01$  & $14.17\pm0.01$  & 2 & $21.09\pm0.04$  & $19.83\pm0.06$  & $21.14^{+0.04}_{-0.04}$  & $0.81\pm0.02$ \\
HD~144470  & $\omega^1$~Sco  & $13.62\pm0.03$   & $14.20\pm0.10$  & $14.31\pm0.08$  & 2 & $21.17\pm0.08$  & $20.05\pm0.11$  & $21.23^{+0.07}_{-0.09}$  & $0.81\pm0.04$ \\
HD~145502  & $\nu$~Sco       & $13.30\pm0.04$   & $<13.00$        & \ldots          & 2 & $21.07\pm0.17$  & $19.89\pm0.15$  & $21.12^{+0.15}_{-0.24}$  & $0.80\pm0.11$ \\
HD~147165  & $\sigma$~Sco    & $13.28\pm0.02$   & $14.40\pm0.08$  & $14.43\pm0.07$  & 2 & $21.34\pm0.15$  & $19.79\pm0.15$  & $21.36^{+0.14}_{-0.22}$  & $0.76\pm0.06$ \\
HD~148184  & $\chi$~Oph      & $14.26\pm0.02$   & $13.62\pm0.08$  & $14.35\pm0.03$  & 2 & $21.07\pm0.10$  & $20.63\pm0.18$  & $21.31^{+0.10}_{-0.13}$  & $0.96\pm0.09$ \\
HD~149038  & $\mu$~Nor       & $14.23\pm0.02$   & $13.79\pm0.13$  & $14.37\pm0.04$  & 2 & $21.00\pm0.08$  & $20.45\pm0.15$  & $21.19^{+0.08}_{-0.10}$  & $0.56\pm0.05$ \\
HD~151890  & $\mu^1$~Sco     & $<11.78$         & $13.20\pm0.06$  & $13.20\pm0.06$  & 1 & $<20.12$        & $<14.26$        & \ldots                & \ldots        \\
HD~157246  & $\gamma$~Ara    & $12.95\pm0.02$   & $13.96\pm0.04$  & $14.00\pm0.04$  & 2 & $20.68\pm0.08$  & $19.23\pm0.23$  & $20.71^{+0.08}_{-0.09}$  & $0.46\pm0.03$ \\
HD~160578  & $\kappa$~Sco    & $<11.53$         & $13.05\pm0.05$  & $13.05\pm0.05$  & 1 & $20.22\pm0.12$  & $<14.23$        & $20.22^{+0.12}_{-0.17}$  & $0.50\pm0.07$ \\
HD~164353  & 67~Oph          & $13.53\pm0.06$   & \ldots          & \ldots          & 2 & $21.00\pm0.15$  & $20.26\pm0.28$  & $21.13^{+0.14}_{-0.21}$  & \ldots        \\
HD~200120  & 59~Cyg          & $12.70\pm0.09$   & $<13.40$        & \ldots          & 2 & $<20.07$        & $19.30\pm0.18$  & \ldots                & \ldots        \\
HD~217675  & $o$~And         & $13.18\pm0.03$   & \ldots          & \ldots          & 2 & \ldots          & $19.67\pm0.18$  & \ldots                & \ldots        \\
HD~218376  & 1~Cas           & $13.98\pm0.10$   & $<13.73$        & \ldots          & 2 & $20.95\pm0.12$  & $20.15\pm0.18$  & $21.07^{+0.10}_{-0.14}$  & $0.60\pm0.06$ \\
\enddata
\tablenotetext{a}{Atomic and molecular hydrogen column densities and sight line depletion factors are obtained from \citet{j09}, except in the case of 67~Oph. For this sight line, H~{\sc i} and H$_2$ column densities are from \citet{s77} and \citet{b78}.}
\tablerefs{(1) \citet{br15}, (2) \citet{m12}.}
\end{deluxetable*}

\section{Fluorine Depletions from the Literature\label{app:fluorine}}
\begin{figure}
\centering
\includegraphics[width=0.44\textwidth]{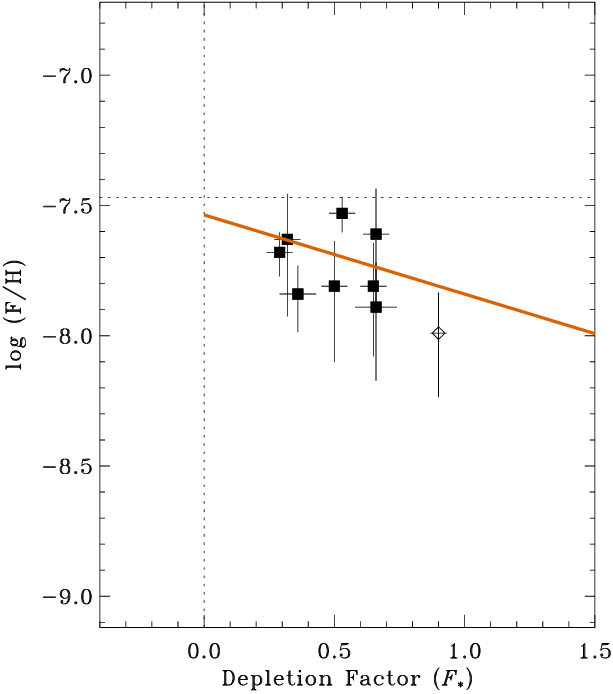}
\caption{Gas-phase F abundances from the literature \citep{f05,sn07} as a function of the sight line depletion factor ($F_*$). Solid squares represent abundances derived from FUSE observations; the open diamond represents the F abundance toward $\delta$~Sco, which is based on Copernicus observations \citep{sy81,f05}. The solid orange line represents a linear fit to the data derived according to the methodology of \citet{j09}. The horizontal dotted line gives the adopted solar system abundance from \citet{l03}.\label{fig:f_depletion}}
\end{figure}

Fluorine abundances were reported by \citet{f05} and \citet{sn07} from analyses of the F~{\sc i} lines at 951 and 954~\AA{} in FUSE spectra. However, fluorine was not included in the depletion study of \citet{j09} (presumably because there are relatively few published detections). From a thermochemical perspective, fluorine is very similar to chlorine. \citep[For example, like Cl$^+$, F reacts exothermically with H$_2$;][]{nw09}. Moreover, the depletions of F and Cl are often assumed to be similar \citep[e.g.,][]{f05}. Thus, in order to compare the depletion behaviors of F and Cl in a more quantitative way, we derived depletion parameters for F using the \citet{j09} methodology. In Figure~\ref{fig:f_depletion}, we plot the gas-phase F abundances from \citet{f05} and \citet{sn07} as a function of the sight line depletion factors from \citet{j09}. (Note that \citet{f05} incorporate a measurement of F~{\sc i} absorption toward $\delta$~Sco into their analysis. This measurement was originally obtained by \citet{sy81} based on Copernicus observations.) A least-squares linear fit yields the following parameters: $A_{\rm F}=-0.304\pm0.288$, $B_{\rm F}=-0.213\pm0.077$, $z_{\rm F}=0.483$, $[{\rm F}/{\rm H}]_0=-0.066\pm0.159$, and $[{\rm F}/{\rm H}]_1=-0.370\pm0.167$. The $\chi^2$ statistic is 13.6 with 7 degrees of freedom, yielding a reduced chi-squared value of 1.9. From these results, we find that the initial depletion of F is consistent with zero (as expected), and the depletion slope for F is very similar to that for Cl. The depletion results for F derived here are compared to those for many other elements in Figure~\ref{fig:depl_tc}.

\begin{thebibliography}{}
\bibitem[Alkhayat et al.(2019)]{a19} Alkhayat, R.~B., Irving, R.~E., Federman, S.~R., Ellis, D.~G., \& Cheng, S.\ 2019, ApJ, 887, 14
\bibitem[Bailer-Jones et al.(2021)]{bj21} Bailer-Jones, C.~A.~L., Rybizki, J., Fouesneau, M., Demleitner, M., \& Andrae, R.\ 2021, AJ, 161, 147
\bibitem[Balashev et al.(2015)]{b15} Balashev, S.~A., Noterdaeme, P., Klimenko, V.~V., et al.\ 2015, A\&A, 575, L8
\bibitem[Blair et al.(2007)]{b07} Blair, W.~P., Ghavamian, P., Long, K.~S., et al.\ 2007, ApJ, 662, 998
\bibitem[Bohlin et al.(1983)]{b83} Bohlin, R.~C., Hill, J.~K., Jenkins, E.~B., et al.\ 1983, ApJS, 51, 277
\bibitem[Bohlin et al.(1978)]{b78} Bohlin, R.~C., Savage, B.~D., \& Drake, J.~F.\ 1978, ApJ, 224, 132
\bibitem[Brown(2015)]{br15} Brown, J.~M.\ 2015, M.S.~Thesis, Univ.~Toledo
\bibitem[Brown et al.(2018)]{b18} Brown, M.~S., Alkhayat, R.~B., Irving, R.~E., et al.\ 2018, ApJ, 868, 42
\bibitem[Brown et al.(2009)]{b09} Brown, M.~S., Federman, S.~R., Irving, R.~E., Cheng, S., \& Curtis, L.~J.\ 2009, \apj, 702, 880
\bibitem[Cartledge et al.(2004)]{c04} Cartledge, S.~I.~B., Lauroesch, J.~T., Meyer, D.~M., \& Sofia, U.~J.\ 2004, ApJ, 613, 1037
\bibitem[Cartledge et al.(2006)]{c06} Cartledge, S.~I.~B., Lauroesch, J.~T., Meyer, D.~M., \& Sofia, U.~J.\ 2006, ApJ, 641, 327
\bibitem[Cartledge et al.(2001)]{c01} Cartledge, S.~I.~B., Meyer, D.~M., Lauroesch, J.~T., \& Sofia, U.~J.\ 2001, ApJ, 562, 394
\bibitem[Dwek(2016)]{d16} Dwek, E.\ 2016, ApJ, 825, 136
\bibitem[Dwek \& Scalo(1980)]{ds80} Dwek, E., \& Scalo, J.~M.\ 1980, ApJ, 239, 193
\bibitem[Federman et al.(2007)]{f07} Federman, S.~R., Brown, M., Torok, S., et al.\ 2007, ApJ, 660, 919
\bibitem[Federman et al.(2005)]{f05} Federman, S.~R., Sheffer, Y., Lambert, D.~L., \& Smith, V.~V.\ 2005, ApJ, 619, 884
\bibitem[Froese Fischer et al.(2006)]{ff06} Froese Fischer, C., Tachiev, G., \& Irimia, A.\ 2006, ADNDT, 92, 607
\bibitem[Gail et al.(2013)]{g13} Gail, H.-P., Wetzel, S., Pucci, A., \& Tamanai, A.\ 2013, A\&A, 555, A119
\bibitem[Gobrecht et al.(2016)]{g16} Gobrecht, D., Cherchneff, I., Sarangi, A., Plane, J.~M.~C., \& Bromley, S.~T.\ 2016, A\&A, 585, A6
\bibitem[Harris \& Bromage(1984)]{hb84} Harris, A.~W., \& Bromage, G.~E.\ 1984, MNRAS, 208, 941
\bibitem[Hibbert(1988)]{h88} Hibbert, A.\ 1988, PhyS, 38, 37
\bibitem[Hobbs(1974)]{h74} Hobbs, L.~M.\ 1974, ApJ, 188, L107
\bibitem[Jenkins(2009)]{j09} Jenkins, E.~B.\ 2009, ApJ, 700, 1299
\bibitem[Jenkins(2019)]{j19} Jenkins, E.~B.\ 2019, ApJ, 872, 55
\bibitem[Jenkins et al.(1986)]{j86} Jenkins, E.~B., Savage, B.~D., \& Spitzer, L.\ 1986, ApJ, 301, 355
\bibitem[Jenkins \& Tripp(2011)]{jt11} Jenkins, E.~B., \& Tripp, T.~M.\ 2011, ApJ, 734, 65
\bibitem[Jensen \& Snow(2007a)]{js07a} Jensen, A.~G., \& Snow, T.~P.\ 2007a, ApJ, 669, 378
\bibitem[Jensen \& Snow(2007b)]{js07b} Jensen, A.~G., \& Snow, T.~P.\ 2007b, ApJ, 669, 401
\bibitem[Jones et al.(2017)]{j17} Jones, A.~P., Kohler, M., Ysard, N., Bocchio, M., \& Verstraete, L.\ 2017, A\&A, 602, A46
\bibitem[Jones et al.(1994)]{j94} Jones, A.~P., Tielens, A.~G.~G.~M., Hollenbach, D.~J., \& McKee, C.~F.\ 1994, ApJ, 433, 797
\bibitem[Jura(1974)]{j74} Jura, M.\ 1974, ApJ, 190, L33
\bibitem[Jura \& York(1978)]{jy78} Jura, M., \& York, D.~G.\ 1978, ApJ, 219, 861
\bibitem[Karovicova et al.(2013)]{k13} Karovicova, I., Wittkowski, M., Ohnaka, K., et al.\ 2013, A\&A, 560, A75
\bibitem[Lebouteiller et al.(2005)]{l05} Lebouteiller, V., Kuassivi, \& Ferlet, R.\ 2005, A\&A, 443, 509
\bibitem[Lodders(2003)]{l03} Lodders, K.\ 2003, ApJ, 591, 1220
\bibitem[Lodders \& Fegley(2023)]{lf23} Lodders, K., \& Fegley, Jr., B.\ 2023, submitted to Geochemistry [arXiv:2301.03674]
\bibitem[Moomey et al.(2012)]{m12} Moomey, D., Federman, S.~R., \& Sheffer, Y.\ 2012, ApJ, 744, 174
\bibitem[Morton(2003)]{m03} Morton, D.~C.\ 2003, ApJS, 149, 205
\bibitem[Neufeld \& Wolfire(2009)]{nw09} Neufeld, D.~A., \& Wolfire, M.~G.\ 2009, ApJ, 706, 1594
\bibitem[Oliver \& Hibbert(2013)]{oh13} Oliver, P., \& Hibbert, A.\ 2013, ADNDT, 99, 459
\bibitem[Paquette et al.(2011)]{p11} Paquette, J.~A., Ferguson, F.~T., \& Nuth, J.~A.\ 2011, ApJ, 732, 62
\bibitem[Press et al.(2007)]{p07} Press, W. H., Teukolsky, S. A., Vetterling, W. T., \& Flannery, B. P.\ 2007, Numerical Recipes, The Art of Scientific Computing (3rd ed.; Cambridge: Cambridge Univ. Press)
\bibitem[Priestley et al.(2021)]{p21} Priestley, F.~D., De Looze, I., \& Barlow, M.~J.\ 2021, MNRAS, 502, 2438
\bibitem[Rachford et al.(2009)]{r09} Rachford, B.~L., Snow, T.~P., Destree, J.~D., et al.\ 2009, ApJS, 180, 125
\bibitem[Rachford et al.(2002)]{r02} Rachford, B.~L., Snow, T.~P., Tumlinson, J., et al.\ 2002, ApJ, 577, 221
\bibitem[Ritchey et al.(2018)]{r18} Ritchey, A.~M., Federman, S.~R., \& Lambert, D.~L.\ 2018, ApJS, 236, 36
\bibitem[Ritchey et al.(2011)]{r11} Ritchey, A.~M., Federman, S.~R., Sheffer, Y., \& Lambert, D.~L.\ 2011, ApJ, 728, 70
\bibitem[Ritchey et al.(2023)]{r23} Ritchey, A.~M., Jenkins, E.~B., Shull, J.~M., et al.\ 2023, submitted to ApJ [arXiv:2301.09743]
\bibitem[Sarangi \& Cherchneff(2015)]{sc15} Sarangi, A., \& Cherchneff, I.\ 2015, A\&A, 575, A95
\bibitem[Savage et al.(1977)]{s77} Savage, B.~D., Bohlin, R.~C., Drake, J.~F., \& Budich, W.\ 1977, ApJ, 216, 291
\bibitem[Savage \& Sembach(1991)]{ss91} Savage, B.~D., \& Sembach, K.~R.\ 1991, ApJ, 379, 245
\bibitem[Schectman et al.(1993)]{s93} Schectman, R.~M., Federman, S.~R., Beideck, D.~J., \& Ellis, D.~J.\ 1993, ApJ, 406, 735
\bibitem[Schectman et al.(2005)]{s05} Schectman, R.~M., Federman, S.~R., Brown, M., et al.\ 2005, ApJ, 621, 1159
\bibitem[Sheffer et al.(2008)]{s08} Sheffer, Y., Rogers, M., Federman, S.~R., et al.\ 2008, ApJ, 687, 1075
\bibitem[Slavin et al.(2015)]{s15} Slavin, J.~D., Dwek, E., \& Jones, A.~P.\ 2015, ApJ, 803, 7
\bibitem[Snow et al.(2007)]{sn07} Snow, T.~P., Destree, J.~D., \& Jensen, A.~G.\ 2007, ApJ, 655, 285
\bibitem[Snow et al.(2008)]{sn08} Snow, T.~P., Destree, J.~D., \& Welty, D.~E.\ 2008, ApJ, 679, 512
\bibitem[Snow \& York(1981)]{sy81} Snow, T.~P., \& York, D.~G.\ 1981, ApJ, 247, L39
\bibitem[Sonnentrucker et al.(2002)]{s02} Sonnentrucker, P., Friedman, S.~D., Welty, D.~E., York, D.~G., \& Snow, T.~P.\ 2002, ApJ, 576, 241
\bibitem[Sonnentrucker et al.(2006)]{s06} Sonnentrucker, P., Friedman, S.~D., \& York, D.~G.\ 2006, ApJ, 650, L115
\bibitem[Sonnentrucker et al.(2007)]{s07} Sonnentrucker, P., Welty, D.~E., Thorburn, J.~A., \& York, D.~G.\ 2007, ApJS, 168, 58
\bibitem[Tayal(2003)]{t03} Tayal, S.~S.\ 2003, ApJS, 146, 459
\bibitem[Toner \& Hibbert(2005)]{th05} Toner, A., \& Hibbert, A.\ 2005, \mnras, 361, 673
\bibitem[Tumlinson et al.(2002)]{t02} Tumlinson, J., Shull, J.~M., Rachford, B.~L., et al.\ 2002, ApJ, 566, 857
\bibitem[Valencic et al.(2004)]{v04} Valencic, L.~A., Clayton, G.~C., \& Gordon, K.~D.\ 2004, ApJ, 616, 912
\bibitem[van Dishoeck \& Black(1988)]{vb88} van Dishoeck, E.~F., \& Black, J.~H.\ 1988, ApJ, 334, 771
\bibitem[Weingartner \& Draine(1999)]{wd99} Weingartner, J.~C., \& Draine, B.~T.\ 1999, ApJ, 517, 292
\bibitem[Welty \& Crowther(2010)]{wc10} Welty, D.~E., \& Crowther, P.~A.\ 2010, MNRAS, 404, 1321
\bibitem[Welty \& Hobbs(2001)]{wh01} Welty, D.~E., \& Hobbs, L.~M.\ 2001, ApJS, 133, 345
\bibitem[Welty et al.(2020)]{w20} Welty, D.~E., Sonnentrucker, P., Snow, T.~P., \& York, D.~G.\ 2020, ApJ, 897, 36
\bibitem[Zhukovska et al.(2016)]{z16} Zhukovska, S., Dobbs, C., Jenkins, E.~B., \& Klessen, R.~S.\ 2016, ApJ, 831, 147
\bibitem[Zhukovska et al.(2018)]{z18} Zhukovska, S., Henning, T., \& Dobbs, C.\ 2018, ApJ, 857, 94
\end{thebibliography}
\end{document}